\documentclass[epj]{svjour}
\usepackage{graphicx}
\usepackage{color}
\usepackage{amssymb,amsmath,amsfonts,graphicx}
\usepackage{bm,color}

\begin{document}

\title{Supernova implosion-explosion in the light of catastrophe theory}
\author{Pierre-Henri Chavanis\inst{1} \and Bruno Denet \inst{2} \and  Martine Le Berre\inst{3}  \and  Yves Pomeau\inst{4} .}

\institute{ 
Laboratoire de Physique Th\'{e}orique (UMR 5152 du CNRS),
Universit\'{e} Paul Sabatier, 118 route de Narbonne, 31062 Toulouse Cedex 4,
France \and
 Universit\'e Aix-Marseille, IRPHE, UMR 7342 CNRS et Centrale Marseille,
Technopole de Ch\^{a}teau-Gombert, 49 rue Joliot-Curie, 13384 Marseille Cedex 13, France.
\and  ISMO-CNRS, Universit\'e Paris-Saclay,  91405 Orsay Cedex, France.
\and Ladhyx, Ecole polytechnique, 91128 Palaiseau, France.
}
\date{\today }

\abstract{
The present understanding   of supernova explosion of massive stars  as a
two-step process, with an initial gravitational collapse toward the center of
the star followed by an expansion of matter after a bouncing on the core, meets
several difficulties. We show that it is not the only possible one: a simple
model based on fluid mechanics, catastrophe theory, and
stability properties of the equilibrium state shows that one can have also a 
\textit{simultaneous} inward/outward motion in the early stage of the
instability of the supernova
 described by a dynamical saddle-center bifurcation. 
The existence of this simultaneous
inward/outward motion is sensitive to the model in such  systems with long-range
interactions. 
If a constant temperature is
assumed 
(canonical ensemble), an overall inward motion occurs, but 
if one imposes with the same equation of state the constraint of
energy conservation (microcanonical ensemble) there is an
 inward velocity field near the center of the star
together with an outward velocity field in the rest of the star. We discuss the
expansion stage of the remnants away from the collapsed
core, and propose a new explanation for the formation of shock waves in the
ejecta which differs from the usual Sedov-Taylor self-similar description.  
}

\maketitle

\section{Introduction}
\label{sec:intro}
In theoretical papers, the sudden death of massive stars is associated to at
least two different processes, depending essentially on their mass (and
secondarily on their composition, rotation speed...). Stars with masses in the
range of $8-40\, M_{\odot}$ die by a supernova phenomenon, which means that they
partially explode. This phenomenon is presently described as an  initial
collapse toward the center of the star, \textit{followed by} a violent expulsion
of the outer layers of the star, leading to the observed supernovae. The death
of
more massive stars, or hypernovae, is believed to be a total collapse of the
star into a black hole, without explosion (or a very faint one) but accompanied
by gamma ray bursts. Although those phenomena are  the most spectacular ones
displayed to us in the Universe, their understanding remains a challenge. Among
the
many unsolved problems, we focus here on core-collapsing supernovae which go with the emission of matter and radiation by explosion. The collapse, which is not  directly observed,
is a phenomenon which has been the subject of many theoretical studies since several decades, including more and more detailed physics, although the observed explosion is still a controversial topic because it requires to explain how to reverse the velocity field of the first stage collapse.
Despite extensive hydrodynamical simulations, the reversal of the motion from inward (collapse or {\it{im}}plosion) to outward (the observed {\it{ex}}plosion of supernovae) is not yet explained because it requires very large outward directed forces to turn the tide. According to most works on core collapse supernovae, this reversal is due to a stiffening of the equation of state at the center,
which stops the collapse and leads to a bounce. An outward propagating shock is
created at this moment but, typically, in numerical studies this shock stalls at
some definite radius except if it is revived by some mechanism (see \cite{Bethe}
and more recently \cite{Burrows} with references herein). Neutrino heating is
often invoked but numerical simulations have shown that this is not generally
sufficient to produce an explosion. More recently, 3D hydrodynamic instabilities
have been discussed but they are still highly controversial. In summary, the
revival of the stalled accretion shock remains an unexplained process since
$1980$ and, as written by Burrows \cite{Burrows}, the understanding of these
phenomena is ``in an unsatisfactory state of affairs" and could remain so until
the state of nuclear matter inside a star can be reproduced on Earth or a close
enough (but not too close!) supernova is observed. 

In the present work we use the same approach as in \cite{epje} (Paper I) where
we definitely do not consider the immensely complex nuclear processes taking
place in a star, but we propose to describe the star as a dynamical system
subject to a loss of stability just before it dies. The starting point of our
theory is the fact that stars die abruptly in a matter of seconds while they
evolve on a very long time, in the billion years range. Similar stability losses
with very different time scales occur in other dynamical systems in nature. The
difference of time scales was recently proposed as a tool for predicting natural
catastrophes before they happen, because it has been shown that one may define
in certain cases a precursor time which stands in between the
very short and
very long time scales \cite{earthquake,creep}.
 It was shown that
this precursor time exists for systems
loosing their stability via a \textit{dynamical} saddle-node bifurcation,
dynamical in the sense that the crossing of the bifurcation results from a slow
sweeping of the bifurcation across a saddle-node, that requires a parameter
changing slowly with time. This time dependence of the parameter can be hidden
into the original equations as in the case of creeping of soft solids and
sleep-wake transitions. The validity of this approach was confirmed
\cite{precursor} by experiments and by mathematical models consisting in coupled
ordinary differential equations (ODE)'s (the nature of the bifurcation in the
case of earthquakes is still an open
question \cite{chua1,chua2}). In these studies, a universal equation was derived
for the order parameter close to the bifurcation, which is first order in time
because these systems are dissipative and reduce to the van der Pol equation in
the relaxation limit. In the case of supernovae we may anticipate that they
likely belong to the class of dynamical catastrophes because of the very
different time scales involved, but we expect that the universal equation
describing the slow-fast transition should be of second order in
time because the system is non-dissipative (we consider compressible inviscid
fluids, at least in the early stage of the dynamics). Therefore, the normal form
should be associated to a saddle-center bifurcation \footnote{A
saddle-center bifurcation occurs when  a center merges with a saddle at the fold
point in Hamiltonian systems, a fairly standard situation as documented in Paper
I.} in place of a saddle-node.
Moreover, we expect to obtain spatial information like density and velocity
profiles at the critical point, in addition to the time evolution of the
amplitude, because our models consist in coupled partial differential
equations (PDE)'s.

To describe the star, we use here, as in Paper I, simple fluid mechanical models
based on the Euler-Poisson equations (with gravity) and a particular equation of
state. We show first that the equilibrium state of such a star may undergo a
saddle-center bifurcation. Then, we study the dynamical solution close to the
critical point in the weakly nonlinear regime where we derive the normal form.
Finally, we describe the strongly nonlinear regime where we show that the
solution displays a self-similar behavior. What differs here from Paper I
concerns the choice of the time dependent parameter. While in Paper I we
considered a fluid with temperature $T(t)$ slowly decreasing with time, here we
consider that the control parameter is the energy $E(t)$. This amounts to going
from a  canonical description (given temperature) to a microcanonical one (given
energy). The interesting result is that this simple change of thermodynamical
ensemble  leads to very different dynamics, as pointed out in previous studies
concerning phase transitions in self-gravitating $N$-body systems (see the
review in \cite{can-microcan}). These studies were launched in view of
applications to astrophysics where galaxies, globular clusters, self-gravitating
dust gas (supposed to be at the origin of planet formation), and fermions gases
(like electrons in white dwarfs, neutrons in neutrons stars, or massive
neutrinos in dark matter models) are examples of self-gravitating systems. Using
tools of thermodynamics and statistical mechanics, it was found that very
different dynamics characterize canonical and microcanonical ensembles,
especially in the vicinity of phase transitions. In general, a single collapsed
core
is formed in the canonical case, whereas a collapsed core surrounded by a halo
is formed in the microcanonical case. Therefore, a question naturally arises:
what should be obtained with the fluid model of Paper I when passing from the
canonical description which leads to a total collapse of the star with a growing
singularity at its core, to the microcanonical one?

Following the same procedure as in Paper I, we show that the microcanonical
Euler-Poisson (MEP) model provides some generic properties that are identical to
those of the canonical Euler-Poisson (CEP) model, but there also exist  very
important differences that drastically change the outcome. We show first that
the loss of equilibrium occurs here via a saddle-center bifurcation, as in Paper
I. In both models, the bifurcating solution reduces to the  Painlev\'{e} I
equation which describes the time dependence of the amplitude of the spatial
mode in the weakly nonlinear regime that we call the Painlev\'{e} regime
although Painlev\'{e} equations were neither derived nor studied in the context
of bifurcation theory before our work to the best of our knowledge. During this
regime, the important difference between the canonical and the microcanonical
models concerns the radial dependence of the neutral mode (in particular the
velocity field) which reflects the loss of  balance between the inward pull of
self-gravity and the outward pull of pressure. This loss of balance (which {\it
a priori} depends on the location in the star) is global. Therefore, if
the spatial profile of the velocity displays different directions, it will
remain so, at least in the early stage of the Painlev\'{e} regime.   In Paper I,
the gravity was found to be dominant everywhere in the star with respect to the
pressure, whereas here the gravity is not dominant everywhere. That gives
different orientations of the radial velocity as a function of the radius. More
precisely, we show that the microcanonical situation turns the all inward-going
velocity field (found in the canonical case, see Paper I) into a velocity
directed inward near the center of the star and outward in the rest of the star.
This shows that in a simple model, fair to study because of the many
uncertainties on what {\it really} happens in supernovae, one somehow gets
rid of the difficulty of reversing an inward collapse  of the star.  Here, the
early stage dynamics does already show a region where an outward going velocity
motion is formed from the very beginning of the supernova process. Moreover, as
soon as matter flows outward, the attraction of the outer shell by the core gets
smaller, and is unable to reverse the outward motion as we numerically observe.

We use, as in Paper I, an equation of state  of the form
$P(r,t)=T(t)g[\rho(r,t)]$, where $P$ is the pressure and $r$ the radial distance
from the center of the star. This equation of state characterizes a barotropic
system and describes  compression and expansion processes including heat
transfer. Such a model is likely unrealistic with respect to the great
complexity of all processes taking place inside a star experiencing supernova
explosion. Nevertheless, we argue that the reality depends on so many
uncontrolled and poorly understood physical phenomena not realizable in
laboratory experiments, and on initial conditions not well defined, that it
seems a better way to try to solve a simple model in a, what we believe,
completely correct way. Moreover, our choice of the function $g[\rho(r,t)]$
gives a finite mass to the star that avoids the box trick encountered in
previous studies \cite{can-microcan}, where the self-gravitating particles are
supposed to be confined within a spherical box, a device proposed by Antonov
\cite{box} for globular clusters. This trick was used  because a stellar system
has the tendency to evaporate under the effect of encounters between stars
\cite{Henon,3corps,YP-pologne,kinaa}, these encounters yielding a huge negative
potential energy which acts as a source for the evaporation of the low energy
stars located in the surrounding halo. The infinite mass problem is also
encountered in the case of purely isothermal stars, and this is why we use a
modified equation of state that is isothermal in the core and polytropic in the
halo (with a polytropic index $n=1$ or $\gamma=2$) so that the density vanishes
at a finite radius.

The paper is organized as follows. In Section \ref{sec:equations}, we present
the MEP model and its equilibrium solutions. We show  that such a
{\it microcanonical} description of a star having a given constant energy $E$
presents a saddle-center bifurcation in its dynamics. In Section
\ref{sec:Painleve}, we show that the  normal form of the MEP model
close to the saddle-center bifurcation takes the form of Painlev\'{e} I equation and we
compare the analytical prediction derived from it with a numerical simulation of
the full MEP model. After the Painlev\'{e} regime, the full numerical study
presented in Section \ref{sec:numerics} displays a self-similar behavior of the
core before the singularity (core collapse) with exponents characterizing the
dominance of gravity over pressure in this region, whereas the outward motion of
the rest of the star continues to accelerate, but with a smaller velocity than
the inward central motion. In Section \ref{sec:post-coll}, we study the
dynamics just after the singularity  where a self-similar solution is given for
both parts of the star,  the core domain which condensates by free fall,  and
the halo supposed to expand freely, these different dynamics being ruled by
the respective importance of gravity and pressure forces. In Section
\ref{sec:free exp},  we consider  the expansion of the remnants supposed to
be well separated from the core. Assuming that the ejecta motion is isentropic,
we show first that no self-similar solution exists when the condition of
conservation of kinetic energy in the halo is imposed, because it makes too many
conditions to be satisfied. We point out that this free expansion stage evolves
naturally toward a non-self-similar solution displaying shocks. This is because
in this regime the velocity field obeys a Burgers-type equation as soon as
gravity and pressure terms become negligible in the Euler equation. This
approach differs fundamentally  from the current description of shocks created
by the collision with the interstellar medium (the Sedov-Taylor regime invoked
in the literature). Here shocks are created by the interactions inside the halo,
not with external matter, as soon as the initial velocity field is maximum
somewhere inside the halo. In Section \ref{sec:can-microcan}, and  also
throughout the paper,  we compare the microcanonical results of the present
study with the canonical results of Paper I, thereby illustrating the notion of
ensembles inequivalence for systems with long-range interactions. Preliminary
results of our study were presented in
Ref. \cite{proceedings}.

\section{Saddle-center bifurcation in the microcanonical
description of a self-gravitating fluid}
\label{sec:equations}

We study the loss of equilibrium of a self-gravitating object (a star) in the
framework of the hydrodynamical Euler-Poisson equations for an inviscid
compressible fluid.

In Paper I, we considered the canonical description: the temperature $T$
of the whole star was assumed to be fixed. This description amounts to
considering the star as a system in contact with a thermostat, its energy $E(t)$
being not fixed. We considered an equation of state $P(\rho)$ presenting a
saddle-center at a critical temperature $T_c$ as the temperature decreases
slowly. At this transition point two equilibrium solutions (one stable, the
other unstable) merge, leading to a loss of equilibrium of the system since no
equilibrium state exists for $T<T_c$. 
We studied the collapse of the star in the weakly nonlinear regime (Painlev\'e regime) near $T = T_c$ and next in the fully nonlinear regime.

Here, we consider the same model but for a closed (isolated, without thermostat)
fluid, namely with fixed energy  $E$, that corresponds to the
\textit{microcanonical} description. In that case, the temperature $T(t)$, which
defines the internal energy $\frac{3}{2}N k_B
T(t) $  in equation (\ref{ae6}), is not a fixed variable. Indeed,  the  fixed
quantity at a given time is the total energy $E$,
including kinetic energy, internal energy and
gravitational energy.
 We consider the same equation of state $P(\rho)$ as in Paper I and show that it presents a saddle-center at a critical energy $E_c$. At that point, two equilibrium solutions (one stable, the other unstable) merge, leading to a loss of equilibrium of the system since no equilibrium state exists for $E<E_c$. Then, we assume that the energy $E(t)$ slowly/adiabatically decreases around the critical value $E_c$ at which the saddle-center bifurcation occurs. The slow decrease of the energy could schematically describe the radiative process of the star burning its matter. We study the collapse of the star in the weakly nonlinear regime (Painlev\'e regime) and in the fully nonlinear regime. We compare the results with those obtained in the canonical ensemble.

\subsection{Description of the microcanonical model}

We use the same notations as in Paper I. The MEP
model presented below differs from the CEP model by an added
equation imposing the conservation of energy in the fluid, although this energy
changes slowly because of added small losses. This constraint modifies the
properties of the equilibrium states with noticeable consequences concerning the
loss of equilibrium of the star.   Some relations  that are needed in our
theoretical study are regrouped in Appendices \ref{app:A} and \ref{app:B}.

\subsubsection{Euler-Poisson equations with conservation of energy}

Let us recall the basic equations written first with the original physical variables.
The Euler-Poisson system is
\begin{eqnarray}
\label{e1}
\frac{\partial\rho}{\partial t}+\nabla\cdot (\rho {\bf u})=0,
\end{eqnarray}
\begin{eqnarray}
\label{e2}
\frac{\partial {\bf u}}{\partial t}+({\bf u}\cdot \nabla){\bf
u}=-\frac{1}{\rho}\nabla P-\nabla\Phi,
\end{eqnarray}
\begin{eqnarray}
\label{e3}
\Delta\Phi=4\pi G\rho,
\end{eqnarray}
where
${\bf u}({\bf r},t)$ is the fluid velocity, $\rho({\bf r},t)$ the mass density, $P({\bf r},t)$ the pressure,  $G$ Newton's constant, and $\Phi({\bf r},t)$ the gravitational potential.
We consider an equation of state of the so-called barotropic
form
\begin{eqnarray}
\label{e4}
P=T(t)g(\rho),
\end{eqnarray}
namely with a uniform temperature. In equation (\ref{e4}),  $P$ and $\rho$
depend on time and space, whereas $T(t)$ is a sort of spatial average of the
temperature which only depends on time.
We assume that the temperature $T(t)$ evolves in time, while remaining
spatially uniform, so as to conserve the total energy (kinetic $+$ thermal $+$
gravitational). We consider a simple energetic constraint of the form
\begin{eqnarray}
E=\frac{1}{2}\int \rho {\bf u}^2\, d{\bf r}+\frac{3}{2}N k_B
T(t)+\frac{1}{2}\int\rho\Phi\, d{\bf r}
\label{ae6}
\end{eqnarray}
which determines the temperature $T(t)$ for a given energy $E$. In doing so, we
are assuming infinite thermal conductivity. This is a rough 
approximation making simpler the theoretical analysis.

\subsubsection{Steady state of the Euler-Poisson equations in physical variables}

It is convenient to introduce the enthalpy per unit mass $h$ defined by $dh=dP/\rho$. For a barotropic equation of state of the form (\ref{e4}), the enthalpy is a function of the density $h(\rho)=\int^{\rho} \lbrack P'(\rho')/{\rho'}\rbrack \, d\rho'$. It is defined up to an additive constant. We impose $h(\rho=0)=0$ which determines the constant. With this choice, the enthalpy vanishes at the edge
of the star. Therefore,
\begin{eqnarray}
\label{he0}
h(\rho)=\int_0^{\rho} \frac{P'(\rho')}{\rho'} \, d\rho'.
\end{eqnarray}
In terms of the enthalpy, the momentum equation can be rewritten as
\begin{eqnarray}
\label{he1}
\frac{\partial {\bf u}}{\partial t}+({\bf u}\cdot \nabla){\bf
u}=-\nabla h-\nabla\Phi.
\end{eqnarray}
The  condition of hydrostatic equilibrium is
\begin{eqnarray}
\nabla h+\nabla\Phi={\bf 0}.
\label{he2}
\end{eqnarray}
Therefore, at equilibrium, $h({\bf r})=-\Phi({\bf r})+C$ where $C$ is a
constant. This is the Gibbs relation. We call $r_0$ the radius of the star at
equilibrium and take $\Phi(+\infty)=0$.  On the boundary of the star, we have
$h(r_0)=0$ and $\Phi(r_0)=-GM/r_0$. Therefore, $C=-GM/r_0$ so that
\begin{eqnarray}
h({\bf r})=-\Phi({\bf r})-\frac{GM}{r_0}.
\label{he2b}
\end{eqnarray}
From equations (\ref{e4}) and (\ref{he0}), we have $\rho=\rho(h,T)$. Taking the
divergence
of equation (\ref{he2}) and using the Poisson equation (\ref{e3}), we obtain the following
differential equation for $h$
\begin{eqnarray}
\Delta h+4\pi G\rho(h,T)=0.
\label{he3}
\end{eqnarray}

\subsubsection{Equation of state : an isothermal core with a polytropic
envelope}

To close the MEP model, we complete equations (\ref{e1})-(\ref{ae6}) by taking
the same equation of state as in Paper I, namely
\begin{eqnarray}
\label{i1}
P(\rho, T)=\rho_* \frac{k_B T}{m}\left (\sqrt{1+\rho/\rho_*}-1\right )^2.
\end{eqnarray}
This equation of state has an isothermal core ($P\sim \rho k_B T/m$ at large
density $\rho\gg \rho_*$) 
and a polytropic halo ($P\sim K  \rho^{\gamma}$ with $K=k_B T/4m\rho_*$ and
$\gamma=2$ at small density $\rho\ll\rho_*$) that confines the system in a
finite region of space. Because of the isothermal core, we infer that the
equation of state (\ref{i1}) should lead to a saddle-center bifurcation
\cite{emden,chandra,aaiso}.
For the equation of state (\ref{i1}), the enthalpy (\ref{he0})  is explicitly given by
\begin{equation}
h(\rho, T)=2 \frac{k_B T}{m} \ln \left ( 1+\sqrt{1+\rho/\rho_*}\right )-2 \frac{k_B T}{m} \ln (2).
\label{i2}
\end{equation}
The inverse relation writes
\begin{equation}
{\rho}(h,T)=4\rho_*\left (e^{m h/k_B T}-e^{m h/2k_B T}\right )
 \mathrm{.}
\label{i3}
\end{equation}

\subsubsection{Dimensionless variables}
\label{sec:tilda}

In the following, it will be convenient to use dimensionless variables. The parameters regarded as fixed are $\rho_*$, $M$, $m$, $k_B$, and $G$. From $\rho_*$ and $M$ we can construct a length $L=(M/\rho_*)^{1/3}$. Then, we introduce the dimensionless quantities
\begin{equation}
{\tilde\rho}=\frac{\rho}{\rho_*},\quad {\tilde r}=\frac{r}{L}, 
\quad{\tilde\Phi}=\frac{\Phi}{G\rho_* L^2}, \quad {\tilde {\bf u}}=\frac{{\bf
u}}{L\sqrt{G\rho_*}},
\label{eq:sc1b}
\end{equation}
\begin{equation}
{\tilde T}=\frac{k_B T}{m G\rho_* L^2},\, {\tilde E}=\frac{E}{\rho_*^2
GL^5},\, {\tilde P}=\frac{P}{GL^2\rho_*^2},\, {\tilde t}=t \sqrt{G\rho_*}.
\label{eq:sc2b}
\end{equation}
Working with the dimensionless variables with tildes amounts to taking
$G=\rho_*=M=m=k_B=1$ in the initial equations, a choice that we shall make in
the following.

\subsection{Equilibrium solutions, energy-radius relation, and caloric curve}
\label{sec:equil}

For a given value of the energy (and therefore of the temperature) the steady
state (equilibrium) is given by equations (\ref{he3}) and (\ref{i3}). The
solutions may be expressed in terms of a second set of scaled variables,
$\hat{r}={r}/{T^{1/2}}$, $\hat{\rho}=\rho$, $\hat{\Phi} = {\Phi}/{T}$, $\hat{h}=
{h}/{T}$, ${\hat p}={p}/{T}$, ${\hat{M}}=M/{T^{3/2}}$ that leads to the
following ODE for the steady state enthalpy
\begin{equation}
 \hat{h}_{,\hat{r}^2}  + \frac{2}{\hat{r}} \hat{h}_{,\hat{r}} + 4 \pi  \hat{\rho}(\hat{h}) = 0
\label{es1}
\end{equation}
with the density-enthalpy relation
\begin{equation}
\hat{\rho}(\hat{h}) =4(e^{\hat{h}}-e^{\hat{h}/2}).
\label{es1b}
\end{equation}
As in Paper I, this equation is solved for a given value of  $\hat{h}(0)=
\hat{h}_0$, the only free parameter, with initial conditions
$\hat{h}_{,\hat{r}}(0)= \hat{h}_{,\hat{r}^3}(0)=0$ and
$\hat{h}_{,\hat{r}^2}(0)=-({2\pi}/{3}) \hat{\rho}(\hat{h}_0)$. The scaled radius
of the star $\hat{r}_0$ corresponds to the smallest root of $
\hat{h}(\hat{r_0})=0$.  The scaled mass  of the star ${\hat M}=\int_0^{{\hat
r}_0} \rho({\hat r},t) 4\pi {{\hat r}}^2\, d{\hat r}$ is related to the
temperature by the relation $\hat{M}  = {T}^{-3/2}$ (we recall that the mass of
the star is $1$ with the units defined in Section \ref{sec:tilda}). Using
$\hat{r}_0=r_0/{T^{1/2}}$ and the relation $\hat{M}(\hat{r}) = -\hat{r}^2
\hat{h}_{,\hat{r}}$ (see Appendix \ref{sec_gauss}) yielding ${\hat h}_{,{\hat
r}}(\hat{r}_0)=-{1}/(\sqrt{T}{r_0^2})$, we obtain
\begin{equation}
\label{es2}
r_0=\left \lbrack \frac{\hat{r}_0}{-\hat{h}_{,\hat{r}}(\hat{r}_0)}\right \rbrack^{1/3},
\qquad
 T=\frac{1}{\left \lbrack -\hat{r}_0^2 \hat{h}_{,\hat{r}}(\hat{r}_0)\right \rbrack^{2/3}}.
\end{equation}
Alternatively, one can compute ${\hat M}=\int_0^{{\hat r}_0} \rho({\hat r},t)
4\pi {{\hat r}}^2\, d{\hat r}$ and determine the temperature by $T={\hat
M}^{-2/3}$ and the radius by $r_0=T^{1/2}\hat{r}_0$. On the other hand, using
equations (\ref{he10}) and (\ref{i1}), the energy writes
\begin{equation}
\label{es3}
E=\frac{3}{2}T -3 T^{{5}/{2}}\int_{0}^{\hat{r}_0} \left( \sqrt{1+{\hat \rho}}-1 \right )^2 4 \pi {\hat r}^2 d{\hat r}.
\end{equation}
We can also compute the energy at equilibrium from the relation (see Appendix
\ref{sec_pot}):
\begin{equation}
\label{es4}
E=\frac{3}{2}T -\frac{1}{2r_0}+\frac{1}{2} T^{{5}/{2}}\int_{0}^{\hat{r}_0} {\hat
h}_{,{\hat r}}{\hat M} d{\hat r},
\end{equation}
where we have used equations (\ref{ae6}), (\ref{he5b}), an integration by parts,
 and ${\hat\Phi}= - {\hat h} + C$ with $C= - {1}/({T^{3/2} \hat{r}_0})=
-{\hat{M}}/{\hat{r}_0}$.

\begin{figure}
\centerline{
\includegraphics[height=1.2in]{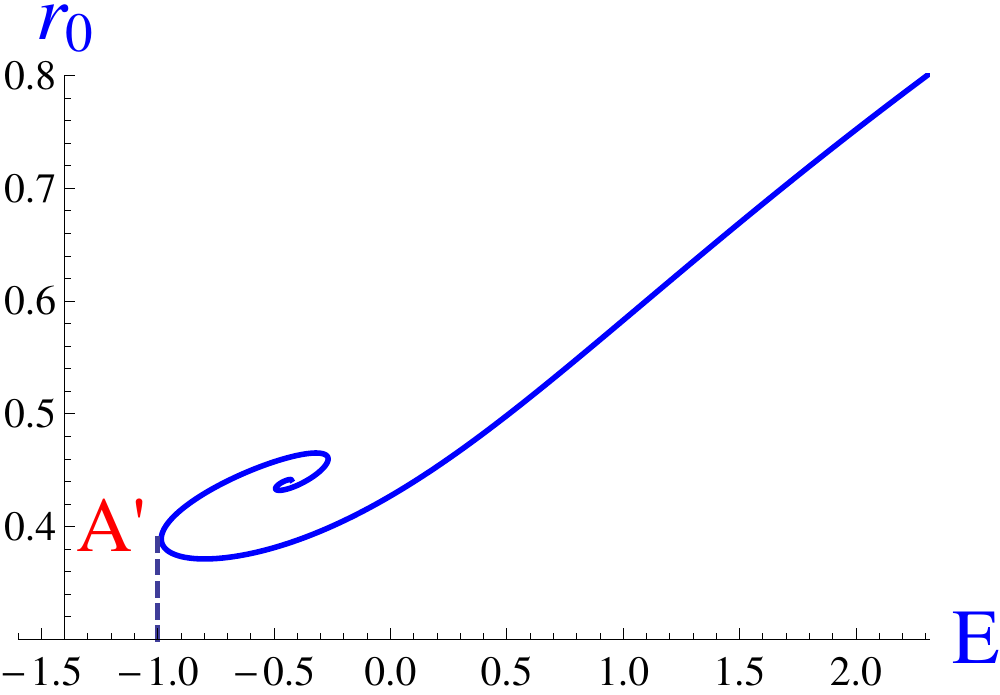}
\includegraphics[height=1.2in]{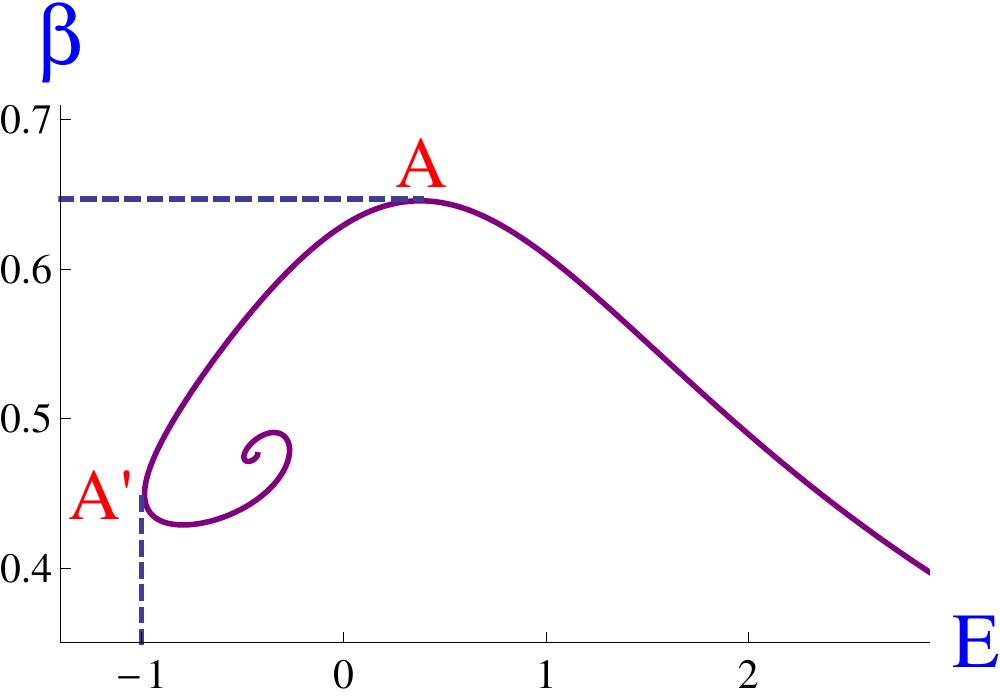}}
\caption{
Series of equilibria:
Left: Radius versus energy $r_0(E)$; Right: inverse temperature versus energy
 $\beta(E)$. The curves are obtained by increasing the parameter $\hat{h}_0$
in the range $[0.5; 20]$ (smaller values of $\hat{h}_0$ are not represented as
they correspond to the right parts of the curves which evolve monotonically).
 }
\label{spi-microcan1}
\end{figure}

Varying $\hat{h}_0$ from $0$ to $+\infty$ allows us to draw spiralling curves,
such as $r_0(E)$ or $\beta(E)$, depicting the steady states. These curves
(series of
equilibria) are
drawn in Figure \ref{spi-microcan1}. At high
energies (and high temperatures), the star is stable since it reduces to a pure
polytrope of index $\gamma=2$ larger than $\gamma_c=4/3$ \cite{chandra}. Using
Poincar\'e's
bifurcation theory \cite{poincare} (see \cite{katz,can-microcan} for an
application of this theory in the case of self-gravitating systems), one can
show that the series of equilibria remains stable in the microcanonical ensemble
until the first turning point of energy (corresponding to $0 \le  \hat{h}_0 <
\hat{h}_0^{(c)}$), and that it becomes unstable afterward (corresponding to
$\hat{h}_0 > \hat{h}_0^{(c)}$). More precisely, a new mode of stability is lost
at each turning point of energy. We focus here on the critical point A' where
the first instability occurs as $E$ decreases ($\hat{h}_0$ increases). It
corresponds to a minimum of the energy, characterized by the following parameter
values
$\hat{h}_0^{(c)}= 6.50655$, $\hat{r}_0^{(c)}= 0.26074$,
$\hat{\rho}_{0}^{(c)}=2574$,   $\hat{M}^{(c)}=0.3012$, $\hat{P}_0^{(c)}=2475$
or in non-hat scalings
$ \rho_{0}^{(c)}=2574$, $r_0^{(c)}= 0.38897$, $P_0^{(c)}= 5508$,  $E_{c}=
-0.984142$
with $T_c=\lbrack \hat{M}^{(c)}\rbrack^{-2/3} = 2.22538$.

\begin{figure}
\centerline{
(a)\includegraphics[height=1.2in]{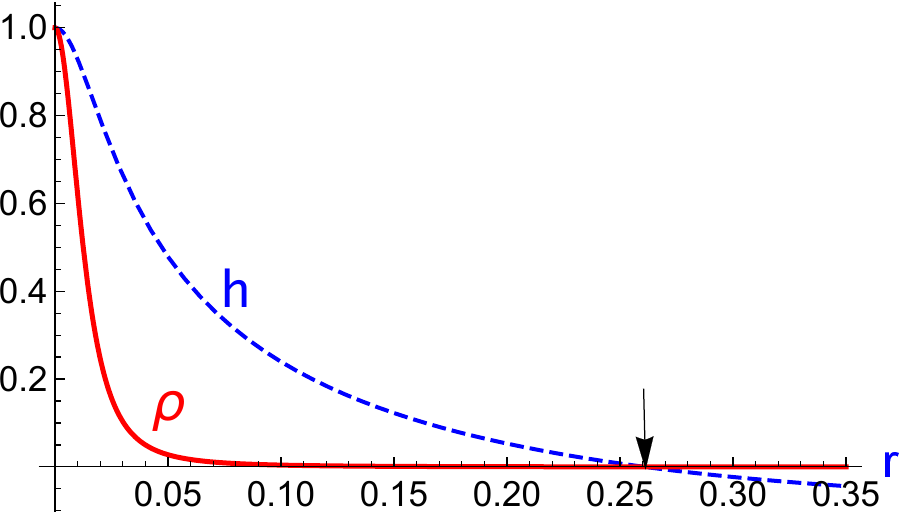}}
\centerline{
(b) \includegraphics[height=1.0in]{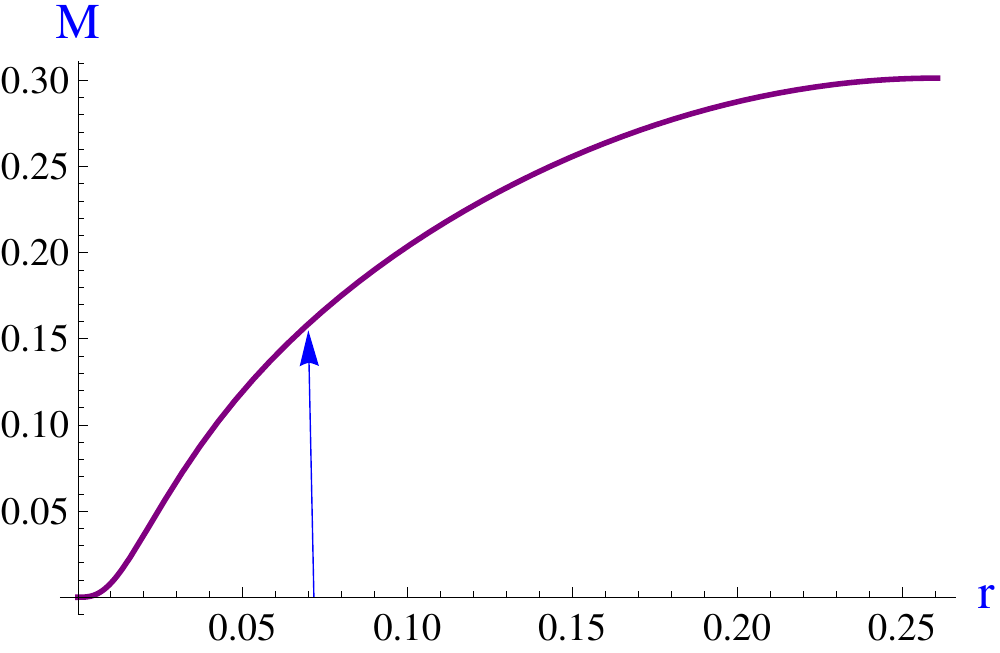}
(c)\includegraphics[height=1.0in]{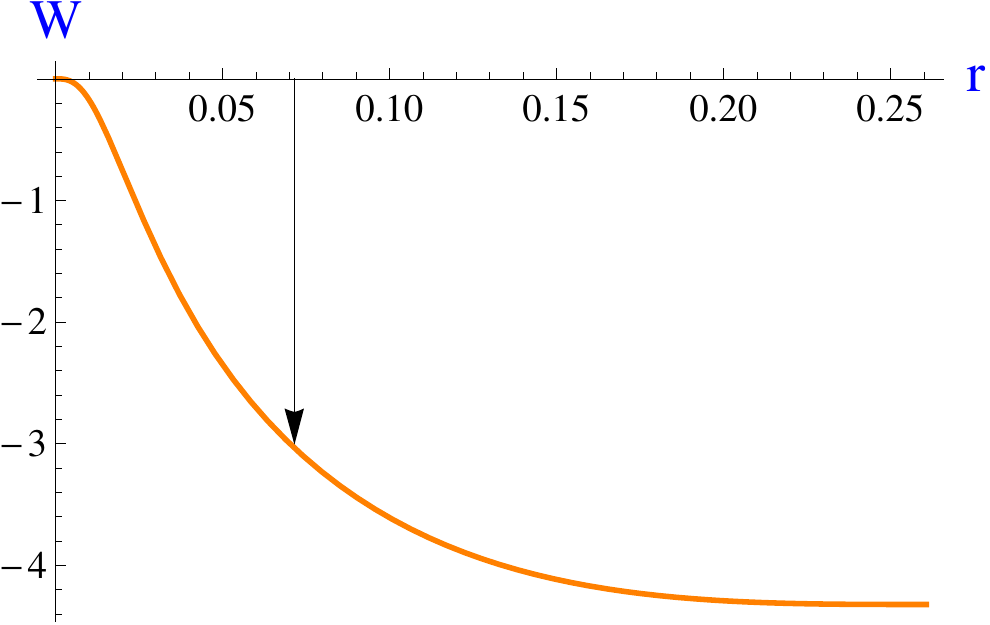}}
\caption{(a)
Density and
enthalpy
profiles $\hat{\rho}(\hat{r})/\hat{\rho}(0)$, $\hat{h}(\hat{r})/\hat{h}(0)$; (b)
mass $\hat{M}(\hat{r})$ and (c) gravitational energy
$W(\hat{r})$  at the microcanonical critical point A' (saddle-center). The
arrow in (a) indicates the radius where the density and the enthalpy vanish,
whereas the arrows
in (b) and (c) display the radius where the velocity sign changes in the
dynamical regime, see Subsection \ref{sec:in-out}.}
\label{fig:equil}
\end{figure}

For these parameter values, the density profile is drawn in Figure
\ref{fig:equil}-(a) where the arrow indicates the equilibrium radius (the radial
distance where the solution crosses zero). Beyond this radius we set $\rho(r)=0$
, as in Paper I, whereas it leads to functions $\hat{\rho}(r)$, $\hat{M}(r)$,
and $\hat{h}(r)$ with discontinuous slopes.

The mass enclosed inside a sphere of radius $\hat{r}$, $ \hat{M}(\hat{r})=\int_0^{\hat{r}}\rho(r') 4\pi {r'}^2\, dr'$, and the gravitational energy (also inside a sphere of radius $\hat{r}$) $W({\hat r})= - 3  T^{5/2} \int_0^{\hat r} (\sqrt{1+{\hat \rho}}-1)^2 4\pi {r'}^2\, dr'$,
 are shown in Figures \ref{fig:equil}-(b) and \ref{fig:equil}-(c) respectively
where the arrows display the radius at which the fluid motion is expected to separate (at critical) between opposite directions (inward and outward) at the saddle-center, as explained below.

\subsection{Saddle-center bifurcation in the microcanonical ensemble and linear stability analysis}
\label{sec:ordre1}

We have seen in the previous section that the equation of hydrostatic
equilibrium can have several solutions with the same energy $E$, but only one
is stable. Close to A', two solutions (one stable and one unstable) merge. This
defines a saddle-center bifurcation. Here, we investigate the structure of the
critical mode.

\subsubsection{Linearized Euler-Poisson system}

The Euler-Poisson set of equations can be rewritten as
\begin{eqnarray}
\label{l1}
\frac{\partial\rho}{\partial t}+\nabla\cdot (\rho {\bf u})=0,
\end{eqnarray}
\begin{equation}
\frac{\partial}{\partial t}(\rho {\bf u})+\nabla \cdot (\rho {\bf u}\otimes {\bf
u})=-\rho\nabla h-\rho\nabla\Phi,
\label{l2}
\end{equation}
\begin{eqnarray}
\label{l3}
\Delta\Phi=4\pi\rho.
\end{eqnarray}
To determine the dynamical stability of a steady state of the Euler-Poisson system (\ref{l1})-(\ref{l3}), we consider a small perturbation about that state and write $f({\bf r},t)=f({\bf r})+\delta f({\bf r},t)$ for $f=(\rho,{\bf u},\Phi)$ with $\delta f({\bf r},t)\ll f({\bf r})$. The linearized Euler-Poisson system writes
\begin{eqnarray}
\label{l4}
\frac{\partial\delta\rho}{\partial t}+\nabla\cdot (\rho \delta{\bf u})=0,
\end{eqnarray}
\begin{equation}
\frac{\partial}{\partial t}(\rho \delta{\bf u})=-\rho\nabla \delta h-\rho\nabla\delta\Phi,
\label{l5}
\end{equation}
\begin{eqnarray}
\label{l6}
\Delta\delta \Phi=4\pi \delta\rho.
\end{eqnarray}
These equations can be combined into a differential equation of the form
\begin{equation}
\frac{\partial^2\delta\rho}{\partial t^2}=\nabla\cdot \left\lbrack \rho(\nabla \delta h+\nabla\delta\Phi)\right\rbrack.
\label{l7}
\end{equation}
Writing the time dependence of the  perturbations as $\delta f({\bf r},t)\propto
e^{\lambda t}$, we obtain the eigenvalue equation
\begin{equation}
\lambda^2\delta\rho=\nabla\cdot \left\lbrack \rho(\nabla\delta h+\nabla\delta\Phi)\right \rbrack,
\label{l8}
\end{equation}
which has to be solved in conjunction with the Poisson equation (\ref{l6}).

\subsubsection{The point of marginal stability}
\label{sec_marg}

We shall now investigate the behavior of the perturbations at the critical
point.
Our aim is to derive the radial profile of the marginal mode which results from
the merging of the stable and unstable equilibrium states. 
 In the case
of the CEP model this amounts to solving an ODE with proper initial conditions
(see Paper I), whereas the MEP model leads to an integro-differential equation,
equations (\ref{marg15})-(\ref{marg17}), that we may solve iteratively by
changing
one of the initial conditions, as explained below.
The neutral mode ($\lambda=0$) which signals the change of stability of the series of equilibria is the solution of the differential equation
\begin{equation}
\nabla\delta h^{(c)}+\nabla\delta\Phi^{(c)}={\bf 0}.
\label{marg1}
\end{equation}
Therefore, at the critical point, we have
\begin{equation}
\delta h^{(c)}({\bf r})=-\delta\Phi^{(c)}({\bf r}).
\label{marg2}
\end{equation}
The constant of integration has been set equal to zero by assuming that the radius does not change at first order (see below). From this relation, and using Newton's law (\ref{he4}) in perturbed form,  we obtain
\begin{equation}
\delta h^{(c)}_{,r}=-\delta \Phi^{(c)}_{,r}=\frac{\delta M^{(c)}(r)}{r^2}.
\label{marg3}
\end{equation}
Taking the divergence of equation (\ref{marg1}) and using Poisson's equation
(\ref{l6}), we get
\begin{equation}
\Delta\delta h^{(c)}+4\pi\delta \rho^{(c)}=0.
\label{marg4}
\end{equation}
The enthalpy  $h$ and  the density $\rho$ are linked by the relation
\begin{equation}
\rho(h,T) =4\left (e^{\frac{h}{T}}-e^{\frac{h}{2T}}\right ).
\label{marg5}
\end{equation}
The first order density deviation is given by
\begin{equation}
\delta\rho= \rho_{,h}\delta h+\rho_{,T}\delta T,
\label{marg6}
\end{equation}
where 
\begin{equation}
 \rho_{,h}= \frac{4}{T}\left (e^{h/T}-\frac{1}{2}e^{h/2T}\right )
 \label{marg7}
\end{equation}
and
\begin{equation}
   \rho_{,T}= -\frac{4h}{T^2}\left (e^{h/T}-\frac{1}{2}e^{h/2T}\right )
    \label{marg8}
\end{equation}
stand for the partial derivatives of $\rho(h,T)$
with respect to $h$ and $T$, respectively. We have
\begin{equation}
   \rho_{,T}= -\frac{h}{T}\rho_{,h}.
    \label{marg9}
\end{equation}
We also note that
\begin{equation}
\delta\rho(r_0)= \frac{2}{T}\delta h(r_0)
\label{marg10}
\end{equation}
since $h(r_0)=0$.

Substituting these relations into equation (\ref{marg4}), we obtain
\begin{equation}
\Delta\delta h^{(c)}+4\pi \rho_{,h}^{(c)}\delta h^{(c)}-4\pi \frac{h^{(c)}}{T_c}\rho^{(c)}_{,h}\delta T^{(c)}=0.
\label{marg11}
\end{equation}
On the other hand, the energetic constraint (\ref{ae6}) writes at first order
\begin{eqnarray}
\label{marg12}
0=\frac{3}{2} \delta T+\int\Phi\delta\rho\, d{\bf r}.
\end{eqnarray}
Substituting equations (\ref{marg6})-(\ref{marg9}) into equation (\ref{marg12}),
we
get
\begin{eqnarray}
\label{marg13}
\delta T=-\frac{\int \Phi \rho_{,h} \delta h\, d{\bf r}}{\frac{3}{2}-\frac{1}{T}\int \Phi h\rho_{,h}\, d{\bf r}}.
\end{eqnarray}
Finally, combining equations (\ref{marg11}) and (\ref{marg13}), we obtain the
integral equation
\begin{equation}
\Delta\delta h^{(c)}+4\pi \rho^{(c)}_{,h}\delta h^{(c)}+
\frac{h^{(c)}}{T_c}\rho^{(c)}_{,h}\frac{4\pi\int \Phi^{(c)} \rho^{(c)}_{,h}
\delta h^{(c)}\, d{\bf r}}{\frac{3}{2}-\frac{1}{T_c}\int \Phi^{(c)}
h^{(c)}\rho^{(c)}_{,h}\, d{\bf r}}=0,
\label{marg14}
\end{equation}
which is the MEP version of the ordinary differential equation (25) of Paper I  obtained for the CEP model free of the energetic constraint.

\begin{figure}
\centerline{
 \includegraphics[height=1.7in]{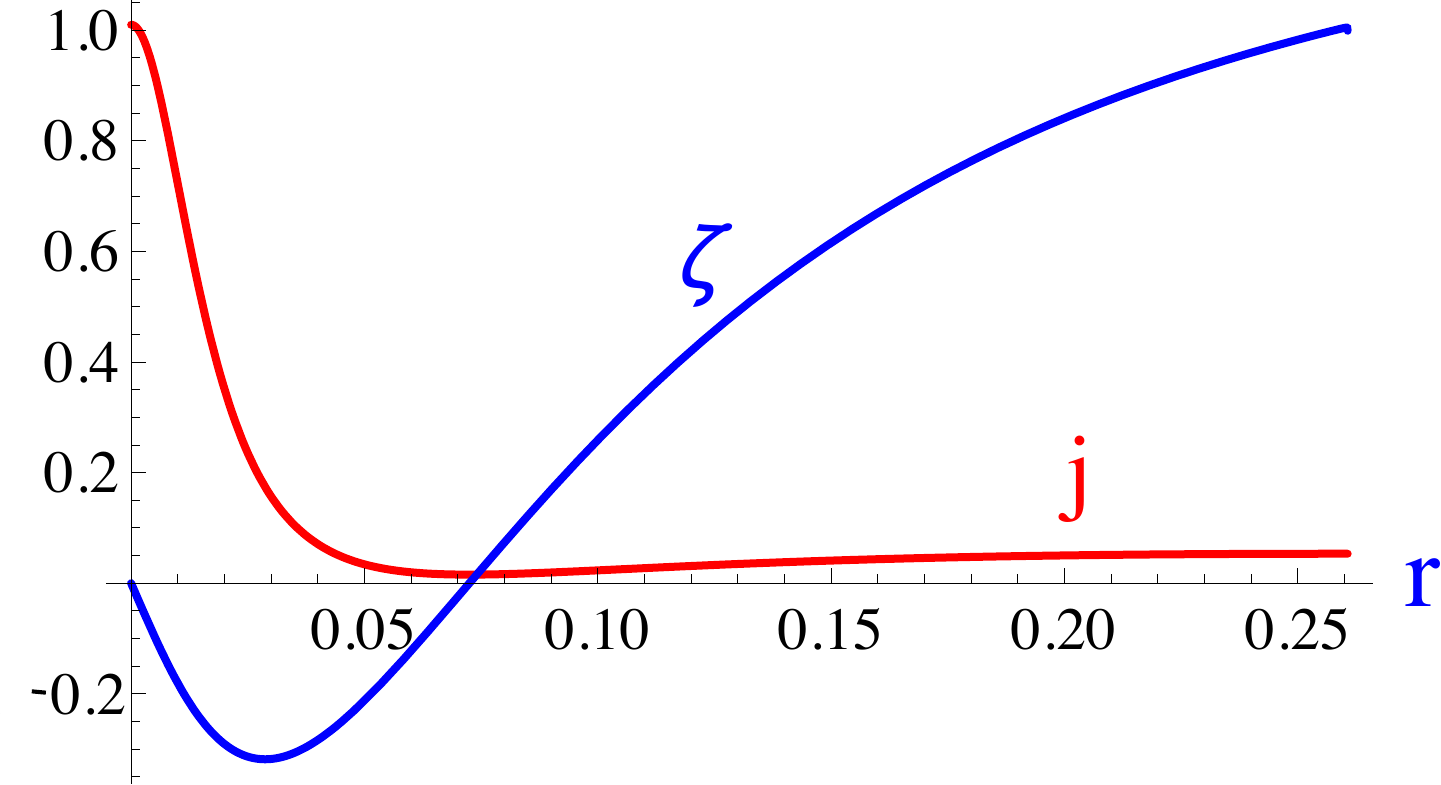}
}
\caption{Neutral mode $j(r)$  (red curve), positive everywhere, and  function
$\zeta(r)$ (blue curve) defined in Section \ref{sec_nfs}, solution of equation
(\ref{n35}). 
}
\label{Figjzeta}
\end{figure}

We now introduce the scaled variables of Section \ref{sec:equil}. Furthermore,
we note $\hat{j}=\delta{\hat h}^{(c)}$ and $\hat{r}_c=\hat{r}_0^{(c)}$. From now
on, we remove the ``hats'' to simplify the expressions. The integral equation
(\ref{marg14}) becomes
\begin{equation}
\Delta j+4\pi \rho^{(c)}_{,h}\, j 
-4\pi h^{(c)} \rho^{(c)}_{,h}\int_0^{r_c} j(r) f(r)\, dr=0,
\label{marg15}
\end{equation}
where
\begin{equation}
f(r)=-\frac{4\pi}{D} \rho_{,h}^{(c)}\, \Phi^{(c)}\, r^2
\label{marg16}
\end{equation}
and
\begin{equation}
D = \, \frac{3}{2T_c^{3/2}} - \int_0^{r_c}  \Phi^{(c)} (r) h^{(c)}(r)\rho^{(c)}_{,h} \, 4\pi r^2d{r}.
\label{marg17}
\end{equation}
This equation has to be solved with the boundary condition $j_{,r}(r_c)=0$  (see
Appendix \ref{app:B})  plus another condition, for example the value of 
$j(0)$. Because of the linearity of the
integro-differential equation with respect to $j(r)$ we can set $\int_0^{r_c}
j(r) f(r)\, d{ r}=1$, and vary $j(0)$ until the resulting solution satisfies
this relation. We find that this occurs for 
$j(0)=20.3$.
The solution $j(r)$ is drawn in Figure \ref{Figjzeta}, red curve. Finally, we
note that the perturbed temperature at the critical point is given by
\begin{equation}
{\cal T}\equiv \frac{\delta T^{(c)}}{T_c}=\int_0^{r_c} j(r)f(r)\, dr=1.
\label{marg18}
\end{equation}

\subsubsection{The inward/outward motion}
\label{sec:in-out}

\begin{figure}[htbp]
\centerline{
(a) \includegraphics[height=1.7in]{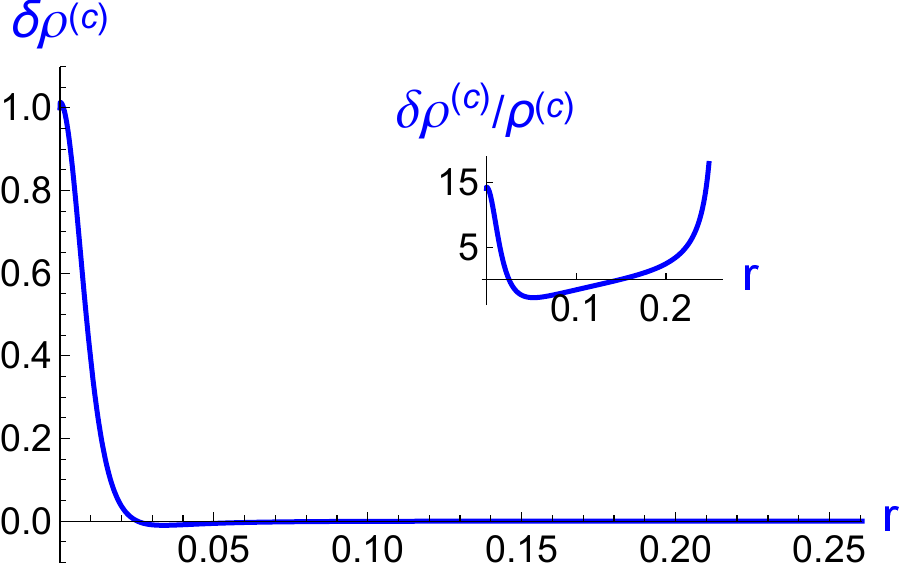}}
\centerline{
(b)\includegraphics[height=1.7in]{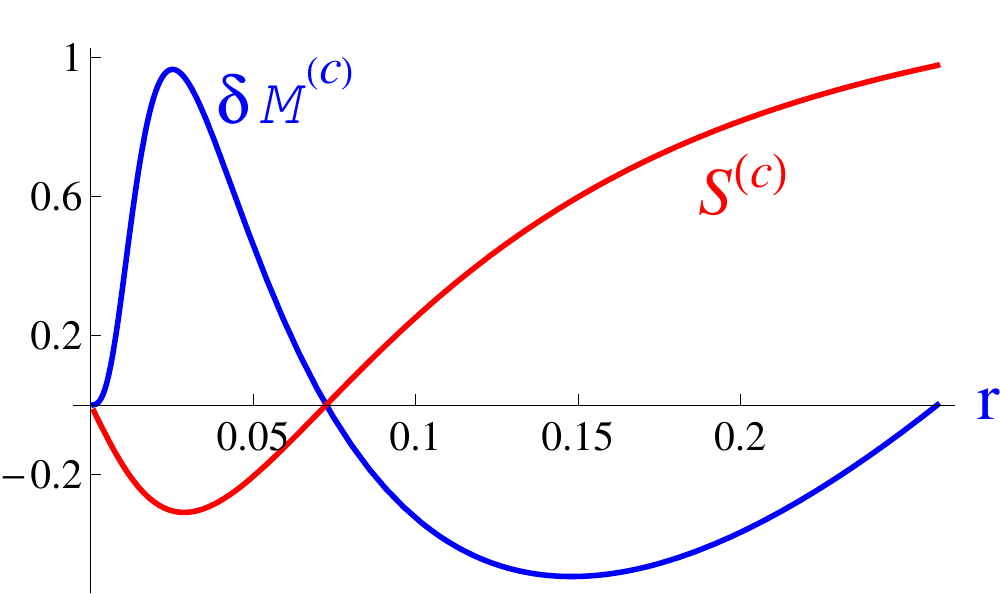}
}
\caption{
Radial profiles of
the first order deviations at the microcanonical critical point: (a) density $
\delta \rho^{(c)}(r)$  (in the insert, the ratio $\delta \rho^{(c)}/ \rho^{(c)}$
displays the two nodes behavior of the density deviation); (b) mass $\delta
M^{(c)}(r)$ and velocity (or displacement) $S^{(c)}(r)$.
}
\label{FigrhM}
\end{figure}

The radial profiles of mass, velocity and density deviations write as $\delta
M^{(c)}(r)=-r^2 j_{,r}(r)$, $S^{(c)}(r)={j_{,r}(r)}/{4\pi \rho^{(c)}(r)}$, and
$\delta\rho^{(c)}(r)={\delta M^{(c)}_{,r}}/{4 \pi r^2}$,  respectively (see
Appendix B). The solution of equation (\ref{marg15}) leads to the radial
profiles shown in Figure \ref{FigrhM}. In Figure \ref{FigrhM}-(a), the insert is
an
enlargement of the lower part of the density deviation normalized by
$\rho^{(c)}(r)$. It displays a small
intermediate region of negative amplitude where the density of mass slightly
decreases with respect to its equilibrium value. Figure \ref{FigrhM}-(b) clearly
displays a simultaneous inward and outward motion of the fluid, the
sign of the velocity profile $S^{(c)}$, or of the mass $\delta M^{(c)}$,
changing at a radius about $28\%$ of the star radius. When this value
is reported on the  curves of Figures \ref{fig:equil}-(b) and
\ref{fig:equil}-(c)
giving the mass $M(r)$ and the gravitational energy $W(r)$, they show that about
$50\%$  of the mass is expected to be expelled at the beginning of the supernova
process, whereas the other half of the total mass begins to move inward.
Concerning the gravitational energy, about $3/4$ of it is concentrated in the
inward-directed core, as indicated by the arrow in Figure \ref{fig:equil}-(c).

The inward/outward motion is illustrated by the radial displacement $S^{(c)}(r)$
shown in  Figure \ref{FigrhM}-(b), red curve. We have to point out that, for the
moment, the sign of the neutral mode is arbitrary since equation (\ref{marg15})
defining the neutral mode profile $j(r)$ is linear. The sign of the radial
profiles $S^{(c)}(r)$, $\delta M^{(c)}$ and $ \delta \rho^{(c)}(r)$ represented
in Figure \ref{FigrhM} is actually derived from higher order terms of the weakly
nonlinear analysis developed in the next Section. Figure \ref{FigrhM}-(b) shows
that any particle located initially in the inner part of the star, $r< 0.073$,
where the density is large (see Figure \ref{fig:equil}-(a)) should move inward, 
whereas any particle located in the outer shell should  move outward.
In the insert of Figure \ref{FigrhM}-(a) we show the ratio $\delta \rho^{(c)}/
\rho$ which displays the two nodes behavior of the density deviation. It
illustrates the formation of a mass close to the center plus a halo further
away, and a decrease of  density in between. In summary, we expect the
formation of a sort of  explosive halo together with a collapse of the inner
part (core) from the first order variations at criticality. This
important point
has to be confirmed by the higher order terms of the weakly nonlinear analysis,
as done in next Section.

\section{Dynamics close to the saddle-center bifurcation:
derivation of the Painlev\'e I equation}
\label{sec:Painleve}

In this Section, we focus on the first stage of the motion, when the system
approaches the critical point A'  in Figure \ref{spi-microcan1} by decreasing
the
energy $E(t)$, following the stable portion of the series of equilibria. Note
that here and in Paper I we call this weakly nonlinear stage ``the Painlev\'{e}
regime", and we call the analysis of the saddle-center bifurcation ``the
Painlev\'{e} analysis", whereas in the work of Painlev\'{e} no connection is
made with bifurcation theory\footnote{Painlev\'{e} found these equations when
searching
 solutions having peculiar properties related to the position of their complex
singularities}. 
In this Painlev\'{e} regime,
because the velocity field has a small amplitude at the beginning of the motion,
we assume that the advection term can be neglected in the Euler equation, an
hypothesis that is justified during a time interval $t_0$ by using the same
arguments as in Section 4.1 of Paper I (the time interval $t_0$
can be defined
in terms of the coefficients appearing in the normal form of the Euler equations
close to the saddle-center, i.e., the Painlev\'e I equation). In this Section,
we use the same procedure and notations as in Section $4$ of Paper I, but this
Section is self-contained.

\subsection{Simplification of the hydrodynamic equations }

Neglecting the advection term in the Euler equation (\ref{l2}) we obtain
\begin{equation}
\frac{\partial}{\partial t}(\rho {\bf u})=-\nabla P-\rho\nabla\Phi.
\label{sim2}
\end{equation}
This equation can be combined with the equation of continuity (\ref{l1}) into a single equation for the density
\begin{equation}
\frac{\partial^2\rho}{\partial t^2}=\nabla\cdot (\nabla P+\rho\nabla\Phi),
\label{sim3}
\end{equation}
where $\Phi$ is given by the Poisson equation (\ref{l3}). The energetic constraint writes
\begin{eqnarray}
\label{sim5a}
E=\frac{3}{2}
T(t)+\frac{1}{2}\int \rho \Phi \, d{\bf r}.
\end{eqnarray}
These equations are  valid during a time interval of order
$t_0$ before the collapse time (see Paper I).

\subsection{The equation for the mass profile}

For a spherically symmetric evolution, using Newton's law (\ref{he4}), we obtain the following partial differential equation for the integrated density
\begin{equation}
\frac{\partial^2 M(r,t)}{\partial t^2}= 4 \pi r^2 P_{,r} +  \frac{1}{r^2}M_{,r}  M.
\label{sim4}
\end{equation}
The energetic constraint writes
\begin{eqnarray}
\label{sim5}
E=\frac{3}{2}
T(t)+\frac{1}{2}\int_0^{R(t)} \Phi M_{,r}\, dr.
\end{eqnarray}
In equation (\ref{sim4}), the term $P_{,r}=P_{,\rho}(\rho) \rho_{,r}$ has to be
expressed as a function of $\rho(r,t)=M_{,r}/(4 \pi r^2)$
 and $\rho_{,r}(r,t)=(M_{,r^2}-2M_{,r}/r)/(4 \pi r^2)$.
 For the equation of state
\begin{eqnarray}
P(\rho)= T(t)\left(\sqrt{1+\rho}-1\right )^2,
\label{mp1}
\end{eqnarray}
we get
\begin{equation}
P_{,\rho}(\rho)=  T(t)\left (1-\frac{1}{\sqrt{1+{\rho}}}\right ).
\label{mp2}
\end{equation}
Introducing this expression into equation (\ref{sim4}), the dynamical equation for $M(r,t)$ writes
\begin{equation}
\frac{\partial^2 M(r,t)}{\partial t^2}=  T(t)\mathcal{L}(M)g(M_{,r}) + \frac{1}{r^2}M_{,r}  M
\label{mp3}
\end{equation}
with
\begin{equation}
\left \{ \begin{array}{l}
\mathcal{L}(M)= M_{,r^2}-\frac{2}{r}M_{,r}\\
  g(M_{,r})=1-\frac{1}{\sqrt{1+\frac{1}{4 \pi r^2}M_{,r}}}
  \mathrm{.}
\end{array}
\right.
\label{mp4}
\end{equation}
The boundary conditions to be satisfied  are
\begin{equation}
 \left \{ \begin{array}{l}
M(0,t)=0  \\
M(R(t),t)= 1 = 4 \pi\int_0 ^{R(t)} {\mathrm{d}} r' r'^2 \rho(r',t)
\mathrm{.}
\end{array}
\right. \label{mp5}
\end{equation}
In the latter relation,  the radius of the star $R(t)$ depends on time. However,
this dependence will be  neglected in this Painlev\'{e} analysis because it can be shown that it plays no role up to order two (with respect to the small parameter $\epsilon$ which characterizes the slow time dependence of $E$), the order considered below. Therefore, we take $R(t)\simeq r_0$.

\subsection{Equilibrium state and neutral mode for the mass profile}

The steady solution of equation (\ref{mp3}) is determined by the partial
differential equation
\begin{equation}
T\mathcal{L}(M)g(M_{,r}) + \frac{1}{r^2}M_{,r}  M=0
\label{esn1}
\end{equation}
with the energetic constraint
\begin{eqnarray}
\label{esn2}
E=\frac{3}{2}T+\frac{1}{2}\int_0^{r_0}\Phi M_{,r}\, d{r}.
\end{eqnarray}
We now consider a small perturbation about a steady state and write
$M(r,t)=M(r)+\delta M(r,t)$ with $\delta M(r,t)\ll M(r)$. Linearizing equation
(\ref{mp3}) about this steady state and writing the time dependence
of the perturbation as $\delta
M(r,t)\propto e^{\lambda t}$, we obtain the eigenvalue equation
\begin{eqnarray}
\lambda^2\delta M=T\left\lbrack {\cal L}(\delta M)g(M_{,r})+{\cal L}(M)g'(M_{,r})\delta M_{,r}\right\rbrack  \nonumber\\
+\delta T {\cal L}(M)g(M_{,r})+\frac{1}{r^2}(M\delta M)_{,r}
\label{esn3}
\end{eqnarray}
with the energy constraint
\begin{eqnarray}
\label{esn4}
\frac{3}{2}
\delta T+\int_0^{r_0}\Phi\, \delta M_{,r}\, dr=0. 
\end{eqnarray}
The neutral mode, corresponding to $\lambda=0$, is determined by the differential equation
\begin{eqnarray}
T\left\lbrack {\cal L}(\delta M)g(M_{,r})+{\cal L}(M)g'(M_{,r})\delta M_{,r}\right\rbrack \nonumber\\
+\delta T {\cal L}(M)g(M_{,r}) +\frac{1}{r^2}(M\delta M)_{,r}=0.
\label{esn5}
\end{eqnarray}

\subsection{Scaled variables}
\label{sec:hatTc}

To study the dynamics close to the critical point A', we introduce the scaled
variables
$\hat{r} =r/\sqrt{T_c}$, $\hat{t}=t$, $\hat{M}=M/T_c^{3/2}$, $\hat{h}=h/T_c$,
${\hat \Phi}=\Phi/T_c$, $\hat{\rho}=\rho$, ${\hat E}=E/T_c$, and ${\hat
T}=T/T_c$. At the critical point, we have ${\hat{ T}}=1$ and all the other
variables coincide with those introduced in Section \ref{sec:equil}. In the
following, we drop the ``hats'' to simplify the notations. With this rescaling,
we obtain
\begin{equation}
\frac{\partial^2 M(r,t)}{\partial t^2}=  T(t)\mathcal{L}(M)g(M_{,r}) + \frac{1}{r^2}M_{,r}  M
\label{sca1}
\end{equation}
with the boundary conditions
\begin{equation}
 \left \{ \begin{array}{l}
M(0,t)=0  \\
M(r_c,t)= T_c^{-3/2} = 4 \pi\int_0 ^{r_c} {\mathrm{d}} r' r'^2 \rho(r',t)
\mathrm{.}
\end{array}
\right. \label{sca2}
\end{equation}
The energetic constraint writes
\begin{eqnarray}
\label{sca3}
E=\frac{3}{2}
T(t)+\frac{1}{2}T_c^{3/2}\int_0^{r_0}\Phi M_{,r}\, dr.
\end{eqnarray}
The steady solution of equation (\ref{sca1}) at the critical point is determined by
\begin{equation}
\mathcal{L}(M^{(c)})g(M^{(c)}_{,r}) + \frac{1}{r^2}M^{(c)}_{,r}  M^{(c)}=0
\label{sca4}
\end{equation}
with the energetic constraint
\begin{eqnarray}
\label{sca5}
E_c=\frac{3}{2}+\frac{1}{2}T_c^{3/2}\int_0^{r_c}\Phi^{(c)} M^{(c)}_{,r}\, d{r}.
\end{eqnarray}
Using Newton's law $\Phi_{,r}={M(r)}/{r^2}$, and the equilibrium relation
$\Phi_{,r}=-h_{,r}$, we can easily check that equation (\ref{sca4}) is 
equivalent to equation (\ref{es1}). On the other hand, at
the critical point, the marginal mode ($\lambda=0$) is determined by the
differential equation [see equation (\ref{esn5})]:
\begin{equation}
\begin{split}
{\cal L}(&\delta M^{(c)})g(M^{(c)}_{,r}) +{\cal L}(M^{(c)})g'(M^{(c)}_{,r})\delta M^{(c)}_{,r} \\
&+\delta T {\cal L}(M^{(c)})g(M^{(c)}_{,r})+\frac{1}{r^2}(M^{(c)}\delta M^{(c)})_{,r}=0
\end{split}
\label{sca6}
\end{equation}
with the energy constraint
\begin{eqnarray}
\label{sca7}
\frac{3}{2}
\delta T+T_c^{3/2}\int_0^{r_c}\Phi^{(c)}\, \delta M^{(c)}_{,r}\, dr=0.
\end{eqnarray}
Using Newton's law in perturbed form $\delta\Phi_{,r}={\delta M(r)}/{r^2}$, and
the relation $\delta\Phi^{(c)}_{,r}=-\delta h^{(c)}_{,r}$ satisfied at the
neutral point (see Section \ref{sec_marg}), we can check that equation
(\ref{sca6}) is equivalent to equation (\ref{marg15}). This implies that the
neutral mass profile is given by
\begin{equation}
\delta M^{(c)}(r)=-r^2 j_{,r},
\label{sca8}
\end{equation}
where $j(r)$ has been determined in Section \ref{sec_marg}.

\subsection{Normal form close to the saddle-center bifurcation}
\label{sec_nfs}

The derivation of the normal form of the hydrodynamic equations close to the
saddle-center bifurcation proceeds by expanding the different
quantities close to their equilibrium value at critical  energy $E_c$  in series
of a small parameter  $\epsilon$ which characterizes a slow variation of the
energy with respect to its value at the saddle-center, supposed to evolve as
$E(t)=E_c-\gamma' t$, with $\gamma' $ small. We set
\begin{equation}
E=E_c-\epsilon^2 E^{(2)},
\label{n1}
\end{equation}
which amounts to defining $\epsilon^2 E^{(2)}=\gamma' t$, and rescaling
the time as $ t=t'/\epsilon ^{{1}/{2}}$.
Equation (\ref{sca1}) is then rewritten as
\begin{equation}
\epsilon\frac{\partial^2 M}{\partial {t'}^2}=  T(t')\mathcal{L}(M)g(M_{,r}) + \frac{1}{r^2}M_{,r}  M.
\label{n2}
\end{equation}
The radial distribution of mass (or radial profile) is expanded as
\begin{equation}
M(r,t')= M^{(c)}(r) + \epsilon M^{(1)}(r,t')+\epsilon^2 M^{(2)}(r,t')+...
\label{n3}
\end{equation}
where $M^{(c)}(r)$ is the equilibrium profile at $E=E_c$ (see  above) drawn in
Figure \ref{fig:equil}-(b).
The expansion of the energy is given in equation (\ref{n1}) and the expansion of
the temperature reads
\begin{equation}
T(t')=1 +\epsilon T^{(1)}(t')  +\epsilon^2 T^{(2)}(t')+...
\label{n4}
\end{equation}
We now substitute the
expansion (\ref{n3}) into equation (\ref{n2}) and consider each order.

\subsubsection{Leading order}

At leading order, we get the equilibrium relation
\begin{equation}
\mathcal{L}^{(c)} g^{(c)} + \frac{1}{r^2}M_{,r}^{(c)}  M^{(c)}=0
\label{n5}
\end{equation}
which  has to satisfy the boundary conditions
\begin{equation}
M^{(c)}(0)= M^{(c)}_{,r}(0)=0, \qquad M^{(c)}(r_c)= M=T_c^{-3/2}.
\label{n6}
\end{equation}
The energy constraint writes
\begin{eqnarray}
\label{n7}
E_c=\frac{3}{2}+\frac{1}{2}T_c^{3/2}\int_0^{r_c}\Phi^{(c)} M^{(c)}_{,r}\, d{r}.
\end{eqnarray}
The mass profile at the critical point is drawn in Figure \ref{fig:equil}-(b).

\subsubsection{First order}

To order $1$ with respect to $\epsilon$, we have
\begin{equation}
T^{(1)}\mathcal{L}^{(c)} g^{(c)} + \mathcal{L}^{(1)}g^{(c)} + \mathcal{L}^{(c)}g^{(1)}+  \frac{1}{r^2}(M^{(1)}  M^{(c)})_{,r}=0,
\label{n8}
\end{equation}
with the energy constraint giving
\begin{equation}
T^{(1)} =- \frac{2}{3}  T_c^{3/2} \int_{0}^{r_c} \Phi^{(c)}(r)  M^{(1)}_{,r} \, dr.
\label{n9}
\end{equation}
Because equation (\ref{n8}) is linear, its solution is of the form
\begin{equation}
M^{(1)}(r,t')= A^{(1)}(t')F(r),
\label{n10}
\end{equation}
\begin{equation}
T^{(1)}(t')= A^{(1)}(t'){\cal T},
\label{n11}
\end{equation}
\begin{equation}
h^{(1)}(r,t')= A^{(1)}(t')j(r),
\label{n12}
\end{equation}
\begin{equation}
S^{(1)}(r,t')= A^{(1)}(t')S^{(c)}(r),
\label{n13}
\end{equation}
\begin{equation}
u^{(1)}(r,t')=\epsilon^{1/2} {\dot A}^{(1)}(t')S^{(c)}(r).
\label{n14}
\end{equation}
This corresponds to the neutral mode multiplied by $A^{(1)}(t')$. In the
foregoing equations $F(r)=\delta
M^{(c)}(r)$, ${\cal T}=\delta T^{(c)}$, $j(r)=\delta h^{(c)}(r)$ and the dot in
equation (\ref{n14}) stands for the time derivative. The neutral mode profiles
(enthalpy, density, mass, and velocity) are plotted in Figures \ref{Figjzeta}
and \ref{FigrhM}. The function $F(r)=\delta M^{(c)}(r)$ was actually derived
from the solution $j(r)$ of equation (\ref{marg15}) thanks to the relation
\begin{equation}
F(r) = - r^2 j_{,r}.
\label{n15}
\end{equation}

\subsubsection{Second order}

To order $2$, equation (\ref{n2}) gives
\begin{equation}
\frac{\partial^2 M^{(1)}}{\partial t'^2} = T^{(2)}\mathcal{L}^{(c)} g^{(c)}+ T^{(1)}(\mathcal{L}^{(1)} g^{(c)}+\mathcal{L}^{(c)} g^{(1)})+
 \mathcal{F}^{(2)},
\label{n16}
\end{equation}
where
\begin{equation}
\mathcal{F}^{(2)}=\mathcal{L}^{(2)} g^{(c)}+\mathcal{L}^{(1)}g^{(1)}+\mathcal{L}^{(c)}g^{(2)}+\mathcal{F}^{(2)}_1
\end{equation}
and
\begin{equation}
\mathcal{F}^{(2)}_1=   \frac{1}{r^2}\left\lbrack (M^{(2)}  M^{(c)})_{,r}+ M^{(1)}  M^{(1)}_{,r} \right\rbrack
\end{equation}
with
\begin{equation}
\mathcal{L}^{(c)}=\mathcal{L}(M^{(c)}), \qquad\mathcal{L}^{(n)}=\mathcal{L}(M^{(n)}),
\end{equation}
\begin{equation}
g^{(1)}=g'^{(c)}M_{,r}^{(1)},\;\;
g^{(2)}=\frac{1}{2}g''^{(c)}(M_{,r}^{(1)})^2+g'^{(c)} M^{(2)}_{,r},
\end{equation}
\begin{equation}
g^{(c)}=g(M_{,r}^{(c)}),\;\;
g'^{(c)}=(\frac{dg}{dM_{,r}})^{(c)}, \;\;
g''^{(c)}=(\frac{d^2g}{dM_{,r}^2})^{(c)}.
\end{equation}
The $r$-dependent quantities can be written in terms of the equilibrium density function $\rho^{(c)}(r)$ as
\begin{equation}
 \left \{ \begin{array}{l}
\mathcal{L}^{(c)}=4\pi r^2\rho_{,r}^{(c)}, \\
g^{(c)} = 1-\frac{1}{\sqrt{1+\rho^{(c)}}}, \\
g'^{(c)}  =\frac{1}{8\pi r^2(1+\rho^{(c)})^{3/2}}, \\
g''^{(c)} = -\frac{3}{4(4\pi r^2)^2(1+\rho^{(c)})^{5/2}}
\mathrm{.}
\end{array}
\right. \label{n17}
\end{equation}
The boundary conditions are
 \begin{equation}
 \left \{ \begin{array}{l}
M^{(n)}(0,t')=0,\quad  M^{(n)}_{,r}(0,t')=0, \\
 M^{(n)}(r_c,t')=0
\mathrm{.}
\end{array}
\right. \label{n18}
\end{equation}
The energetic constraint writes
\begin{eqnarray}
\label{n20}
-E^{(2)}=\frac{3}{2} T^{(2)}+\frac{1}{2}T_c^{3/2}\int_0^{r_c} \Phi^{(1)}M^{(1)}_{,r}\, dr \nonumber\\
+T_c^{3/2}\int_0^{r_c} \Phi^{(c)}M^{(2)}_{,r}\, d{r}.
\end{eqnarray}
This determines $T^{(2)}$. In terms of $M^{(n)}$ the resonant and non-resonant
parts of the second order temperature deviation
$T^{(2)}={T}^{(2)}_{\rm res.}+{T}^{(2)}_{\rm n.res.}$ are given by the relations
\begin{eqnarray}
{T}^{(2)}_{\rm res.}=  - \frac{2}{3} T_c^{3/2}\int_0^{r_c} \Phi^{(c)}
M^{(2)}_{,r} \, dr \nonumber\\
= - \frac{2}{3} T_c^{3/2}\int_0^{r_c}  h^{(c)}_{,r} M^{(2)} \,
dr
\end{eqnarray}
and
\begin{eqnarray}
{T}^{(2)}_{\rm n.res.}=  -\frac{1}{3} T_c^{3/2} \int_0^{r_c} \Phi^{(1)}
M^{(1)}_{,r} \, dr  - \frac{2}{3}E^{(2)} \nonumber\\
=  -\frac{1}{3} T_c^{3/2} \int_0^{r_c}
h^{(1)}_{,r} M^{(1)} \, dr  - \frac{2}{3}E^{(2)}.
\end{eqnarray}
Using equations (\ref{n10}) and (\ref{n12}), we get
\begin{equation}
{T}^{(2)}_{\rm n.res.}=   -\frac{1}{3} T_c^{3/2} \lbrack
A^{(1)}\rbrack^2\int_0^{r_c} j_{,r}(r) F(r) \, dr  - \frac{2}{3}E^{(2)}.
\label{n21m1}
\end{equation}
After splitting the resonant and non-resonant terms in equation (\ref{n16}), we
obtain
\begin{equation}
F(r)\ddot{A}^{(1)}(t') = T^{(2)}_{\rm n.res}\mathcal{L}^{(c)} g^{(c)}
+\mathcal{K}(F) [A^{(1)}]^2 + \mathcal{C}(M^{(2)}),
\label{n21}
\end{equation}
where the non-resonant contribution to the quadratic term is
\begin{eqnarray}
 \mathcal{K}(F)=\frac{1}{r^2}FF_{,r} +\frac{1}{2}\mathcal{L}^{(c)} g''^{(c)} F_{,r}^2 +g'^{(c)}{\cal L}(F)F_{,r} \nonumber\\
 +\mathcal{T}\left ( {\cal L}(F)g^{(c)}+\mathcal{L}^{(c)} g'^{(c)} F_{,r}\right ),
 \label{n22}
\end{eqnarray}
whereas the resonant term is
\begin{equation}
\begin{split}
\mathcal{C}(M^{(2)})= &\mathcal{L}^{(2)} g^{(c)} +\frac{1}{r^2}(M^{(2)}M^{(c)}
)_{,r} \\
+&\mathcal{L}^{(c)} g'^{(c)}  M^{(2)}_{,r} +T^{(2)}_{\rm res}\mathcal{L}^{(c)} g^{(c)}.
\end{split}
\label{n23}
\end{equation}
Substituting equation (\ref{n21m1}) into equation (\ref{n21}), introducing the
slow decrease of the energy versus time, $E^{(2)}\sim \gamma' t/\epsilon^2$, and
making the rescaling $A=\epsilon A^{(1)}$ to eliminate $\epsilon$ (we note that
$A(t)$ is the true amplitude of the mass profile $M^{(1)}(r,t)$), we get
\begin{equation}
\begin{split}
&F(r)\ddot{A} =  - \frac{2}{3}\gamma' t \,\mathcal{L}^{(c)} g^{(c)}  +\epsilon^2 \mathcal{C}(M^{(2)})  \\
+&\left\lbrack \mathcal{K}(F)
-\frac{1}{3} T_c^{3/2} \mathcal{L}^{(c)} g^{(c)}\int_0^{r_c} j_{,r}(r) F(r) \, dr \right\rbrack  A^2 .
\end{split}
\label{n24}
\end{equation}

\subsubsection{Solvability condition}

To write the dynamical equation for $A(t)$ in a normal form, we  multiply
equation (\ref{n24}) by a function $\zeta(r)$ and integrate over $r$ for $0 <r
<r_c$. We are going to derive the function $\zeta(r)$ so that the term ${\cal
C}(M^{(2)})$  disappears after integration. By definition, the function $\zeta$
must satisfy, for any function $M^{(2)}(r)$,
the integral relation
\begin{equation}
\int_0 ^{r_c}  \mathcal{C}(M^{(2)})(r) \zeta(r) \, dr=0.
\label{n28}
\end{equation}
Let us expand  $\mathcal{C}$ as
\begin{equation}
\mathcal{C}(M^{(2)})=
g^{(c)} M^{(2)}_{,r^2} +b M^{(2)}_{,r}+c M^{(2)}+\mathcal{I}[M^{(2)}],
\label{n29}
\end{equation}
where
\begin{equation}
\mathcal{I}[M^{(2)}]=\delta (r)\int_0^{r_c} h^{(c)}_{,r} M^{(2)} dr
\end{equation}
with
\begin{equation}
\delta(r)= - \frac{2}{3} T_c^{3/2}\mathcal{L}^{(c)}(r) g^{(c)}(r).
\end{equation}
We have also introduced $b(r) =
-{2g^{(c)}}/{r}+{M^{(c)}}/{r^2}+\mathcal{L}^{(c)}g'^{(c)}$  and $c(r) =
{M^{(c)}_{,r}}/{r^2}$. In terms of the equilibrium values of the density and
potential functions at the saddle-center, we have
\begin{equation}
 \left \{ \begin{array}{l}
 g^{(c)}(r) = 1-\frac{1}{\sqrt{1+\rho^{(c)}}}, \\
b(r) = -\frac{2g^{(c)}}{r} -h_{,r}^{(c)}+\frac{\rho_{,r}^{(c)}}{2(1+\rho^{(c)})^{3/2}}, \\
c(r) = 4\pi \rho^{(c)}
\mathrm{.}
\end{array}
\right.
\label{n30}
\end{equation}
Integrating the first three terms of $\mathcal{C}(M^{(2)})(r)$ in equation (\ref{n28}) by parts, using  $M^{(2)}(r)=0$ on the boundaries $r=0$ and $r=r_c$, and using $M^{(2)}_{,r}(0)=0$ and $g^{(c)}(r_c)=0$,  gives
\begin{equation}
\begin{split}
\int_0^{r_c}dr\, & M^{(2)}\mathcal{D}\lbrack \zeta\rbrack  \\
+& \int_0^{r_c}dr\, \zeta(r) \delta(r) \int_0^{r_c}dr\, M^{(2)}(r) h_{,r}^{(c)}(r)=0,
\end{split}
\label{n31}
\end{equation}
where the action of the differential operator $ \mathcal{D}[.] $ on a function $\zeta(r)$ is such that 
 \begin{equation}
 \mathcal{D}[\zeta] = (g^{(c)}\, \zeta)_{,r^2}-(b\, \zeta)_{,r}+ c\,\zeta.
\label{n32}
\end{equation}
It can be written equivalently as
\begin{equation}
 \mathcal{D}[\zeta] =  g^{(c)}(r)\zeta_{,r^2}+ a_1(r) \zeta_{,r}+  a_0(r)\zeta,
\label{n33}
\end{equation}
where the coefficients
\begin{equation}
 \left \{ \begin{array}{l}
a_1(r) = 2g^{(c)}_{,r} -b(r), \\
a_0(r) = c(r)+ g^{(c)}_{,r^2}(r) -b_{,r}(r),
\end{array}
\right.
\label{n34}
\end{equation}
can be expressed in terms of the radial density by using equations
(\ref{n17}) and (\ref{n30}). The function $\zeta$ is the solution of the integro-differential equation
\begin{equation}
\mathcal{D}[\zeta] + h_{,r}^{(c)}(r)\int_0^{r_c}\zeta(r)\delta(r)\,dr =0.
\label{n35}
\end{equation}
The solution  $\zeta(r)$ is drawn in Figure \ref{Figjzeta}, blue curve. This
solution is obtained by solving the integro-differential equation (\ref{n35}) 
with two initial conditions. Close to the center, it can be shown that the
solution of equation (\ref{n35}) writes $\zeta(r)= z_1 r \;+\;z_3 r^3+...$.
Therefore we set $\zeta(0)=0$ and $\zeta'(0)=z_1$, an a priori unknown parameter
proportional to $G\equiv\int_0^{r_c} \zeta(r)\delta(r)\,dr$ which may be taken
as unity since the integro-differential equation is linear with respect to
$\zeta$.  The value $z_1$ of the slope of $\zeta(r)$ at the center is determined
numerically by increasing $z_1$ step by step. At step $n$, for a given
$z_1^{(n)}$, we solve the ordinary differential equation
\begin{equation}
\mathcal{D}[\zeta^{(n)}] + h_{,r}^{(c)}(r) =0,
\label{n36}
\end{equation}
calculate the value of $\int_0^{r_c}\zeta^{(n)}(r)\delta(r)\,dr$, and increase the slope $z_1$ until we obtain the expected  result $\int_0^{r_c}\zeta(r)\delta(r)\,dr\; =\; 1$.

\subsubsection{Painlev\'e I equation}

\begin{figure}
\centerline{
(a)\includegraphics[height=1.7in]{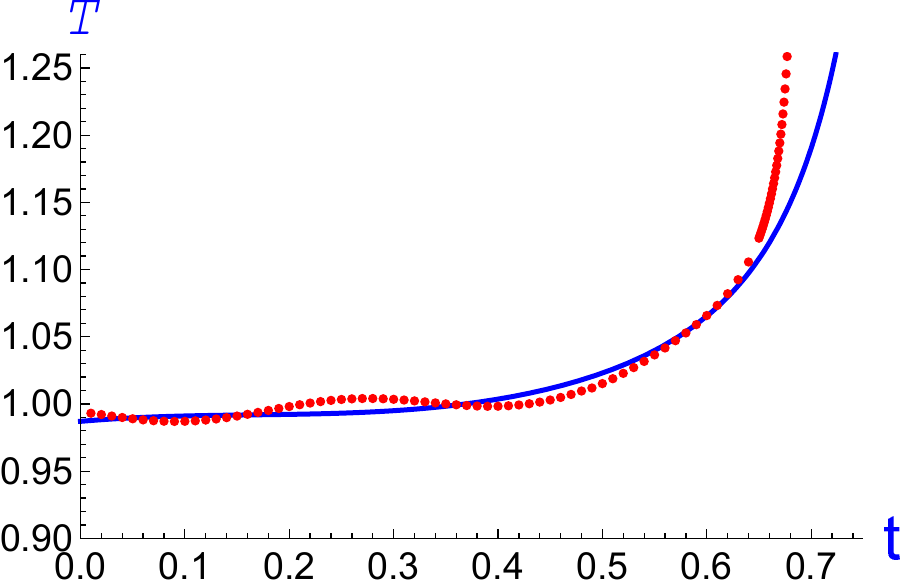}}
\centerline{
(b)\includegraphics[height=1.7in]{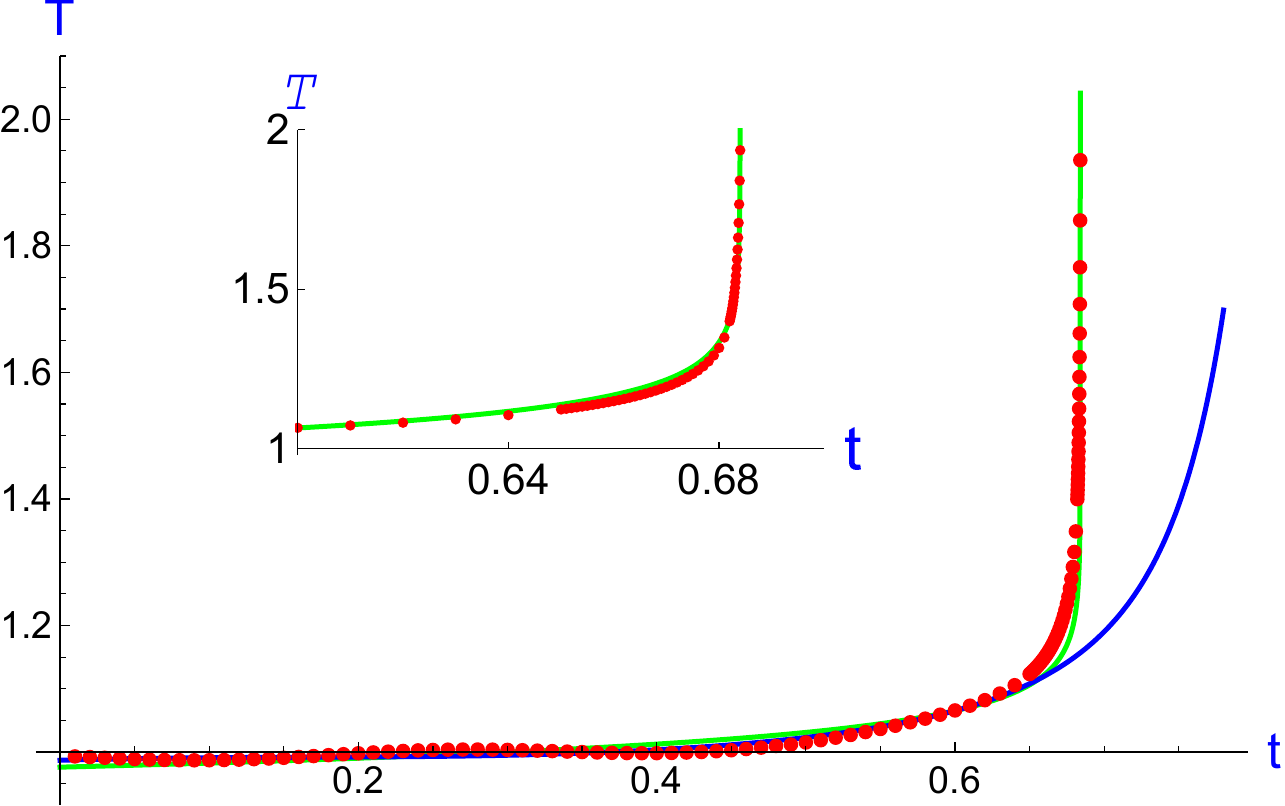}
}
\caption{
Evolution of the temperature (a) in the Painlev\'e regime, $0
\lesssim  t \lesssim 0.6 $, (b) in the whole time
interval before collapse, $0 \lesssim  t \lesssim 0.684$. The
numerical solution (dots) of the
MEP model with the energetic constraint $E=E_c-\gamma' (t-t_0)$, where
$\gamma'=0.1$, is compared in (a) with the solution $A(t)$ of the
Painlev\'{e} I
equation (\ref{n25}) with the initial condition $A(t_0)=-0.008$ and
$\dot{A}(t_0)=0.01$, where $t_0=0.18$.
  In (b) we add the self-similar solution $T(t) \sim (t-t_*)^{a(k)}$ of
Section \ref{sec_cc} where
$k=8$ ($a=-1/24$) in the green portion of the principal curve, and $k=8.4$
($a=-0.074$) in the insert. 
}
\label{Fig:Painl}
\end{figure}

Now that we have obtained the function $\zeta(r)$ satisfying the integral
relation (\ref{n28}), we find that equation (\ref{n24}) multiplied by $\zeta(r)$
and integrated over $r$ for $0 <r<r_c$ takes the form of the
Painlev\'{e} I equation 
\begin{equation}
\ddot{A}(t)= \tilde{\gamma} t+K A^2.
\label{n25}
\end{equation}
The first coefficient in equation (\ref{n25})  is given
explicitly as a function of the parameters at the critical point by the
expression
\begin{equation}
\tilde{\gamma}= -\frac{2}{3}\gamma' \frac{\int_0^{r_c} {\cal L}^{(c)}(r)
g^{(c)}(r)\zeta(r)\, {\mathrm{d}}r}{\int_0^{r_c}F(r)\zeta(r)\, {\mathrm{d}}r}
\label{n26}
\end{equation}
which is found to be equal to $\tilde{\gamma}= 46.63...\gamma'$ . Moreover, the
second coefficient  in equation (\ref{n25})  is given by
\begin{equation}
K= \frac{\int_0^{r_c} \mathcal{G}(r) \zeta(r)\,
{\mathrm{d}}r}{\int_0^{r_c}F(r)\zeta(r)\, {\mathrm{d}}r}
\label{n27}
\end{equation}
with
\begin{equation}
\begin{split}
\mathcal{G}(r)=&{\cal L}^{(c)}F_{,r}\left (\frac{1}{2}g''^{(c)} F_{,r} +
\mathcal{T}g'^{(c)} \right ) \\
+ &\left (g'^{(c)}F_{,r} +  \mathcal{T}g^{(c)}\right
) (F_{,r^2}-\frac{2}{r}F_{,r}) + \frac{1}{r^2} F\; F_{,r}\\
 + &\frac{1}{3}{\cal L}^{(c)} g^{(c)} T_c^{3/2}
\int_0^{r_c} j(r)F_{,r}(r) dr.   
\end{split}
\end{equation}
It is found  to have the numerical value $K=1055.98...$

As noted in Section \ref{sec:in-out}, an important point to
make clear is the sign of the neutral mode which was unknown at first order.
With the choice we made in the previous Section, we obtain at second order two
positive coefficients $\tilde{\gamma}$ and $K$  in the Painlev\'{e} equation
(\ref{n25}). This result confirms that we made the good choice at first order,
because it leads to the acceleration of the velocity field initially chosen. A
change of sign of the neutral mode amounts to changing $A$ into $-A$ in the
Painlev\'{e} equation (\ref{n25}), or to changing the sign of the nonlinear
coefficient $K$ (this change of sign being formal because it is just a
consequence of the sign chosen for the neutral mode). {\it In fine}, this
imposes us to reverse $-A$ into $A$ because we want to look at a growing
perturbation. In summary, the weakly nonlinear analysis provides the time
evolution of the perturbation \textit{and} the sign of the growing mode. This is
an intrinsic property of saddle-center bifurcations which is absent in the case
of ``classical" transitions from a linearly stable to a linearly unstable
situation, where the unstable mode may have either positive or negative
amplitude. Such a fair property of saddle-center bifurcations comes from the
fact that the stable and unstable equilibrium states are merging at the critical
point, so that no equilibrium state exists beyond that point.

\begin{figure}
\centerline{
 (a)\includegraphics[height=2.0in]{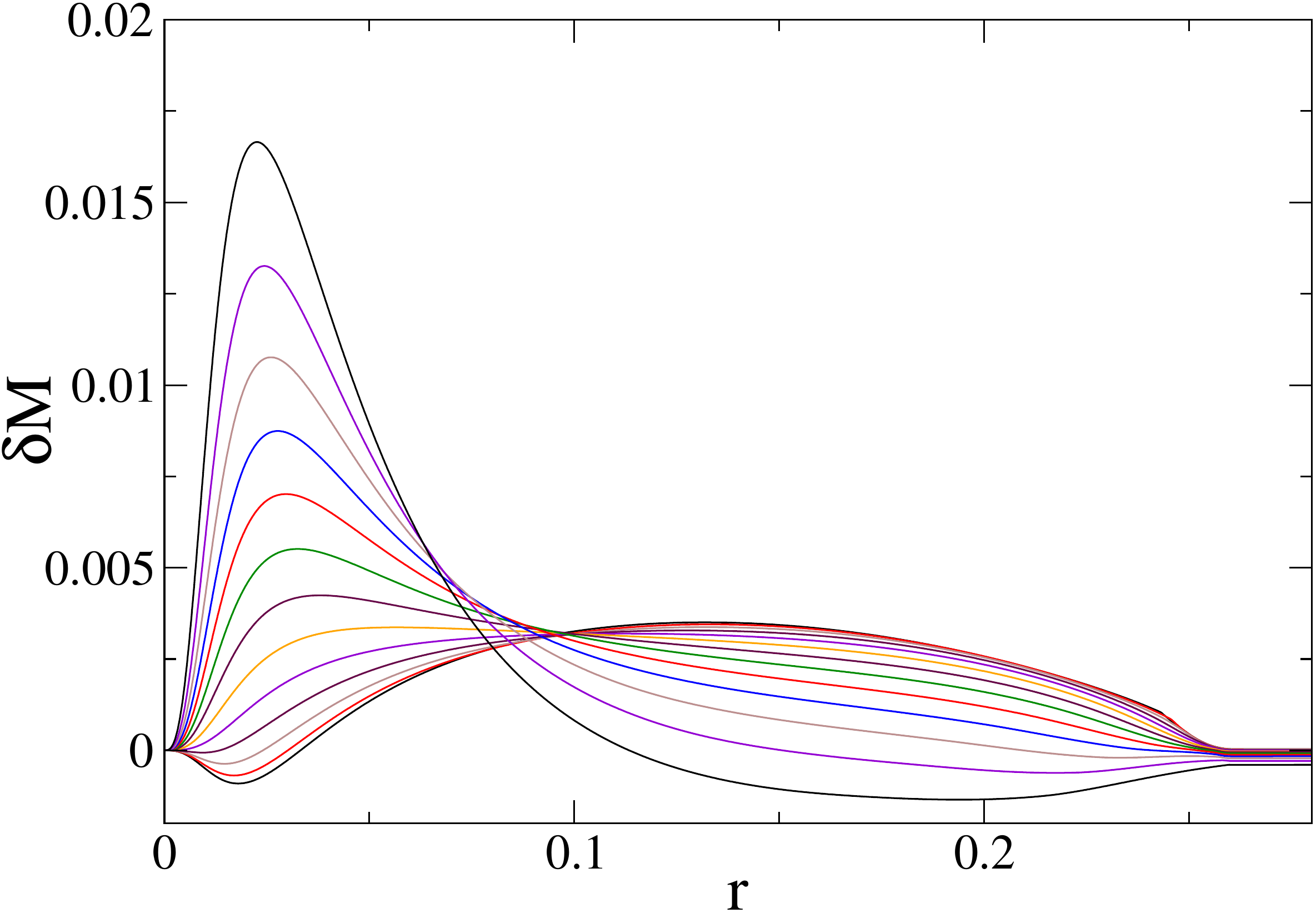}}
\centerline{ 
 (b)\includegraphics[height=2.0in]{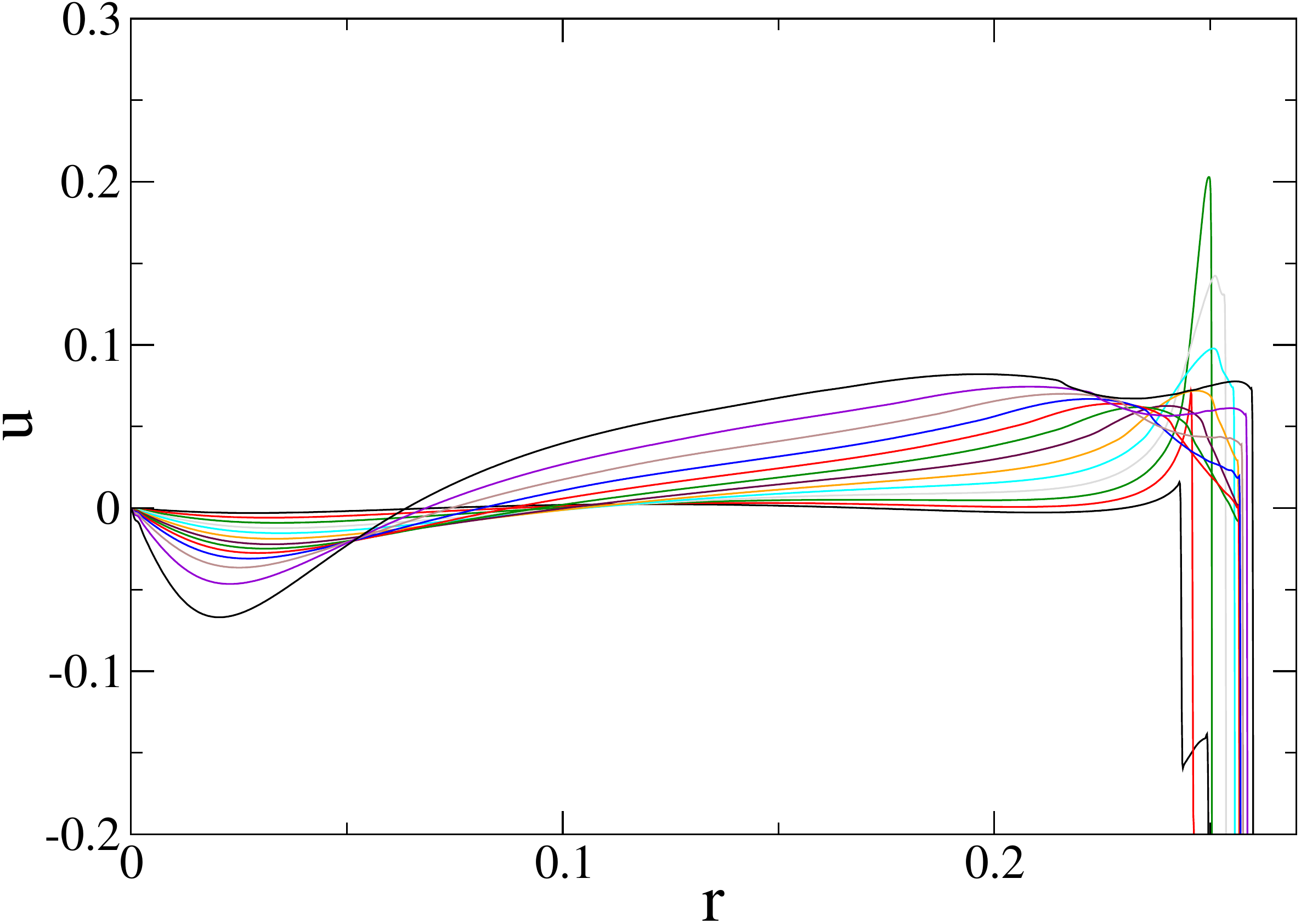}
}
\caption{Fluctuations of mass and velocity of the MEP model showing the back and
forth motion concomitant with the oscillations of temperature $T(t)$ during the
Painlev\'{e} regime. We also see that the velocity is negative in the core and
positive in the halo. Therefore, these direct numerical simulations confirm the
inward/outward motion of the star predicted by our linear or weakly nonlinear
analysis close to the critical point. Furthermore, the numerical profiles
$\delta M(r,t)$ and $\delta u(r,t)$ are in qualitative agreement with the radial
profiles of
the first order deviations at the microcanonical critical point plotted
in Figure \ref{FigrhM}. 
}
\label{Fig:motion-Painl}
\end{figure}

We now compare the prediction of the weakly nonlinear analysis derived here with
the solution of the full MEP model. The numerical solutions of the full MEP
model were obtained using a variant of the CentPack Software
\cite{progbalbas1,progbalbas2}
by Balbas and Tadmor, with  a spatial mesh of $3000$ points and adaptative time
increments. In Figure \ref{Fig:Painl}-(a), we plot in solid line the temperature
$T(t)=\mathcal{T} A(t)$  resulting from the above Painlev\'{e} analysis and show
in dotted line the temperature calculated with the full numerical MEP model for
the early stage of the explosion-implosion process. In the numerical study of
the MEP model, our aim was to take as initial condition the equilibrium state at
the critical energy $E_c$ defined theoretically in Section \ref{sec:equil}, and
let the energy slowly decrease. However, the numerical value of the equilibrium
state in the MEP solution is not exactly the one predicted by the theory (point
A' in Figure \ref{spi-microcan1}) because of finite mesh effects, as already
observed in the canonical case (Paper I). Here, the density in the core is about
two orders of magnitude larger than in Paper I, causing rapid fluctuations
of the MEP solution around an average value. These fluctuations are 
clearly visible on the temperature $T(t)$ of Figure \ref{Fig:Painl}-(a) which
displays a few oscillations before increasing strongly. Such rapid oscillations
are not observed  in solving Painlev\'{e}'s equation: with
off-equilibrium initial conditions close to the critical point, we would get
oscillations with a long period as described  in Paper I (see Figure 2 and
equation (10)). The rapid oscillations observed here  could be attributed to
acoustic waves formed because of the stiffness of the density. They are
characterized  by a back and forth motion of matter in the  star, as illustrated
in Figure \ref{Fig:motion-Painl} which reports the deviations of mass and
velocity
in the whole star versus $r$ at various times in the weakly nonlinear regime.
Due to this back-and-forth motion, the MEP profiles shown in this figure agree
only qualitatively with the neutral mode profiles shown in Figure
\ref{FigrhM}-(b).
Note that the fluid mechanical equations we solve are without any damping term,
so spurious time oscillations are easily generated in the numerics.  
 
To compare the temperature of the MEP model with the weakly nonlinear analysis,
we
take as initial conditions of the Painlev\'{e} equation the ones of the MEP
solution averaged over the oscillations at a given time
$t_0=0.18$  which is chosen close to the point where
$T(t)=T_{c}=1$ (in normalized variables).
The agreement between the MEP model and the weakly nonlinear
analysis (see Figure \ref{Fig:Painl}-(a)) is very good until $t_s=0.6$ which
characterizes the
end
of the Painlev\'{e} regime where nonlinear terms of higher order come into play.
After $t=t_s$, the two curves separate, the solution of the full equations
increasing much more strongly than the weakly nonlinear one. The solution of the
full equations has a self-similar behavior leading to a finite time singularity
at $t_*\simeq 0.684$ illustrated by the green
portion of the curve (see the next Section). The solution of
the Painlev\'e equation also displays a divergence but it occurs later
(at $t_P \simeq 0.8$),
in a regime where the Painlev\'e equation is not valid anymore (see below).

The duration of the Painlev\'{e} regime is expected to be a few times the
precursor time (intermediate time scale) which
stands between the short and long time scales, defined in Section 2 of Paper I,
and given by the relation
\begin{equation}
t_0= (\tilde{\gamma} K)^{-1/5}.
\label{n27b}
\end{equation}
Introducing the numerical values of $K$ and $\tilde{\gamma}$ in equation
(\ref{n27b}), we obtain $t_0=0.18$ which is about $1/3$ the full
Painlev\'{e} regime duration illustrated in  Figure 
\ref{Fig:Painl}-a, as expected. 

In the framework of Painlev\'{e}'s equation, the collapse time
is given by the
relation $t_P \simeq 3.4 t_0$, or $t_P \simeq 0.4\vert
T_c/\dot{T}\vert^{1/5}$,  where
$\dot{T}/T_c=\gamma'$ (see Paper I).
Numerically,
this gives $t_{P }^{\rm approx}=  0.63$. This approximate
value agrees  
well with the exact Painlev\'{e} diverging
time  $t_{P}=  0.8$ when taking the origin at $t_0=0.18$.
Note that this
``Painlev\'{e}'' collapse time is not reached by the solution of the MEP model
which diverges before at $t_*\simeq 0.684$. On the other
hand, we have shown in
Paper I that the amplitude of the Painlev\'{e} solution (for example the
temperature drawn in solid line) diverges close to $t_P$ as
\begin{equation}
A(t)= \frac{0.0063}{(t_P-t)^{2}},
\label{n27bb}
\end{equation}
a solution different from the MEP model solution as discussed in the next
Section [see equation (\ref{Tself})].

\section{The post-Painlev\'e regime before explosion
(pre-collapse regime)}
\label{sec:numerics}

After the weakly nonlinear Painlev\'{e} regime, the full numerical MEP model
displays a solution which ultimately diverges at $t_*$ defined as the collapse
time. This divergence occurs in the
core domain whose radius shrinks to zero while the density and the velocity
increase up to infinity there. Simultaneously, the temperature also diverges
as shown in  Figure \ref{Fig:Painl}-(b). In the halo, the outward velocities
continue to grow, but more slowly than in the core, so that at the
collapse time $t_*$ the outward motion of matter is still at an early stage. The
solution of the MEP model will be described separately in the two regions.

\subsection{Core collapse}
\label{sec_cc}

\begin{figure}
\centerline{
(a)\includegraphics[height=2in]{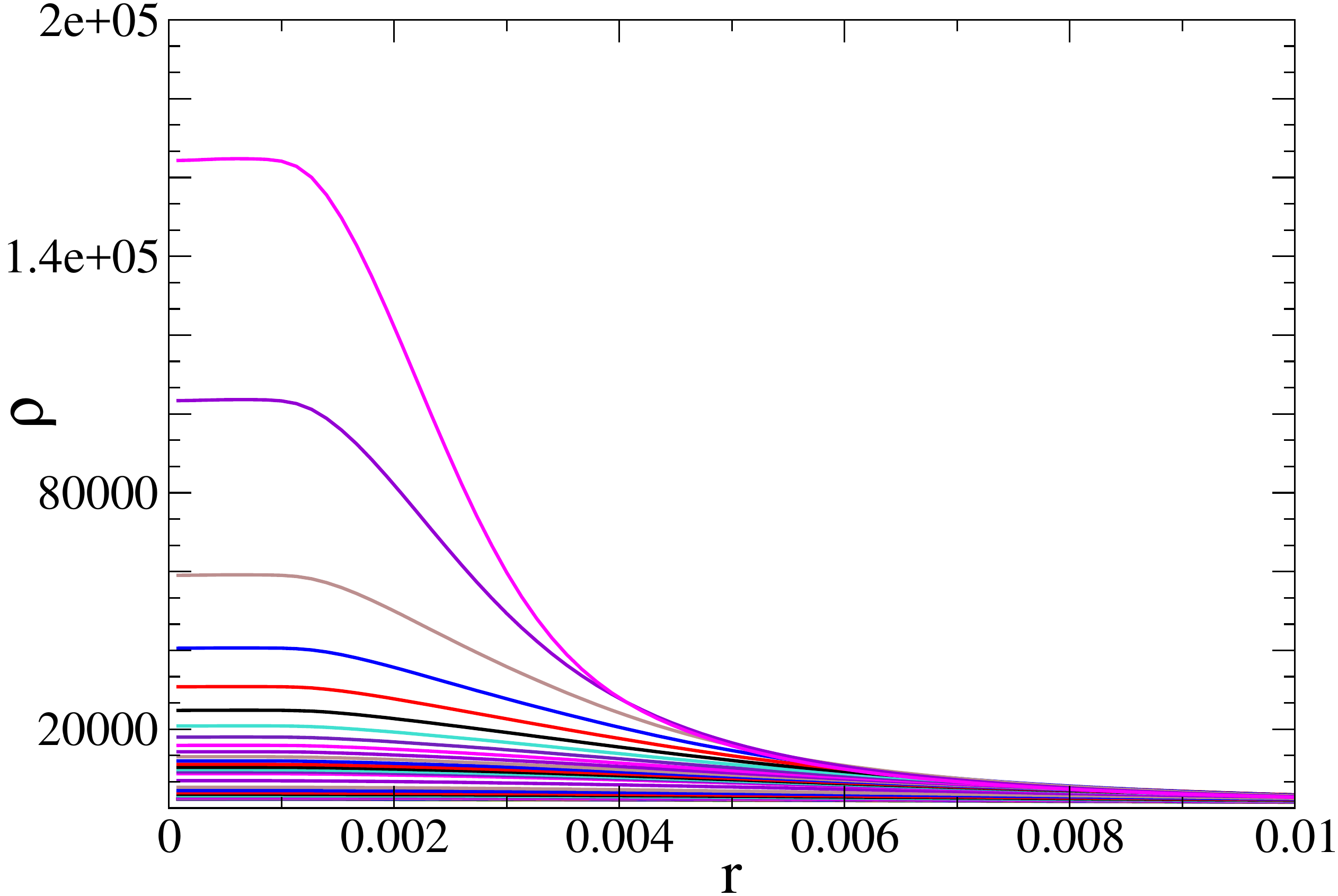}}
\centerline{
(b) \includegraphics[height=1.8in]{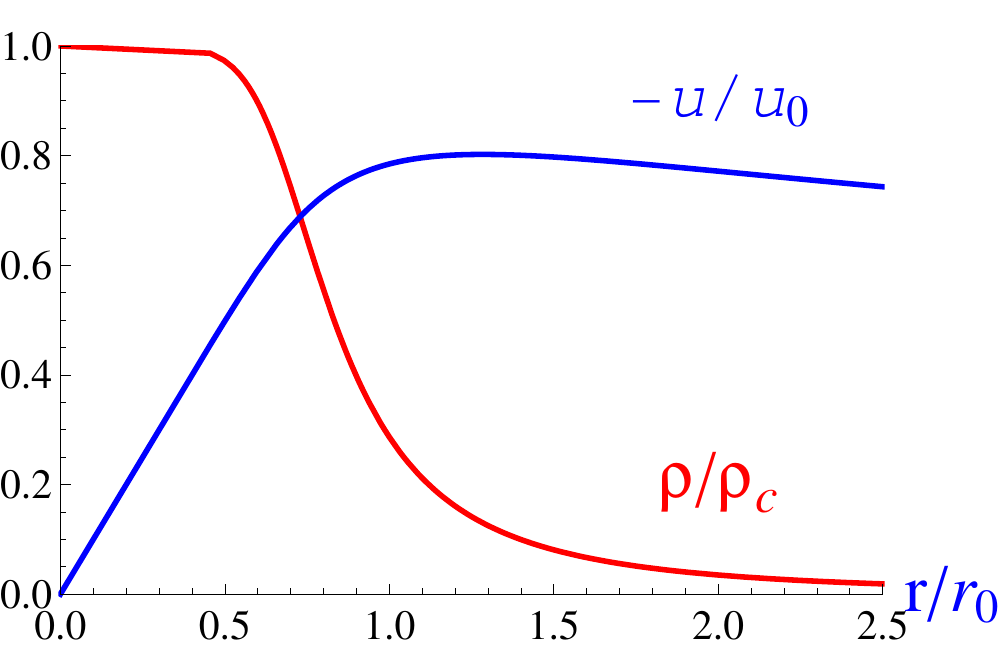}
}
\caption{ Density in the core domain before collapse:
(a)  Numerical solutions  of the MEP model ($\rho (r,t)$ increases with time);
(b) Invariant profiles  (\ref{ff5})-(\ref{ff7})  versus $r/r_{0}$
of the
self-similar pressureless Penston-type solution  with $k=8$.
}
\label{Fig:dens1}
\end{figure}

The increase of density and velocity close to the center of the star, which are
well visible in linear scale in Figures \ref{Fig:dens1}-(a) and
\ref{Fig:post-Painl-u}-(a), deserves to be specified. The numerical study
displays a solution which  becomes self-similar in the core after the
Painlev\'{e} regime, with a singularity of the second kind in the sense of
Zel'dovich \cite{Zeldo}. This property was already found in the
canonical case (Paper I) where the whole star collapses. In both cases, in the
collapsing domain, the values of the exponents characterizing the self-similar
regime show that gravity dominates over pressure forces. However, direct
numerical simulations show that the exponents of the MEP model are different
from those of the CEP model. Recall that for the gravity-dominated case, using
the notations of Paper I, the self-similar density is of the form
 \begin{equation}
\rho(r,t) = (-t)^{-2} R(r(-t)^{-2/\alpha})
\mathrm{,}
\label{eq:rhoa}
\end{equation}
and the self-similar velocity is of the form
 \begin{equation}
 u(r,t) = (-t)^{-1+\frac{2}{\alpha}} U(r(-t)^{-2/\alpha})
 \mathrm{,}
\label{eq:ua}
\end{equation}
where $R$ and $U$ are invariant profiles, $\xi=r(-t)^{-2/\alpha}$ is the scaled
radial distance, and the exponent $\alpha$ is larger than two. We have taken
the origin of time at the collapse time $t_*$. These self-similar solutions
require that $R(\xi)\sim \xi^{-\alpha}$ and $ U(\xi)\sim \xi^{-(\alpha/2-1)}$
for $\xi\rightarrow +\infty$ in order to have a steady profile at large
distances, as necessary. The  exponent $\alpha$ is not free; it is related to
the behavior of the self-similar solution as $\xi\rightarrow 0$ \cite{epje}.
More precisely, expanding $R$ as $R = R_0 + R_2 \xi^2  + ... + R_k \xi^k+ ... $
and $U$ as  $U= U_1 \xi  + ... + U_k \xi^{k+1}  + ...$, one finds
\begin{equation}
\alpha(k)=\frac{6k}{2k+3},
\label{eq:alfak}
\end{equation}
where $k$ is an even number because we consider solutions with spherical symmetry.

The behavior of the temperature
\begin{eqnarray}
\label{exp3}
T(t)=\frac{2}{3}E-\frac{1}{3}\int\rho\Phi\, d{\bf r}-\frac{1}{3}\int\rho {\bf u}^2\, d{\bf r},
\end{eqnarray}
can be deduced from the above scalings in the core domain. This can be done if
one neglects the contribution of the halo to the energy in equation
(\ref{exp3}), an assumption justified because the kinetic and gravitational
energies are much smaller in the halo than in the
core.\footnote{These quantities are of order $M u_{i}^{2}$ and
$M^{2}/r_{i}$, where the index $i=1,2$ refers to the core and halo domains (here
$M_{1}\sim M_{2} \sim M/2$). Their relative values depend on the mean
velocity
and on the size of the fluid in each region. We observe numerically that the
velocity in the halo is much smaller than in the core (and the inverse  for the
size $r_{i}$).} In the core, the potential energy
behaves as $W\sim \rho_c \Phi_c r_0^3$, where $\Phi_c\sim -\rho_c r_0^2$
according to Poisson's equation. Using $\rho_c\sim (-t)^{-2}$ and 
$r_0\sim (-t)^{2/\alpha}$, we get
 \begin{eqnarray}
T \sim - W \sim (-t)^{10/\alpha-4} \;\;\; \sim  
(-t)^{\frac{15-2k}{3k}} 
 \label{Wself}
 \end{eqnarray}
which diverges  for $\alpha$ larger than $5/2$, or  for the even number $k$
larger than $6$.\footnote{The kinetic
energy behaves as $E_{\rm
kin}\sim \rho_c u_0^2 r_0^3$. Using $\rho_c\sim (-t)^{-2}$,
$r_0\sim (-t)^{2/\alpha}$ and
$u_0\sim (-t)^{-1+2/\alpha}$, we get $E_{\rm kin}\sim  (-t)^{10/\alpha-4}\sim
-W$. Therefore, the divergence of the kinetic energy $E_{\rm kin}\rightarrow
+\infty$ in Eq. (\ref{exp3}) could compensate the divergence of the
gravitational energy $W\rightarrow -\infty$. However, numerical simulations
show that $W$ dominates. Therefore, as a result of the conservation of the
energy, the collapse of the core ($W\rightarrow -\infty$) is associated with an
increase of the temperature of the system ($T\rightarrow +\infty$).}
 In
order to investigate whether a self-similar solution of the
form (\ref{eq:rhoa}), (\ref{eq:ua}) and (\ref{Wself}) agrees with the numerical
results, let us first look at the temperature behavior in the post-Painlev\'{e}
regime before the divergence,  for $t_{s}< t < t_{*}$, where $t_{s}\simeq 0.60$
and $t_{*} \simeq 0.684$. Restoring the initial
notations, equation 
(\ref{Wself}) writes
\begin{equation}
T (t)= T(t_s) \left (\frac{t_*-t}{t_*-t_s}\right )^{a(k)}
\quad {\rm
with} \quad  a(k)=\frac{15-2k}{3k}.
 \label{Tself}
 \end{equation}
The  best fit  with the numerical results occurs for
$a= -0.074$ which is chosen in the insert (green line) of Figure
\ref{Fig:Painl}-(b). This value corresponds to $k=8.4$,  or  $\alpha=2.5454$,
indicating that the integer value $k=8$ is a possible candidate. The
corresponding exponent  
\begin{equation}
\alpha(8)=\frac{48}{19}\qquad {\rm implying}\qquad
a(8)=-\frac{1}{24}
\label{eq:alpha}
\end{equation}
chosen to draw the green line  superposed to the full curve $T(t)$ gives a good
fit with the numerical curve. For the value $k=8$, the temperature diverges at
the collapse time as
\begin{equation}
T(t)\propto (t_*-t)^{-1/24}.
\end{equation}

\begin{figure}
\centerline{
(a) 
\includegraphics[height=2.0in]{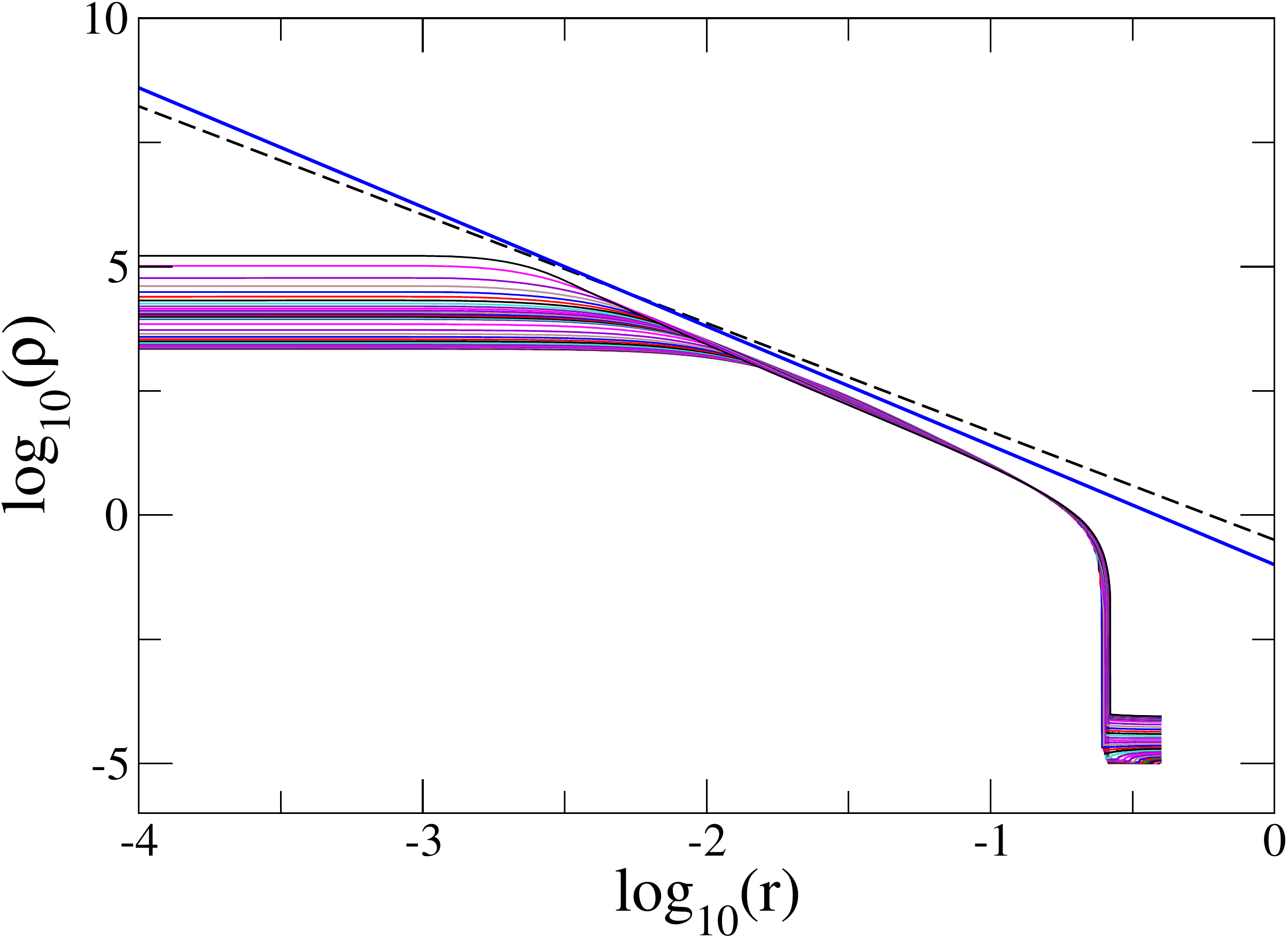}}
\centerline{
(b) \includegraphics[height=2.0in]{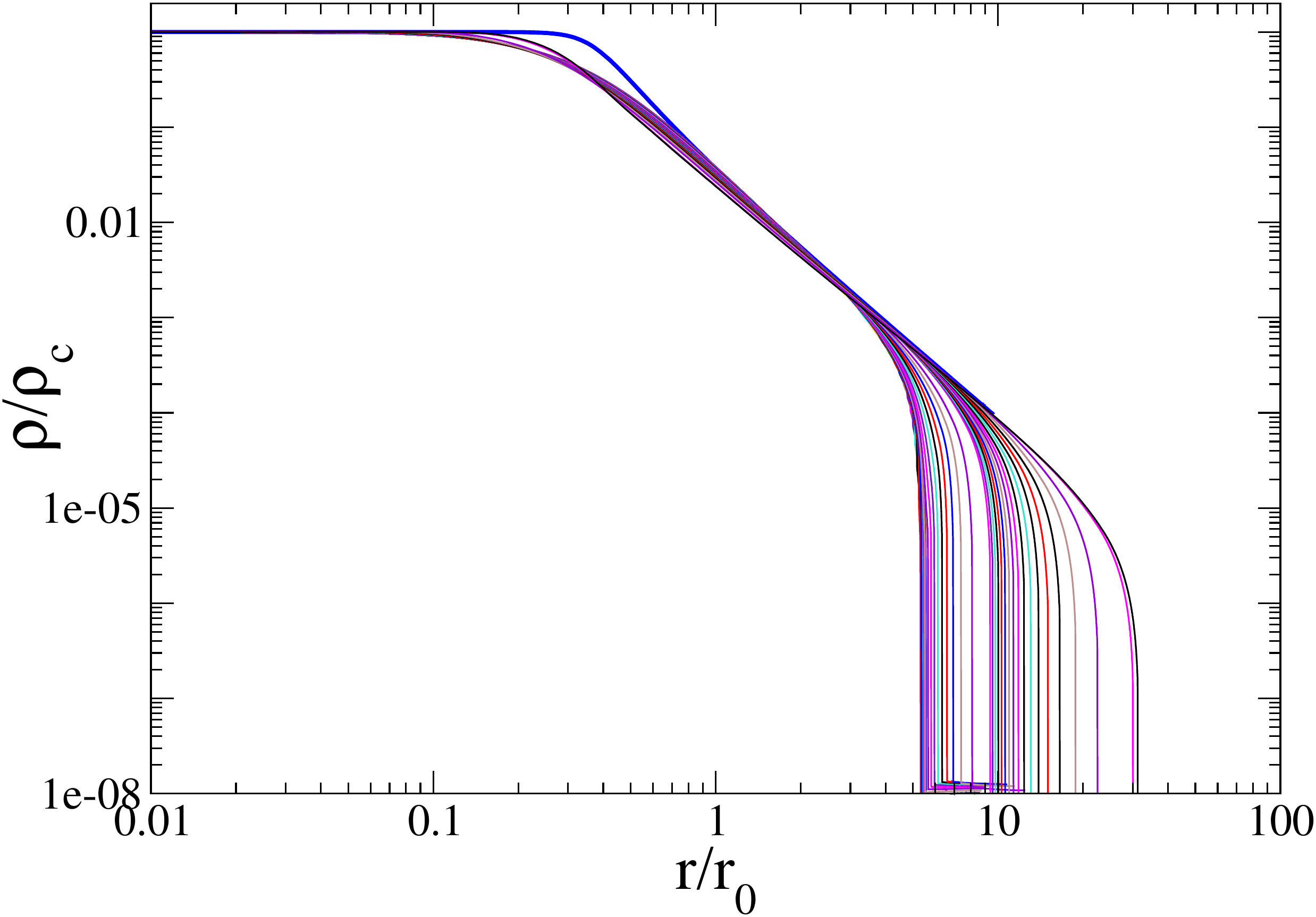}
}
\caption{Density $\rho (r,t)$ before core collapse in $\log_{10}$
scale (numerical solution of the MEP equations). In (a)
the thick straight lines display the slopes
 $\alpha(8)=-48/19$ (solid blue) and  $\alpha(4)=-24/11$ (dashed black) respectively;  in (b) the numerical solutions
   $\rho (r,t)/\rho (0,t)$ versus $r/r_0(t)$   for $\alpha(8)$  are 
superposed to the invariant function (\ref{ff5})-(\ref{ff6}) in blue thick
line.
}
\label{Fig:dens-log}
\end{figure}

Let us now check and see if the radial solutions $u(r,t)$ and  $\rho(r,t)$
also display a self-similar behavior with invariant functions corresponding to 
the value $k=8$ suggested by the temperature behavior. The numerical density
curves $\rho(r,t)$ for increasing time values are shown in linear and
logarithmic scales in Figures \ref{Fig:dens1}-(a) and  \ref{Fig:dens-log}-(a)
respectively. The latter shows an  asymptotic behavior $r^{-\alpha}$ which
agrees with the the  slope $\alpha(8)=48/19$ reported above the
curves (blue straight
line) of Figure \ref{Fig:dens-log}-(a). For comparison,  the slope
$\alpha(4)=24/11$ of
the CEP model is plotted in dashed (black) line. Note that in both cases  (CEP
and MEP models) the value of $\alpha$  is larger than $2$ contrary to
Penston's isothermal solution \cite{Penston} deduced by assuming that pressure
and gravity forces keep
the same order of magnitude until the collapse.
 
The whole self-similar solution ($U(.),R(.)$)  can be derived either  by solving
the two coupled differential equations (102) and (103) of Paper I, or by using
the parametric solution given in Appendix B.1 of Paper I  which
generalizes Penston's pressureless solution \cite{Penston},
namely
\begin{equation}
\frac{\rho(r,t)}{\rho_c(t)}=\frac{3}{3+2(3+k)y+(3+2k)y^2},
\label{ff5}
\end{equation}
\begin{equation}
\frac{r}{r_0(t)}=y^{1/k}(1+y)^{2/3},
\label{ff6}
\end{equation}
\begin{equation}
\frac{u(r,t)}{u_0(t)}=-\frac{y^{1/k}}{(1+y)^{1/3}},
\label{ff7}
\end{equation}
where $y(t)\propto t_{*}/(t_{*}-t)$ goes from $0$ to
$+\infty$.

The exponent of the MEP model, $\alpha(8)=48/19$, corresponds 
to the
on-axis behavior of the function $R = R_0 + R_8 \xi^8+ ...$, whereas  with  the
CEP model we found $k=4$, $\alpha(4)=24/11$  and $R = R_0 + R_4
\xi^4+ ...$. The invariant functions ($U(.),R(.)$) are drawn in
Figure \ref{Fig:dens1}-(b) in linear scale. 

To compare the numerical curves with the functions ($U(.),R(.)$) we 
define a time dependent core radius $r_{0}(t)$ by the relation $\rho(0,t)
r_{0}(t)^{\alpha}=1$  and plot $\rho
(r,t)/\rho (0,t)$ and $u(r,t)/u_{0}(t)$ 
versus $r/r_0(t)$.  Using this procedure, the density curves merge quite well
(in the core domain) into the theoretical solution of equations (\ref{ff5}) and
(\ref{ff6}) plotted in blue line for $k=8$, as illustrated in  Figure
\ref{Fig:dens-log}-(b). 

\begin{figure}
\centerline{
 (a)\includegraphics[height=2.0in]{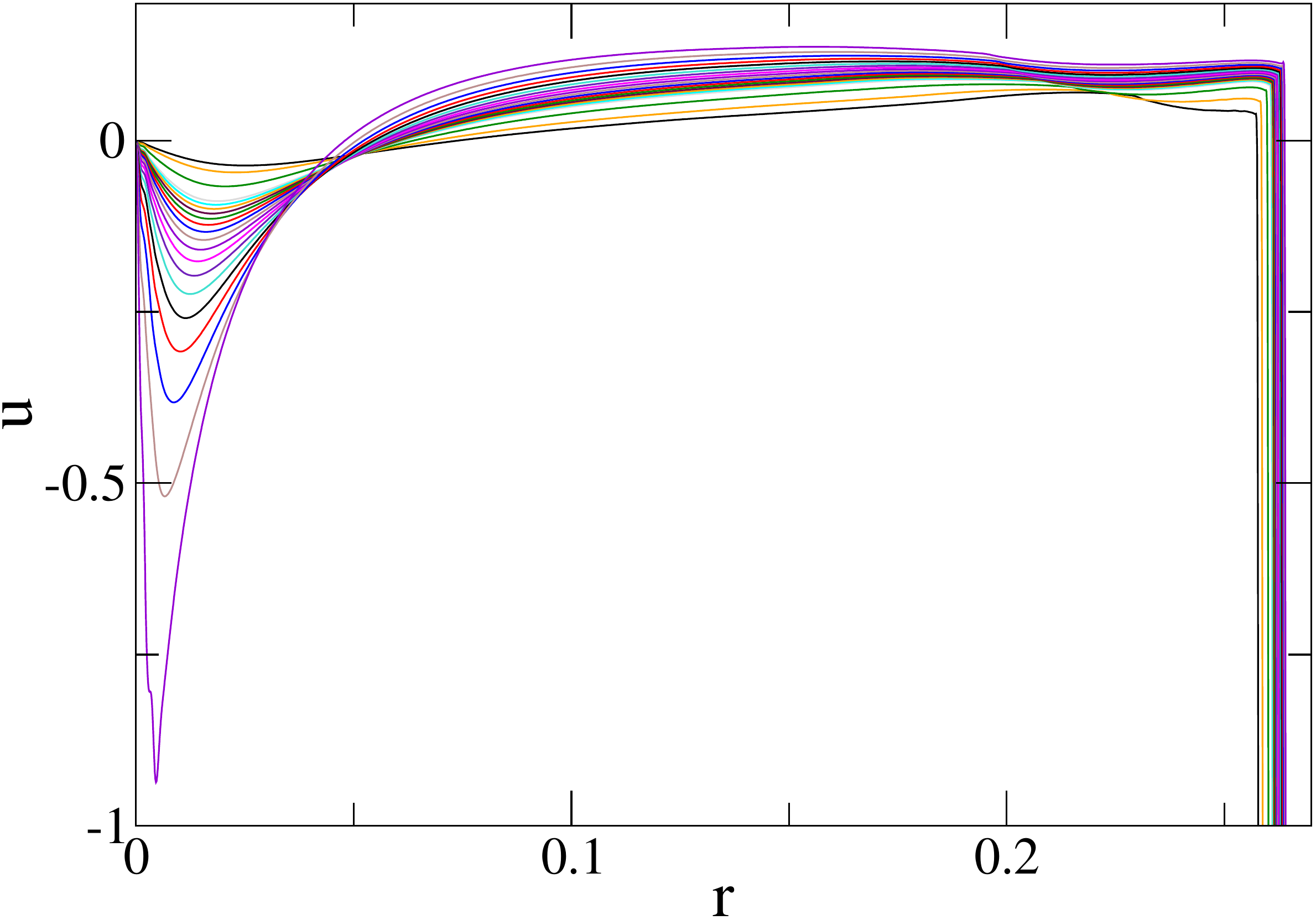}}
 \centerline{
 (b)\includegraphics[height=2.0in]
{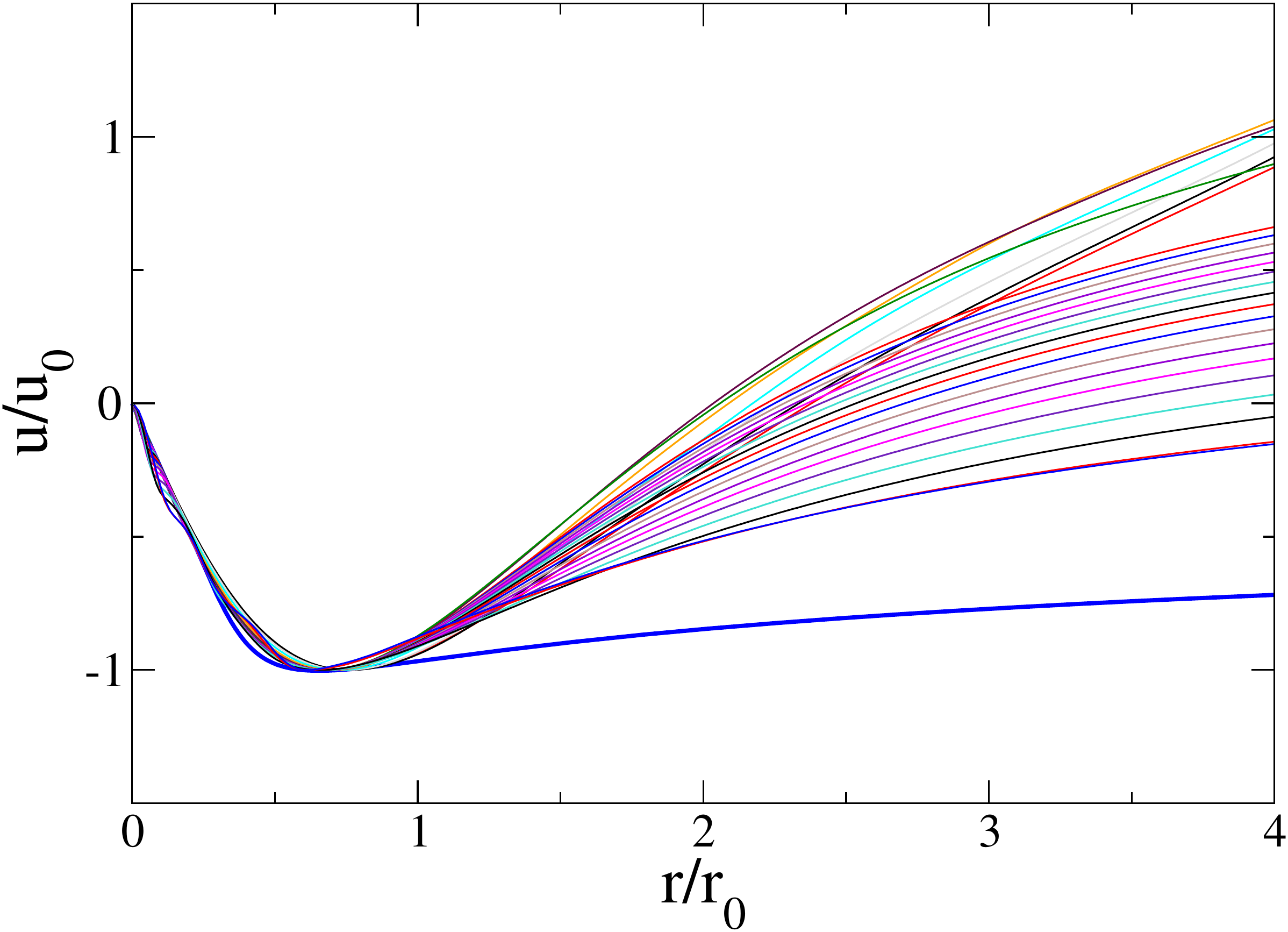}
}
\caption{(a) Velocity $u(r,t)$  for $t_s< t < t_*$ in the whole
star (the
modulus of the extremum increases with time); (b) Velocity ratio 
$-u(r,t)/u_{0}(t)$  versus $r/r_{0}(t)$  compared to the invariant
parametric solution (\ref{ff5})-(\ref{ff7}) applying to the
core for $k=8$ (blue thick line).
 }
\label{Fig:post-Painl-u}
\end{figure}

The merging of the velocity curves into a single one $U$ is not as good, except 
close to the center, for $r \leq r_{0} (t)$, see Figure
\ref{Fig:post-Painl-u}-(b).
At larger radii, the numerical curves separate, approaching asymptotically the
solution $U$ (blue line) in the core as time tends to $t_{*}$.  Compared to the
CEP results,
where the self-similar behavior was also better for the density than for the
velocity,  we note that here the non-merging region concerns the right part of
the curve only (compare  Figure \ref{Fig:post-Painl-u}-(b) with Figure 14 of
Paper I). We attribute this effect (at large radii) to the fact that the
velocity has
to change its sign at the internal surface of the halo $r_{h}$, which
enforces the slope of the velocity at the beginning of the self-similar regime
when the ratio $r_{0}/r_{h}$ is not small.

\subsection{Halo: No self-similar solution}
\label{sec:precoll-halo}

 The velocity curves presented in
Figure \ref{Fig:post-Painl-u}-(a) clearly illustrate the simultaneous 
inward/outward motion of matter in the time interval $t_s < t <
t_{*}$, where  the velocity is negative in the core, and
\textit{positive} in the halo, with a 
modulus increasing  with time  in both parts. During the self-similar growth of
density and inward velocity in the core, what
happens in the halo? Is the solution self-similar there? We shall see that the
answer is NO. 

\begin{figure}
\centerline{
\includegraphics[height=2.0in]{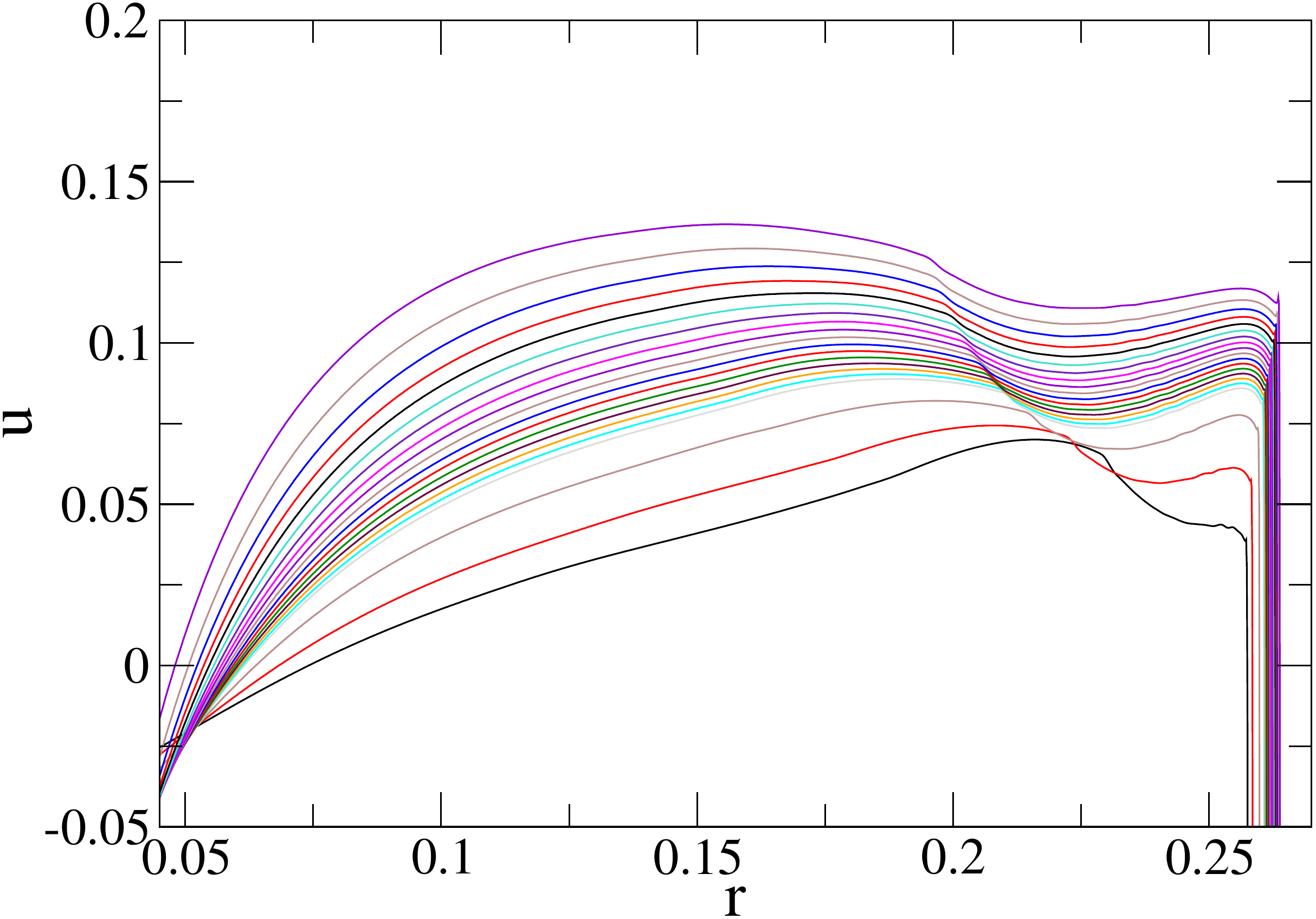} }
\caption{
Velocity versus radius  in the halo  for $t_s < t < t_*$ (the
maximum increases
with time).
 }
\label{Fig:post-Painl-uhalo}
\end{figure}

First, we note that the velocity and density diverge only in
the core while they remain
finite in the halo, as  illustrated in Figure \ref{Fig:post-Painl-uhalo}
which
zooms in Figure \ref{Fig:post-Painl-u}-(a) in the halo region. Moreover,
during the time interval  $t_s < t < t_{*}$, the halo gains
a radial extension
of about $20\%$, a small evolution compared to the strong shrinking of the
core. These two observations seem to indicate that the
expansion of the halo is still in a
preliminary stage when the core collapses.  
However, one may ask if the  solution  is self-similar in the halo before the core
collapse (or if it will become self-similar  after the collapse,  a property
investigated  in the next section, while not  studied numerically). 
Looking at this possibility, we search for a self-similar solution for a dilute 
medium by neglecting the self-gravity, so the Euler equations (\ref{e1}) and
(\ref{e2}) reduce to 
\begin{eqnarray}
\label{exp1}
\frac{\partial\rho}{\partial t}+\frac{1}{r^2} \frac{\partial}{\partial r}  (r^2
\rho u)=0
\end{eqnarray}
and
\begin{eqnarray}
\label{exp2}
\frac{\partial u}{\partial t}+ u\frac{\partial u}{\partial r}=-
\frac{1}{\rho}\frac{\partial P}{\partial r}.
\end{eqnarray}
We assume that the equation of state is purely isothermal\footnote{Since the
initial
density is larger than unity in the main part of the halo,  we shall consider that 
the equation of state (\ref{mp1})  may be approximated by equation (\ref{exp4}).
This was a
problem to build equilibrium solutions with a finite mass but this is not a
problem if we consider dynamical solutions.} 
 so that
\begin{eqnarray}
\label{exp4}
P=\rho T(t).
\end{eqnarray}
The pressure  increases
because the temperature $T(t)$ increases with time 
when the core shrinks self-similarly, as described just above 
if $k$, an 
even number, is larger than $6$ in equation (\ref{Tself}). We take
\begin{eqnarray}
\label{exp4b}
T(t)\propto (-t)^a\qquad (a<0).
\end{eqnarray}
We neglect the size of the core as compared to the size of the halo
(this is marginally valid since $r_h=0.05$ and
$r_c=0.26$).

We first 
look for
a
self-similar solution of the form
\begin{equation}
 \rho(r, t) = (- t)^{\beta} R\left(r(-t)^{\beta/\alpha}\right)
\mathrm{,}
\label{selfrho}
\end{equation}
\begin{equation}
u(r, t) = (- t)^{\delta} U\left(r (-t)^{\beta/\alpha}\right)
\mathrm{.}
\label{selfu}
\end{equation}
The exponents are linked by the
relations
\begin{eqnarray}
\label{eq:sc1}
\beta/\alpha + \delta +1=0,
\end{eqnarray}
\begin{eqnarray}
\label{eq:sc2}
\delta -1= a + \beta/\alpha.
\end{eqnarray}
Assuming that the mass in the halo $M_h=4\pi\int
\rho(r',t) r'^2\, {\mathrm{d}}r'$
is approximately constant during this short time interval (an approximation not
really
fulfilled here where matter comes from the inside layer,
see  Figure \ref{Fig:post-Painl-uhalo}   
where   the zero-velocity radius 
$r_{h}(t)$ decreases from $0.075$ to $0.05$), we
obtain
\begin{eqnarray}
\label{eq:sc4}
\alpha= 3.
\end{eqnarray}
Then we find
\begin{eqnarray}
\label{eq:sc4b}
\beta= -3\left (\frac{a}{2}+1\right ),\qquad  \delta=
\frac{a}{2}.
\end{eqnarray}
Using the value $a=-1/24$ from equation
(\ref{eq:alpha}), we obtain a self-similar
solution which diverges at $t_{*}$, the density  increasing as $
(-t)^{-47/16}$, the velocity as  $(-t)^{-1/48}$, and the radius shrinking  to
zero as $ (-t)^{47/48}$. This clearly 
disagrees with the numerical results
where the dimension of the halo increases by a factor $1.2$,   the velocity 
barely increases and the density decreases.  To explain the irrelevance of the
above scalings for our model, we
have to notice that they are derived within the hypothesis that the pressure
dominates over gravity, an hypothesis that could be irrelevant at this stage
because the halo is not  yet dilute enough. Precisely, the density in the inner
part of the halo is  approximately equal to $30$  (which is the initial  density
at $r=0$  in the canonical case where gravity dominates)  and the gravitational
attraction by the core is not negligible with respect to the self-gravity forces
in the halo because the mass of the core is approximately equal to the mass in
the halo, see Figure \ref{fig:equil}-(b),  and the halo is still close to the core:
its internal radius ($r_{h} \sim 0.05$) is noticeably smaller than its size
($r_{c} - r_{h }= 0.21$).

To go further in the investigation 
of self-similar solutions in the dilute gas, and  motivated by the linear
behavior of the velocity profile for $t \simeq t_{s}$  in Figure
\ref{Fig:post-Painl-uhalo}, we 
 perform a more precise study of the 
self-similar solutions of equations (\ref{exp1})-(\ref{exp2})
 of the form
(\ref{selfrho}) and (\ref{selfu}) 
under the assumption
that the velocity increases linearly with the distance, namely with the ansatz
\begin{eqnarray}
\label{ansatz}
\rho({\bf r},t)=\frac{M}{R(t)^3} f\left\lbrack \frac{\bf r}{R(t)}\right
\rbrack,\qquad
{\bf u}({\bf r},t)=H(t){\bf r}.
\end{eqnarray}
This study is reported  in  Appendix \ref{sec_a} for an equation of state
$P=K(t)\rho^{\gamma}$. There, we show  that the radius obeys the differential
equation 
\begin{eqnarray}
\label{radius}
\ddot R R^{3\gamma-2}  =(t_{*}- t)^a.
\end{eqnarray}
 In Appendix \ref{sec_c1}  we consider first solutions of the form
\begin{eqnarray}
\label{7ab}
R(t)=A(t_{*}-t)^{q}.
\end{eqnarray}
For $\gamma=1$,   we show
 that
  a solution of the form (\ref{ansatz}) exists only for $a<-2$. 
Therefore  this solution
with $u\propto r$ is incompatible with our numerics where we found $a=-1/24$.
This result confirms that a gravity-free self-similar solution  of the form
(\ref{selfrho})-(\ref{selfu}) is not appropriate to
describe the dynamics of the halo during the strongly nonlinear regime before
the core collapse.

On the  other hand, assuming that gravity dominates pressure
does not fit the numerics as well (because this hypothesis gives  the same
exponents as the ones found for the collapse). Finally, assuming that gravity
and pressure forces are of the same order of magnitude leads to the exponents
values $\alpha= - \beta=2$ and $\delta=a/2$ which do not fit our numerical
results. 
We conclude that the halo does not follow a self-similar evolution of the form
of equations (\ref{selfrho}) and (\ref{selfu}) in 
the pre-collapse regime, whatever is the ratio between the
gravity  and the pressure forces.


Secondly, in Appendix \ref{sec_c2}, we consider another
type of  gravity-free
self-similar solution which also obeys (\ref{radius}),  but  we replace
equation (\ref{7ab})  by the condition that the radius of the star $R(t)$  
  increases
and tends to a
constant $B$ at $t_*$, i.e., 
\begin{eqnarray}
\label{Rcst}
R(t)=B+\epsilon(t)  \qquad   \textrm{with} \qquad \epsilon(t)_{t\to t_{*}} \to
0.
\end{eqnarray}
In that case,  we show that the solution exists and is valid provided that $-2 <
a <0$. The 
velocity of expansion  $\dot R$ takes  a finite value at  time $t_{*}$ when
$a>-1$, and the density decreases as $t\to t_{*}$. These two properties agree
with our numerical results, Figure \ref{Fig:post-Painl-uhalo}, although the 
starting hypothesis   $u=H(t)r$ is definitely not fulfilled. 
In summary a  gravity-free self-similar solution associated to diverging 
temperature and non-diverging radius exists, but it is irrelevant  to describe
the  dynamics of  the halo in the pre-collapse regime because it supposes that
the velocity increases linearly with the radius for $t_{s} <t<t_{*}$, which is
not observed in  our simulation.


\section{Self-similar dynamics just after the
singularity (post-collapse regime)}
\label{sec:post-coll}

In this Section, we present self-similar solutions for the core and the halo
just
after the singularity time $t_{*}$ (as in the previous Section, we take it as
the origin of time). In the core region, we assume  that
gravity forces overcome pressure forces, as it was stated in the pre-collapse
regime. On the other hand, we propose a
self-similar solution for the halo which is based on the opposite assumption
(pressure overcoming gravity).  This solution may
be valid soon after the singularity,  when the star is very hot and the
pressure in the halo is larger than the gravity because the expansion already
took place. We did not perform any
numerical simulation
to check whether
these self-similar  solutions agree with the MEP model, particularly because of
the formation of a singularity at $r=0$ (Dirac peak) in the core plus
the lack of knowledge on  the temperature evolution  in the halo (assumed
here of the form  $T(t)\propto t^{a}$ where the exponent $a>0$ is unknown).

\subsection {Solution in the core domain }
\label{sec_scd}

We outline here the derivation of the  solution in the core which is similar to
the solution of the CEP model but with a different value of the exponent
$\alpha$ ($\alpha_{\rm MEP}=48/19$ instead of $\alpha_{\rm CEP}=24/11$). We
emphasize that the density does not
write as a Dirac distribution at the singularity time $t_{*}$, but as a power law $\rho(r,0) \propto
r^{-\alpha}$ which yields  a mass equal to zero at the center because the mass integral converges at $r = 0$ for $\alpha< 3$. 

At very short times after the collapse, we assume that the inward motion follows
a free fall dynamics in the core region. The situation is then qualitatively the
same as in the CEP model, and  looks (mathematically) like the one of the
dynamics of the Bose-Einstein
condensation where the mass of the condensate begins to grow from zero
{\it{after}} the time of the singularity
\cite{BoseE,bosesopik}.  A self-similar
solution exists which is the one derived in \cite{epje} but for the value of the
exponent $\alpha$ found here, equation (\ref{eq:alpha}). 
We recall that the main change with respect to the pre-collapse study amounts to
adding to the equations of density and momentum
conservation, an equation for the mass at the center $M_c(t)$ (with $M_c(0) = 0$).  The mass flux across a sphere of radius $r$  being $ J = 4 \pi r^2 \rho(r) u(r)$, the
equation for $M_c(t)$ is
\begin{equation}
\frac{d{M}_{c}}{dt} = \left[-4 \pi r^2 \rho(r) u(r)\right]_{r \to 0}
\mathrm{.}
\label{eq:Mc}
\end{equation}
Therefore, the equations one has to solve now are the same as before,
\begin{equation}
\frac{\partial \rho}{\partial t} + \frac{1}{r^2} \frac{\partial}{\partial
r}\left( r^2 \rho u \right) = 0
\mathrm{,}
\label{eq:Euler.1+}
\end{equation}
\begin{equation}
\frac{\partial u}{\partial t} + u \frac{\partial u}{\partial r} = - \frac{G
M(r,t)}{r^2}
\mathrm{,}
\label{eq:Euler.2+}
\end{equation}
plus the mass inside a sphere of radius $r$ 
 \begin{equation}
M(r,t)  = 4 \pi \int_0^r {\mathrm{d}}r' r'^2 \rho(r',t) + M_c(t)
\mathrm{.}
\label{eq:Euler.2.1+}
\end{equation}
These equations after singularity include the whole set of
equations leading to
the singularity. Moreover, the solution at $t_{*}$ has the same asymptotic
behavior on both sides of the singularity. It follows that the scaling laws are
the same before and after $t_{*}$. At very short
times after $t_{*}$ taken as
the origin of time, only the solution very close to $r = 0$ is changed by the
occurrence of a finite mass at $r = 0$ which is very small. We look for a
self-similar solution of equations (\ref{eq:Euler.1+})-(\ref{eq:Euler.2.1+}) 
with $\rho(r, t)$
and $u(r,t)$ having the same exponents as before
collapse:
 \begin{equation}
\rho(r,t) = t ^{-2} R_+(r t^{-2/\alpha}) \mathrm{,}
\end{equation}
\begin{equation}
u(r,t) =
t^{-1+\frac{2}{\alpha}} U_+(r t^{-2/\alpha}) \mathrm{,}
\end{equation}
plus another scaling for $M_c(t)$:
\begin{equation}
M_c(t) = K_M t^b
\mathrm{,}
\end{equation}
where $\alpha=48/19$ and $b$ is a positive
exponent to be found.
The two terms on the right-hand side of equation
(\ref{eq:Euler.2.1+}) are of the same order of magnitude with respect to $t$ if
\begin{equation}
b = \frac{6}{\alpha} - 2,
\end{equation}
a positive exponent as it should be (recall the condition
that $\alpha$ is less than 3). For $\alpha=48/19$, we get $b=3/8$.
Therefore,  the mass at $r=0$ and  the core radius evolve with
time $t$ (positive) as
\begin{equation}
M_c(t) = K_M t^{\frac{3}{8}},  \qquad R_{c}(t) \sim  t^{\frac{19}{24}}, 
\label{eq:singsc}
\end{equation}
in this self-similar post-singularity regime. We refer the reader to
\cite{epje} 
for additional information about this self-similar solution (see, in particular,
the explicit analytical solution
given in Appendix B of \cite{epje}).

The evolution of the temperature of the system in the
post-collapse regime
where the core is a mathematical singularity (Dirac peak) is not clear.
Indeed, the divergence of the potential energy of the core would imply
an infinite temperature (for global energy conservation). However,
if we replace the singular core by a relativistic compact object
such as a neutron star, we can get an estimate of the temperature by the
relation $k_B T\sim M_c\, c^2$ leading to the scaling
\begin{equation}
T(t) \propto  t^{{3}/{8}}.
\label{tev}
\end{equation}
We shall consider this law of evolution of the temperature in
the following section and
in Appendix \ref{sec_b}.

\subsection{A self-similar solution for the halo}
\label{subsec:phcsol}

We assume that the energy released during the collapse of the core heats the halo and provides its expansion. Indeed, as the
gravitational energy $W$ of the core decreases (and becomes
very negative), the
temperature $T$ of the halo and its macroscopic kinetic energy $E_{\rm kin}$
increase and become very large
($T\sim E_{\rm kin}\sim -W$) as a result of energy conservation.
Therefore, the pressure inside the halo can be high enough to accelerate its
expansion. More
precisely, we assume that  the pressure forces in the halo are stronger than the
gravity forces, a condition which will be checked {\it in fine}. For the sake of
generality, we assume that the halo has a
polytropic equation of state of the form
\begin{eqnarray}
P=K(t)\rho^{\gamma},
\end{eqnarray}
where $K(t)$ is an increasing function of time.
The isothermal case is recovered for $\gamma=1$ and $K(t)=T(t)$. 

In Appendix \ref{sec_a} we show that a self-similar
solution exists within such a frame, a question 
which is interesting from a mathematical point of view in addition to its
potential applicability to the expansion of the halo in the supernova problem. 
The self-similar solution has
a Tsallis \cite{tsallis} invariant density profile (reducing to a Gaussian for
$\gamma=1$) with
a typical radius $R(t)$ and a velocity field
which increases linearly with the radius $r$. The halo expands with
time,
as expected, its size $R(t)$ evolving according to the second order differential
equation
\begin{eqnarray}
\label{eq:phc2}
\ddot R R^{3\gamma-2}  = K(t).
\end{eqnarray}
Equation (\ref{eq:phc2}) shows that the  expansion rate $\dot R(t)$
is time dependent, contrary to what is generally
admitted in the first stage of the expansion, an important point which is
discussed below. The case where $K(t)=K$ is independent of time corresponds
to 
\begin{eqnarray}
\label{eq:phc2b}
\ddot R  = \frac{K}{R^{3\gamma-2}}.
\end{eqnarray}
This equation is similar to Newton's law for a particle in a
potential of the form $V(R)=[K/3(\gamma-1)]R^{-3(\gamma-1)}$. It is studied in
Ref. \cite{prep} by analogy with cosmological models (the isothermal case
$\gamma=1$ is treated specifically in Appendix \ref{sec_k} of the present
paper and asymptotic results valid for an arbitrary index $\gamma$ are also
given in that Appendix). Here, we
assume that $K(t)=t^{a}$, where $a$ can be of any sign for the sake of
generality. In that case, 
the radius of the halo obeys the differential equation
\begin{eqnarray}
\label{eq:phc3}
\ddot R R^{3\gamma-2}=t^{a}.
\end{eqnarray}
The asymptotic behavior of the solution of this equation is studied in
Appendix \ref{sec_b}. Below we illustrate some particular behaviors
numerically.

\begin{figure}[htbp]
\centerline{
(a) \includegraphics[height=1.5in]{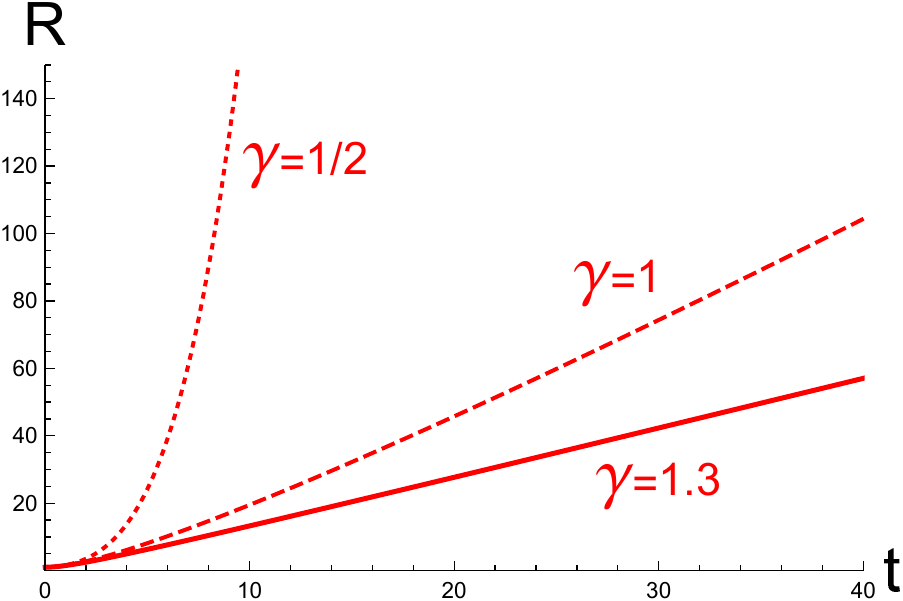}}
\centerline{
(b)\includegraphics[height=1.5in]{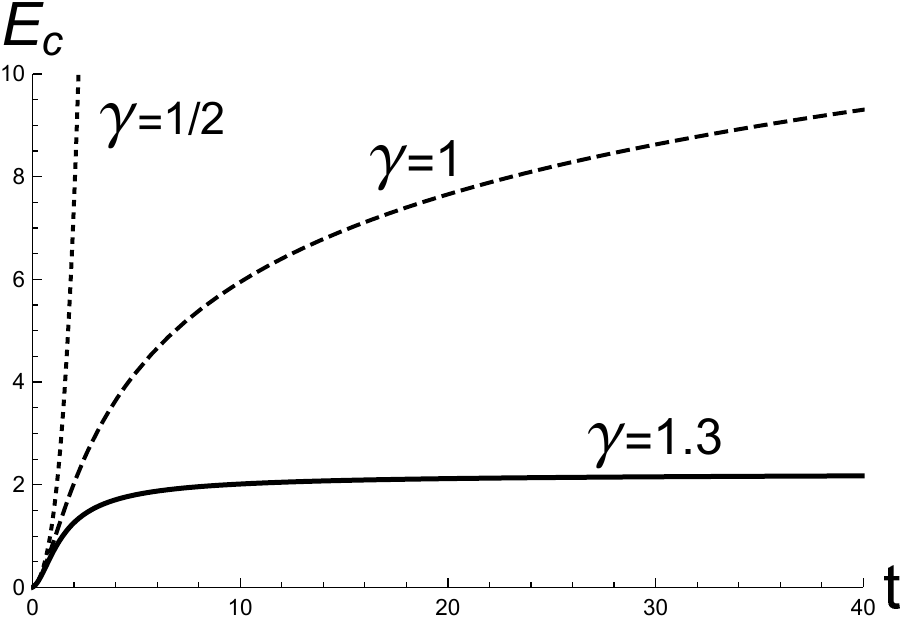}
}
\caption{Solution of equation (\ref{eq:phc3}) for the case of an adiabatic
expansion ($a=0$) with initial conditions $R(0)=1$ and $\dot R(0)= 0.1$.   (a)
Radius of the halo $R(t)$;  (b) Kinetic energy $E_{\rm kin}(t)\sim \dot
R(t)^{2}$. In (a) the solid line
for $\gamma=1.3$ shows an asymptotic constant rate,
the dashed line for $\gamma=1$ displays an
expansion rate increasing with time as $ t \log(t)^{1/2}$, and the dotted line
for $\gamma=1/2$ evolves as $R(t)\sim t^{4}$ asymptotically.  Same legend for
curves (b).}
\label{Fig:phcs}
\end{figure}

\begin{figure}
\centerline{
(a)\includegraphics[height=1.5in]{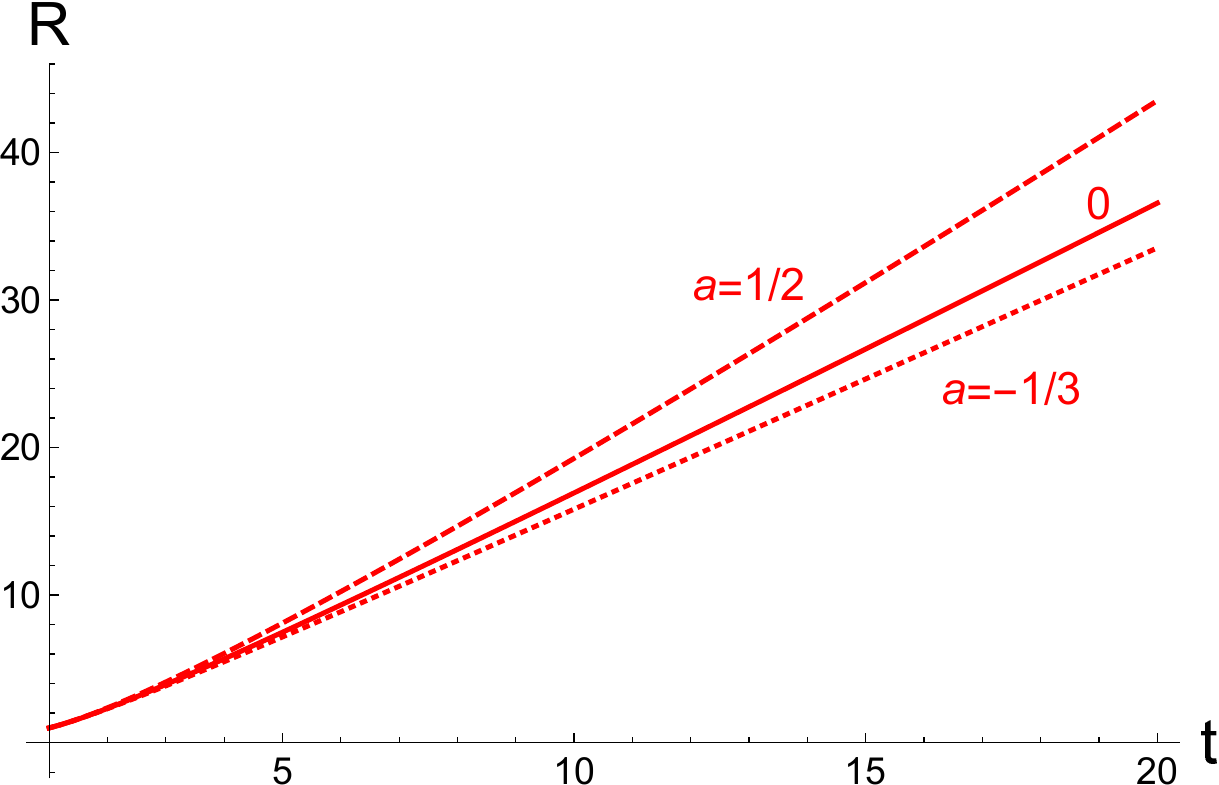}}
\centerline{
(b)\includegraphics[height=1.5in]{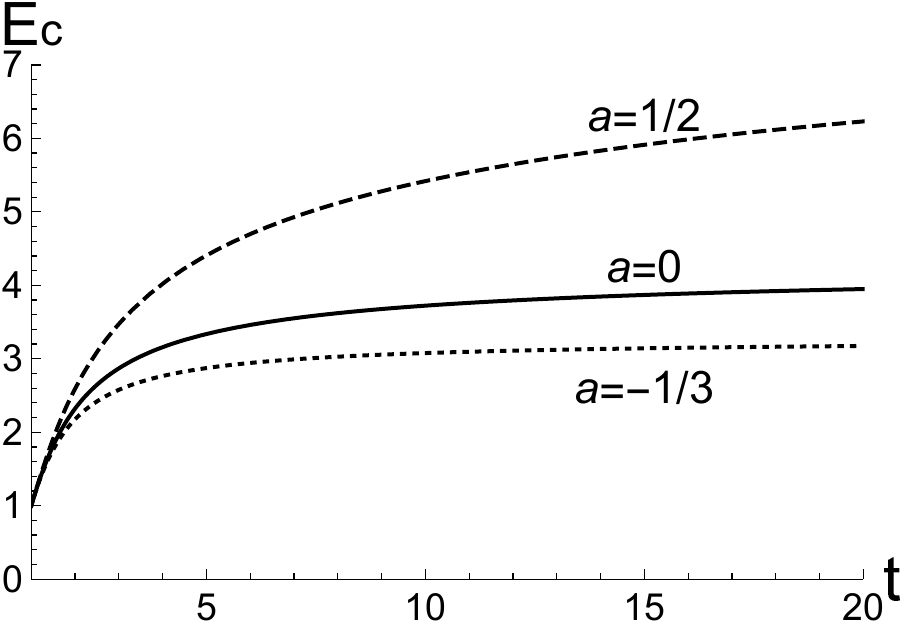}
}
\caption{Solution of equation (\ref{eq:phc3}) versus time with
initial conditions $R(1)=\dot{R}
(1)=1$ 
 for  $\gamma=1.2$.
(a) Radius of the halo $R(t)$; (b) 
Kinetic energy $E_{\rm kin} \propto \dot{R}(t)^{2} $. We have taken 
$a$ equal to $-1/3$ (dotted), $0$ (solid), and $+ 1/2$ (dashed), respectively.
In each case the kinetic energy tends to a constant for $t\rightarrow +\infty$.}
\label{Fig:PHCsol}
\end{figure}

Numerical solutions of equation (\ref{eq:phc3})
are presented  in Figures \ref{Fig:phcs}-(a)  and \ref{Fig:PHCsol}-(a)
respectively for several values of the exponents $a$ and $\gamma$ chosen for
their role in the late time dynamics.  In all cases the expansion rate $\dot R$
is clearly time dependent, see the curves of Figures \ref{Fig:phcs}-(b) and
\ref{Fig:PHCsol}-(b) which display the kinetic energy
$E_{\rm kin}\propto \dot R ^{2}$. This
result differs from the common
description of the
remnant motion just after the explosion (supposed to expand with a constant
velocity due to the conservation of kinetic energy). We shall return to this
so-called ``free expansion regime'' in the next Section. Here, we look if
there is a range of parameter  values ($a$ and $\gamma$) such that the
self-similar
solution has an expansion rate which tends \textit{asymptotically} to a constant
value. In Appendix \ref{sec_cst} we show that such an asymptotic
solution exists, and fulfills our assumptions that pressure is stronger than
gravity, provided that 
\begin{eqnarray}
\label{range}
\frac{3+a}{3} < \gamma <  \frac{4+a}{3}.
\end{eqnarray}
In that case, the  asymptotic behavior  of the velocity is
given by 
\begin{eqnarray}
\label{drt}
\dot{R}(t)\simeq v + \frac{1}{v^{3\gamma-2}}
\frac{t^{-3(\gamma-1)+a}}{a-3\gamma+3}+...\;\; (t\rightarrow +\infty).
\end{eqnarray}
Other solutions with a different asymptotic behavior, that are valid for values
of $\gamma$ in a range different from equation (\ref{range}), are given in
Appendix \ref{sec_b}. Assuming an adiabatic expansion and an homogeneous entropy
inside
the halo,
amounts to considering the case $a=0$ (see the next Section). In
that case, the condition of validity of the solution (\ref{drt}) corresponding
to $R\sim vt$ is
$1<\gamma<4/3$. For example, for $\gamma=1.3$, we can check on  Figure
\ref{Fig:phcs} that the asymptotic expansion rate is
constant. By contrast, for $1/3<\gamma<1$ the radius increases as $R(t)\propto
t^{2/(3\gamma-1)}$ (see Appendices \ref{sec_kb} and \ref{sec_ut}) and for
$\gamma=1$ it increases
as $R(t)\propto t\sqrt{\ln t}$ (see Appendix \ref{sec_ka}). More
generally, for $\gamma=1$, the asymptotic expansion rate is constant when
$-1<a<0$ while the radius increases as $R(t)\propto
t^{(a+2)/2}$ when $a>0$. In
particular, for $\gamma=1$ and
$a=3/8$ (see the end of Sec. \ref{sec_scd}), the radius increases as
$R(t)\propto
t^{19/16}$ (see Appendix \ref{sec_ut}) which is not very far from  a linear
behavior.\footnote{The linear behavior characterizes  what is
generally called  
free expansion, understood as the propagation of the remnant with constant
kinetic energy, at the very beginning of the expansion  when the pressure of the
interstellar gas is negligible, before accumulated mass of this gas affect the
expansion.}
Note that for
an ideal gas the exponent $\gamma$ is
equal to
unity for an isothermal transformation only, otherwise one has $\gamma>1$. This
is because an adiabatic transformation implies
$\gamma=c_{p}/c_{v}$ with $c_{p}$ larger than $c_{v}$,  and other
transformations (called ``polytropic'') are intermediate between adiabatic and
isothermal. 
Therefore we chose a value of $\gamma$ larger than unity to illustrate the role
of the exponent $a$ on the dynamics of the solution. The solution of equation
(\ref{eq:phc3})  is shown in Figure \ref{Fig:PHCsol}  for $\gamma=1.2$ and 
various values of the exponent $a$ which satisfy the condition
(\ref{range}). 
In addition to the adiabatic case ($a=0$), we have chosen a positive ($a=1/2$)
and a
negative ($a=-1/3$) value of the
exponent $a$.
In all cases, the late time dynamics displays the condition of asymptotic
constant kinetic energy, but with different time scales. In
Figure \ref{Fig:PHCsol}-(b), the
velocity tends to a constant plus a term decreasing as $t^{-0.93}$, $t^{-0.6}$,
and $t^{-0.1}$ for the dotted, solid and dashed line respectively.
Note that if the self-similar solution proposed here could describe the first
stage of the expansion of the halo,  we expect that it  merges ultimately with
the non self-similar Burgers solution suggested in
Section \ref{sec:free exp}
which displays shocks.

\section{Isentropic expansion of the halo and  shock formation}
\label{sec:free exp}

\subsection{Physics of the free expansion stage}
\label{sec:general-post}

Here, we consider the expansion of the remnants, namely the gas ejected by the
explosion of a supernova. We shall assume that the remnant is a dilute
gas, although much denser than the interstellar medium. We consider a
general equation of state but,
 as we show below, the pressure  is irrelevant
for the expansion of a dilute gas. The study of the remnants is of interest
because it represents the last stage of the evolution following a supernova
explosion which should be matched with the previous stage of the supernova. In
addition,  those remnants have been observed both on SN1987A and for a number of
supernovae having exploded in our Galaxy not too long ago.  Lastly, this study
is of interest also because it is one of the few instances where a lab-model
could be looked at on Earth with some relevance for an astrophysical
problem. 

We shall neglect a set of perhaps crucial phenomena, namely plasma effects due
to the finite electric conductivity of the expanding gas (this yields Laplace
forces which could supersede inertia and gravity in the expanding gas). We
shall also assume spherical symmetry, not displayed by the observed
remnants, except for their large scale structure. As shown later, asphericity is
not so crucial from the point of view of the present analysis.  

We consider what is called sometimes in the literature the ``free expansion
stage'' following the supernova explosion, when the ejecta are believed to expand freely in space and cool adiabatically because of their expansion. 
This stage begins after the formation of the neutron star (if
any is formed) and when the temperature, pressure and density of the remnants
are  small enough, see below. It should last until the density of the remnants
becomes of the order of magnitude of the density of the interstellar gas. 
It seems that this is considered
as a rather uneventful stage of the expansion, although we
believe  that this stage of adiabatic cooling is of primary interest.
 Here, we point out 
that  in order to create structures in an expanding gas volume, as observed in
the remnants, there is no need to have interaction with an outside interstellar
gas because shocks can occur even within the dynamics of the expanding gas. As we shall
show, this occurrence of shocks depends on the initial distribution of the fluid
velocity \emph{inside} the remnants: if the radial velocity is larger for a
given radius than for a larger one, a shock forms because the larger velocities
overcome the slower ones. This shock has its own dynamics related to the
conservation relations. Lastly, this early stage of the expansion is, by far,
the one that is the best known experimentally because it is a stage where the
remnants are still far more luminous than the rest of the Galaxy. Based on the
existence of such internal shock waves, we suggest an explanation for the very
sharp luminous rings observed in the remnants of SN1987A. Therefore, we believe
it is of interest to try to understand this ``free expansion stage". 

A basic question concerning the free expansion is the validity of its fluid
mechanical description. This question concerns the expanding matter, but could
also concern the interstellar medium. 
The validity of such a picture requires that the mean free path of molecules,
ions, electrons and atoms in the 
expanding gas  (or in the interstellar medium) is much smaller than the length scale of the structure under consideration. If this is not the case, one has what is called a dust gas, without interaction between the particles other than possibly gravitation. In such a dust gas, the conservation of entropy does not occur, velocity fields with more than one value at a given point are perfectly possible,  and shock waves are absent. 
Here,  we assume that the expanding matter is dilute, but not infinitely dilute,
with a mean free path far smaller than its size. Therefore, it  can be described
as a fluid at very large Mach number, in the sense of regular fluid mechanics,
namely with a single valued velocity field.
The constraint of single valued 
velocity field explains the formation of shock waves inside this expanding gas, before any  interaction with the interstellar medium, as discussed below. 

For the interstellar medium, a fluid mechanical description is questionable.
According to Spitzer \cite{Spitzer1,Spitzer2}, its mean free path  in the usual
sense is
of the order of the size of the Galaxy. It is about $500$ pc for a $2$ MeV
proton moving in a gas of neutral hydrogen atoms of density of order one atom
per cubic centimeter, that should forbid to consider it as a fluid. However,
Spitzer notices that, because of the existing magnetic field in the Galaxy, the
gyration radius of protons is far smaller than their mean free path, which could
reduce by orders of magnitude the mean free path of protons. This could well be,
but we must notice that charged particles move freely along the lines of the
magnetic field making this reduction of the mean free path not so efficient.
Moreover, as noticed by Spitzer, the energy density of the galactic magnetic
field is well below the one of the expanding gas so that it is not clear that
the electric currents due to the galactic magnetic field are able to slow down
this expanding gas.

Another question is the interaction between the molecules of the expanding
bubble. As it expands, its density decreases and it should enter in the
so-called
Knudsen regime where the mean free path of atoms and molecules becomes of the
same order or bigger than its radius $R$.  This stage comes quite late in the
expansion, as shown by the following simple estimate: the radius of the blob is
of order $(M_h/ \rho)^{1/3}$ although the mean-free path $l$ is of order $ l = m
/
(\rho\sigma)$ where $m$ is the mass of the atoms making the gas and  $\sigma$ is
the cross section for the collisions. Therefore, as the density decreases, the
mean free path should become larger than the radius of the gas blob, forbidding
to describe this gas by the equations of fluid mechanics supplemented by the
thermodynamic relations. To put this condition in a dimensionless form, let us
introduce the quantity 
 \begin{equation}
 \rho_{\rm Kn}= m \sigma^{-\frac{3}{2}}
\textrm{,}
\label{eq:kn1}
\end{equation}
which is the mass density of dense matter, the inter-particle distance of which
is of order of $\sigma^{1/2} $. The mean free path becomes of the order of the
radius of the cloud when its density is such that 
 \begin{equation}
 \rho =  \rho_{\rm Kn} \left(\frac{m}{M_{h}}\right)^{\frac{1}{2}}
\textrm{,}
\label{eq:kn2}
\end{equation}
a very low density compared to usual densities of condensed matter,  $m/M_{h}$
being like the inverse number of atoms in the expanding cloud, surely a very
small number.

In summary,  we consider below the post-explosion regime where the expanding gas
is dense enough to be interacting with itself (the attraction by the
core of the exploded star will also be considered), but not with the
interstellar medium, and such that it makes a continuous fluid, not a Knudsen
gas.

\subsection{Difference with the Sedov-Taylor problem}
\label{sec:self-sim no g}

In this regime, we first point out that  a self-similar expansion with  gravity and pressure forces included in the equation has to be rejected.   Many authors
invoke the Sedov-Taylor solution (see Section 106, p. 403-406, of \cite{ll}
and \cite{sedov2,sedov3,sedov4}) to describe the second phase of
the expansion of the remnant. But the Sedov-Taylor solution was derived
for the expansion of
an explosion releasing energy in another gas.  This solution  conserves energy only, while  mass
conservation does not enter into the solution because the initial mass is mixed
with the infinite mass around. Therefore Sedov-Taylor
 is not suitable here because, in our description of 
the free expansion stage,  the remnant is an entity  which  exchanges neither
energy nor mass with the interstellar medium. Then we have to impose the
constraints of conservation of mass and energy for the remnant.
 In our case, self-similarity of the \textit{ free
expanding bubble}, if it exists, requires to neglect some physical effects, as
it was done for the description of free fall of dense molecular clouds where the
pressure forces were assumed to be negligible with respect to the gravity
forces. 

Below,  we consider two cases, first when gravity is negligible compared to
pressure, secondly when both gravity and pressure are negligible. In the former
case, we show that the conservation of mass and energy is compatible
with a self-similar solution only for the case $\gamma=1$  which corresponds to
an isothermal process, not to an adiabatic one. This solution must
be rejected since $c_{p}=c_{v}$ is unphysical
for a dilute gas. In the latter case, we  derive rather straightforwardly 
 an approximation of the fluid equations of Burgers-type, which is well-known to
yield shocks. This equation has a simple solution which can be extended beyond
situations of perfect spherical symmetry.  This solution  differs from the one
of  free fall of a dust gas in a few points. First, the Burgers solution is not
self-similar and 
depends  on the initial conditions contrary to the free fall solution.
Secondly, if those conditions are such that a shock wave is
created, the subsequent evolution couples mass, energy and momentum
conservation
by the Rankine-Hugoniot relations which link the flux across the shock waves and
yield ultimately their trajectory. Somehow, dynamics of mass and energy (the two
being linked by the
adiabatic condition) are enslaved to the velocity field when this one is
smooth (before the shock), but actively enters into the dynamics when
shock waves are formed.

The equations for an inviscid compressible ideal fluid in
spherically symmetric situations, including self-gravitation, read (see
Section 6 of \cite{ll}):
\begin{equation}
r ^2 \frac{\partial \rho}{\partial t} + \frac{\partial (\rho u r^2) }{\partial r} = 0  
\textrm{,}
\label{eq:densite}
\end{equation}
\begin{equation}
\frac{\partial u}{\partial t} + u \frac{\partial u }{\partial r} = - \frac{1}{
\rho} \frac{\partial P}{\partial r}  -  \frac{4\pi G}{r^2} \int_0^r 
\mathrm{d}r' r'^2 \rho (r')
\textrm{,}
\label{eq:moment}
\end{equation}
\begin{equation}
r ^2 \frac{\partial}{\partial t} \left ( \frac{1}{2} \rho u^2 +  \rho
\varepsilon\right ) + \frac{\partial}{\partial r}\left\lbrack \rho u r^2 
\left (\frac{1}{2} u^2 + w\right ) \right\rbrack = 0  
\textrm{,}
\label{eq:energy}
\end{equation}
where $\rho$ is the mass density, $u$ the radial velocity, $ \varepsilon $ the
internal energy per unit mass, $w =  \varepsilon +p/\rho$ the enthalpy per unit
mass and $G$ is Newton's constant. 
Equation (\ref{eq:energy}) can be transformed into the condition that the flow
is isentropic 
\begin{equation}
 \frac{\partial (r ^2 s \rho)}{\partial t} + \frac{\partial}{\partial r}(u r^2 s \rho) = 0  
\textrm{,}
\label{eq:entropy}
\end{equation}
where $s$ is the entropy per unit mass such that $d\varepsilon = T ds
+(p/\rho^2) d\rho$. To take advantage that the flow is isentropic, one uses as
thermodynamic variables the density $\rho$  and the entropy $s$. Therefore, the
pressure $P$ is a function of those two quantities, $P(\rho, s)$. If the heat
capacities of the gas are independent of temperature, Laplace's relation between
pressure and density reads
\begin{equation}
 P = K(s)  \rho^ {\gamma} 
 \textrm{,}
\label{eq:pressure}
\end{equation}
where $\gamma = c_p/c_v$ is the ratio of heat capacities at constant pressure
and constant volume (for an ideal gas undergoing an adiabatic process). For the
sake of generality, we did introduce a constant $K(s)$ depending explicitly on
the entropy $s$ because it is possible to have an initial state of non-uniform
entropy. However, we shall assume later that the entropy is initially uniform,
and try to find a possible self-similar solution of the above set of equations
for a gas bubble expanding {\it in vacuo}. To that purpose, we look for
solutions depending on time and radius as $F(r, t) = r^a f(r t^b)$, where
$a$ and $b$ are exponents to be derived from the equations, with $a$ depending
on the field $F$ under consideration (that is either $ \rho$, $u$ or $s$)
although $b$ is the same for all fields. Moreover $f(.)$ is a numerical function
with values of order one when its argument is of order one. 

In the frame of a perfect gas expanding adiabatically, 
equation (\ref{eq:pressure}) requires that the pressure tends to zero as the
density does.
 Moreover the pressure depends on the temperature as $P  \sim
T^{c_p/(c_{p}-c_{v})}$ and $c_{p }> c_{v}$,
which implies that the absolute temperature of the
gas tends to zero as well. Therefore, the internal energy of the gas, being
proportional to $T$, must also tend to zero.  Moreover, the
gravitational energy in equation (\ref{ae6}) tends also to zero because the
typical dimension of the halo becomes very large (see below the condition for
neglecting the gravitation with respect to pressure forces).   We shall assume
that the isentropic expansion starts after this (short) transient, when
the internal and  gravitational energies are  almost wholly converted into
kinetic energy, so that
the  density  of energy per unit mass
is of order $ \rho u^2$  after this conversion has been
achieved.  In summary, we assume that the condition of conservation
of total energy amounts to imposing that the order of magnitude of the velocity
$u$ is constant during the isentropic  expansion.

\subsection{No self-similar expansion without gravity}
\label{witug}

Let us insert the self-similar solution 
\begin{eqnarray}
\label{eq:sc-halo}
\rho(r, t) = r^a R(r t^b),\qquad u(r, t) = r^c U(r t^b), 
\end{eqnarray}
in the equations of perfect fluids with the pressure-density relation
(\ref{eq:pressure}) and neglect the gravitational term. For an expanding
solution, one must have $b \leq 0 $. From mass conservation, the integral of $
\rho(r, t) $ over the whole space must be constant, which implies 
\begin{equation}
 a=-3
  \textrm{.}
\label{eq:a3}
\end{equation}
The two terms on the left-hand side of equation (\ref{eq:moment}) are of the
same order of magnitude if the condition
 \begin{equation}
 b(1-c)=-1
\label{eq:bc1}
\end{equation}
is fulfilled. As noted above, the (reasonable) condition of conservation of
kinetic energy requires $c=0$ which ensures a constant order of magnitude of the
velocity field. In this case, the relation (\ref{eq:bc1}) reduces to
\begin{equation}
b=-1
\label{eq:bc}
\end{equation}
which means that the  size of the shell expands linearly with time,
as $R(t)=v t$. This solution corresponds to what is generally
called the ``free  (or Joule) expansion stage" in the literature
\cite{Joule}. 
Lastly, imposing that the pressure gradient
divided by $ \rho$, with $ P = K(s)  \rho^ {\gamma} $ and a constant entropy, is
minus the acceleration yields 
 \begin{equation}
b= - \frac{1 }{3\gamma - 2},
\label{eq:bgam}
\end{equation}
 which agrees with the condition that $b$ is negative if $\gamma$ is  larger
than $2/3$ .
 Equations (\ref{eq:bc}) and (\ref{eq:bgam}) are compatible for the particular
value 
 \begin{equation}
\gamma=1
\end{equation}
only. However, as mentioned in the previous Section, $\gamma$ 
 is always larger than $1$ in the case of adiabatic or polytropic processes.
Therefore, \textit{no physically meaningful self-similar solution of the
expanding halo exists for an adiabatic process}. By physically meaningful, we
refer to a solution  which conserves the kinetic energy and the entropy as
assumed above (then $\gamma$ is larger than unity).

Let  us now discuss the range of validity of our hypothesis of negligible
gravity forces for self-similar solutions  of the form  (\ref{eq:sc-halo}).
Keeping the relation (\ref{eq:bc1}) with equation (\ref{eq:bgam}), one finds
that the gravitation term scales like 
\begin{equation}
\frac{4\pi G}{r^2} \int_0^r  \mathrm{d}r'  r'^2 \rho (r') \sim r^{-2}  \sim t^{2b}
 \textrm{.}
\label{eq:scalinggrav}
\end{equation}
This is to be compared with the other terms, for exemple with the scaling of the
acceleration $ u_{,t} \sim t^{-1}$
 just derived.  For any negative value of $b$ the exponent  $-1$ of the
acceleration as a function of time is  larger than the exponent $2b$ of the
gravitational force  if  $  \gamma < 4/3$. Therefore, when looking at the limit
of large $t$, the gravitational force is negligible when $\gamma$ fulfills
the relation
 \begin{equation}
 \frac{2}{3} < \gamma < \frac{4}{3}
 \textrm{.}
\label{eq:condgam}
\end{equation}
 If $\gamma$ is bigger than $4/3$,  the gravitational attraction dominates at
large time, 
then any self-similar  solution is expected to evolve toward a collapse.

\subsection{Expansion without pressure and without gravity}
\label{sec:Burgers}
The alternative to the self-similar solution considered above is to assume that one of the terms in the dynamical equations is  negligible with respect to the others, an assumption
which changes the scaling laws.
Because the pressure tends to zero by adiabatic
expansion, it is natural to assume that the pressure term in
equation (\ref{eq:moment}) becomes small with respect to the other terms as well
as the gravitational interaction in this late stage of the expansion {\it in
vacuo}. This reduces the momentum equation to  the simple form
\begin{equation}
\frac{\partial u}{\partial t} + u \frac{\partial u }{\partial r} = 0
\textrm{,}
\label{eq:momentmod}
\end{equation}
an equation well-known since Poisson to have the implicit solution
\begin{equation}
u(r, t) = u_0 (r - u t)
\textrm{,}
\label{eq:momentmodsol}
\end{equation}
where $u_0(r)$ is the initial radial velocity. This solution conserves
the order of magnitude of $u$ in the course of time, which is consistent with
the conservation of energy. 
The solution in equation (\ref{eq:momentmodsol}) is also well-known to become
multi-valued after a finite time for a wide range of initial conditions.
However, it is easy to find initial conditions remaining single-valued forever
by choosing an initial velocity field growing uniformly as $r$ goes from $0$ to
$\infty$, as assumed in the peculiar form (\ref{ansatz}). 
An initial condition
leading to a multi-valued solution after a finite time yields actually shock
waves regularized by viscosity and heat conduction, which could well be what is
observed in the remnants of supernovae. Once the velocity field is known as well
as the initial distribution of mass density, one can find, at least by implicit
relations, the distribution of mass at any later time. 

The solution for the density takes the form  
\begin{equation}
  r^2 \rho(r, t) = \frac{\partial r_0 }{\partial r}   r_0^2 \rho_0(r_0(r, t)) 
\textrm{,}
\label{eq:momentmodsoldens}
\end{equation}
where $ \rho_0(r_0) $ is the initial condition for the density, and where $r_0$
is the function of $(r, t)$ such that $r_{,t} = u(r, t) $ with the initial
condition  $r(t = 0) = r_0$, and where $u(r, t)$ is given by
equation (\ref{eq:momentmodsol}).

At this stage, one should check that the neglected terms are actually negligible
in this limit of long times compared to what has been kept. Let us look at
equation (\ref{eq:moment}) and compare the terms on the left- and right-hand
side in the late stage of the expansion of the gas, that is when $r$ becomes
large. The term $u u_{r}$,  called dynamical pressure term, of order $u^{2}/r$,
is proportional to $1/r$ because $u$ keeps constant order of magnitude to ensure
the conservation of energy. The term $P_{,r}/\rho$ involving the
thermodynamical pressure $P \propto \rho^{\gamma}$  is of order
$\rho^{\gamma-1}/r$  with $\rho\sim M_{h}/r^{3}$, where $M_{h}$ is the mass of
the expanding cloud. That gives a term of order $1/ r^{3\gamma-2}$ decreasing
more rapidly with time (as the radius size $r$ increases) than the left hand
side term in equation (\ref{eq:moment}) because  $\gamma $ is larger than
unity. The gravitational term, namely $-  (4\pi G/r^2) \int_0^r  \mathrm{d}r' 
r'^2 \rho (r')$,  scales like  $G M / r^2$, where $M$ is the total mass, a
constant. Therefore, it decays faster than the term of dynamical pressure, by a
factor $1/r$ as $r$ tends to infinity. This shows that, as assumed, the
dynamical pressure is dominant in the regime of a dilute gas. This analysis was
based on the fluid equations for a compressible inviscid gas. If this gas
becomes highly diluted, it enters the so-called Knudsen regime discussed in
Subsection \ref{sec:general-post}.

Let us notice again that even though the equations of motion do not
include explicitly the temperature (or the entropy) this one is known from the
constraint of conservation of entropy (\ref{eq:entropy}) where the velocity
field is given by the implicit equation (\ref{eq:momentmodsol}) and the density
derived from the equation of transport of mass (\ref{eq:densite}) or
(\ref{eq:momentmodsoldens}). Somehow, one could say that the velocity field of
the expanding gas acts a little bit like a piston with an imposed motion such
that the gas expands. A related physical phenomenon is the Ranque effect
\cite{Ranque} where a gas  injected at high pressure tangentially in a cylinder
makes a very strong vortex and cools down spontaneously when extracted near the
axis of the cylinder where the pressure is low. In the Ranque effect, the
cooling occurs  not just by expansion as in the theory presented above, but
because of the centrifugal force due to the rotation of the gas inside the
cylinder.   

Because we have in mind the expansion of the mass of a star after it exploded as
a supernova, we must consider also the possible effect of a mass remaining at
the center of the star, being the rest of the core after the explosion, even
though no such dense core has been observed (yet?) in SN 1987A, the best known
supernova. This adds another gravity term in equation (\ref{eq:moment}) which
becomes 
\begin{equation}
\frac{\partial u}{\partial t} + u \frac{\partial u }{\partial r} = - \frac{1}{ \rho} \frac{\partial p}{\partial r}  -  \frac{4\pi G}{r^2} \int_0^r  \mathrm{d}r'  r'^2 \rho (r') - \frac{G M_c}{r^2} 
\textrm{,}
\label{eq:momentcore}
\end{equation}
where $M_c$ is the mass of the core, a point mass located at $r = 0$. Contrary
to the term of self gravitation, the last term, representing the attraction by
the core, diverges near $r = 0$ and so cannot be neglected anymore, at least for
$r$ small.  Let us suppose that the solution remains like the one derived
before, that is with a density decreasing to zero with time, and a radial
velocity keeping a constant order of magnitude. Comparing the kinetic energy of
a unit mass and the gravitational energy due to the attraction by the core, one
finds a critical radius $r^*$ such that if $r > r^*$ the velocity is positive
while  it is negative otherwise. This radius is $r^* = G M_c/u^2$ where $u$ is
the order of magnitude of the velocity. 
At $ r = r^*$ the velocity $u(r)$ changes sign. Because of the scalings, $r^*$
remains of the same order of magnitude, in particular because for  $r$ larger
than $r^*$ the velocity  is directed outward and so no mass is added to the
core. Therefore because there is no feeding of mass coming from outside, the
radius $r^*$ should stay constant, and the local density tends to zero as it
does for $r$ much larger than $r^*$. For $r \gg r^*$, the blob keeps expanding
as explained above because the attraction by the core becomes negligible
compared to the dynamic pressure, and the density inside the core tends to
collapse on the center, but with a negligible attraction on the expanding gas at
radii much larger than  $r^*$, where most the mass is located.  

Depending on the initial conditions $u_0(r)$ for the velocity field, the
solution of equation (\ref{eq:momentmod}) may or may not lead to a finite time
singularity. If it does not, the assumptions leading to this equation remain
correct for all positive times. If this solution becomes singular at finite
time, there is the question of the evolution after the singularity time. As
well-known since Riemann, the finite time singularity of the solutions of
equation (\ref{eq:momentmod}) is physically transformed into a solution with a
propagating discontinuity, a shock wave, once molecular transport (heat
conductivity and viscosity) is taken into account. Notice that such a shock wave
is neither the one derived from the Sedov-Taylor model of an expanding gas (the
remnant) inside an exterior medium (the interstellar medium) which is supposed
to occur at the boundary between the two media, nor  the one often referred to
in theories of supernovae, which is supposed to occur inside the star just after
the  core collapse
and is believed to play a role in the emission of matter outside. In the present
case, the discontinuity propagates also inside the medium where it was born (as
in the latter case), but the propagation occurs through an expanding rarefied
gas.  In our case, we have neglected the pressure term $- P_{,r}/\rho$  in the
momentum equation (\ref{eq:moment}), that adds  complexity to the standard
theory of shock waves. This approximation was based on the fact that the order
of magnitude of the thermodynamic pressure $P$ becomes negligible compared to
the one of the dynamic pressure (giving rise to the term $ u u_{, r}$ 
in equation (\ref{eq:moment})).

To neglect the thermodynamic pressure with respect to the dynamical pressure
amounts to taking the limit where the velocity of sound,
$c_{s}=\sqrt{p'(\rho)}$, is much less than the actual fluid velocity $u$,
equivalent to the limit of a very large Mach number 
 \begin{equation}
\mathcal{M}=\frac{u}{c_{s}}\gg 1
\textrm{.}
\label{eq:Mac}
\end{equation}
In this limit, one can use the known relations giving the ratio between the
thermodynamic parameters on both sides of a shock wave. In the present case, we
shall be concerned with the ratio of number densities. As shown in Section 89
of \cite{ll}, this ratio is, for shock waves of arbitrary Mach number in
polytropic gases, given by 
 \begin{equation}
\frac{\rho_2}{\rho_1} = \frac{(\gamma + 1) \mathcal{M}_1^2}{(\gamma - 1) \mathcal{M}_1^2 + 2}
\textrm{,}
\label{eq:ratiodeb}
\end{equation}
where the index $1$ refers to the upstream part of the shock, and $2$ to the
downstream part, both being located inside the expanding remnant. In the case of
a shock propagating outward, the index $1$ refers to the outside and $2$ to the
inside, while $u_1$ is the fluid velocity near the shock front on the upstream
side in
the frame of reference of the shock. Its order of magnitude is the one of the
fluid velocity in the expanding gas, much bigger in the low density limit than
the speed of sound  $c_1$ on the upstream side. Therefore, as already mentioned,
neglecting the thermodynamic pressure is valid in the limit of large Mach
number. In this limit, the ratio of densities across the shock takes the finite
value $(\gamma + 1)/(\gamma - 1)$, which shows that the accumulation of matter 
on the shock is limited to a finite ratio. Notice that this ratio is obviously
larger than 1 because it is the ratio of the density on the downstream side
(index 2) to the upstream side (index 1). It is equal to $7$ for a gas such
that $\gamma = 4/3$.  It would be interesting to know if a larger effect of mass
concentration happens on manifolds where the velocity is more singular than on
shock waves, like for instance near the line of merging of two shock surfaces or
at points where three shock surfaces meet.  An interesting possibility is that
such an accumulation of mass and energy increase could explain the observation
of rings in SN1987A, with a fair axial symmetry, likely due to the initial
rotation of the star. At sufficiently long time after the initial explosion, the
shock waves due to the initial conditions for the velocity field likely get an
axisymmetrical shape which could result in lines of intersection having this
symmetry and so be circles in planes perpendicular to the same axis, the
symmetry between the two thin circles being due to a symmetry with respect to
the lid plane perpendicular to the axis of rotation of the star. 

It is of interest to remark that a shock wave 
occurring in the expanding gas at decreasing density and temperature is a manner
for the system to increase its temperature. In strong shocks (see equation
(89.10) of \cite{ll}) propagating through polytropic
gases, there is a very large increase of temperature on the downstream side. The
ratio of the downstream temperature $T_2$ to the upstream temperature $T_1$  is
given by
 \begin{equation}
 \frac{T_2}{T_1} = \frac{2 \gamma (\gamma - 1)}{(\gamma + 1)^2} \mathcal{M}_1^2 
\textrm{,}
\label{eq:ratiodebb}
\end{equation}
where $ \mathcal{M}_1 $ is the large upstream Mach number. Such a large
temperature increase could well explain the observation of a light emitting part
of the remnants, particularly near their edge where the effect of an initial
velocity difference is more likely to yield a shock wave because of the
structure of the solution of the equation (\ref{eq:momentmod}). 

\section{Comparison with the canonical description}
\label{sec:can-microcan}

In this Section, we summarize the main results 
obtained in this paper, which are valid in the microcanonical ensemble (fixed
energy $E$), and we compare them with those obtained in Paper I, which are valid
in the canonical ensemble (fixed temperature $T$). As discussed in the
Introduction, the CEP model gives some results that are identical to those found
here for the MEP model, but there are also important differences.

\subsection{Series of equilibria}

First of all, we recall that the series of equilibria are the same in the
canonical and microcanonical ensembles.  They are made of all the solutions of
equations (\ref{es1}) and (\ref{es1b}), stable or unstable, corresponding to the
condition of hydrostatic equilibrium. This leads to the spiralling curve of
Figure
5 in Paper I and to the spiralling curves of Figure \ref{spi-microcan1} in this
paper. However, the stability of the solutions is different in the
microcanonical and canonical ensembles. Using the Poincar\'e theory
\cite{poincare,katz,can-microcan}, one can show that the series of equilibria is
stable in the canonical ensemble before the first turning point of temperature
and that it becomes unstable afterward. The instability occurs when the
specific heat $C=dE/dT$ becomes infinite, passing from positive to
negative values.
Furthermore, a new mode of stability
is lost at each turning point of temperature as the series of equilibria
$\beta(E)$ rotates anticlockwise. The critical point A where the
first instability occurs as $T$ decreases ($\hat{h}_0$ increases) corresponds to
a minimum of the temperature. This canonical critical point (saddle-center) has
been fully characterized in Paper I. It corresponds to $T_c^{\rm cano}=1.546$
and
$E_c^{\rm cano}=0.378$. Similarly, one can show that the series of equilibria is
stable in the microcanonical ensemble before the first turning point of energy
and that it becomes unstable afterward. The instability occurs when the
specific heat vanishes, passing from negative to positive values. Furthermore, a
new mode of stability
is lost at each turning point of energy as the series of equilibria
$\beta(E)$ rotates anticlockwise. The critical point A' where the first
instability occurs as $E$ decreases ($\hat{h}_0$ increases) corresponds to a
minimum of the energy. This microcanonical critical point (saddle-center) has
been fully characterized in Section \ref{sec:equil} of the present paper. It
corresponds to $E_c^{\rm micro}=-0.984142$ and $T_c^{\rm micro}=2.22538$. The
fact that
the onset of instability differs in microcanonical and canonical ensembles
(A $\neq$ A') is a manifestation of ensembles inequivalence for systems with
long-range interactions \cite{can-microcan}. Considering the caloric curve of
Figure \ref{spi-microcan1}-(b), we note that the region of ensembles
inequivalence
(between points A and A') occurs in the region of negative specific heats
$C=dE/dT<0$. This is natural
because we know from general arguments of
thermodynamics that the specific heat must be positive in the canonical ensemble
while there is no {\it a priori} constraint on its sign in the microcanonical
ensemble. These results regarding the caloric curve $\beta(E)$ and the notion of
ensembles inequivalence are similar to those obtained in the context of
box-confined isothermal spheres (see \cite{box,lbw,aaiso}). We also note that
the 
curves which depict the succession of equilibrium states lead to spirals
spinning inversely when the equilibrium radius of the star is plotted versus $T$
or versus $E$: compare $r_0(T)$ shown in Figure 5 of Paper I with the curve
$r_0(E)$ shown in Figure \ref{spi-microcan1}-(a) of the present paper. In the
canonical ensemble, the series of equilibria is stable until point A, so that
the radius always decreases as the temperature decreases (see Figure 5 of Paper
I). As a result, we
anticipate that the canonical description should give a contraction of the
radius of the star (collapse/implosion) after the instability point A. In the
microcanonical ensemble, the series of equilibria is stable until point A'.
The radius first decreases as the energy decreases, then, after the turning
point of radius $r_0=0.371$, energy $E=-0.80$, and temperature $T=1.97$, the
radius increases as the energy keeps decreasing (conjointly, in the region of
negative specific heat, the temperature increases as the energy decreases). As a
result, we anticipate that
the microcanonical description should give an expansion of the radius of the
star and an increase of temperature
(explosion) after the instability point A' (see Figure
\ref{spi-microcan1}-(a)). 

\subsection{Neutral mode}

\begin{figure}[htbp]
\centerline{
(a) \includegraphics[height=1.5in]{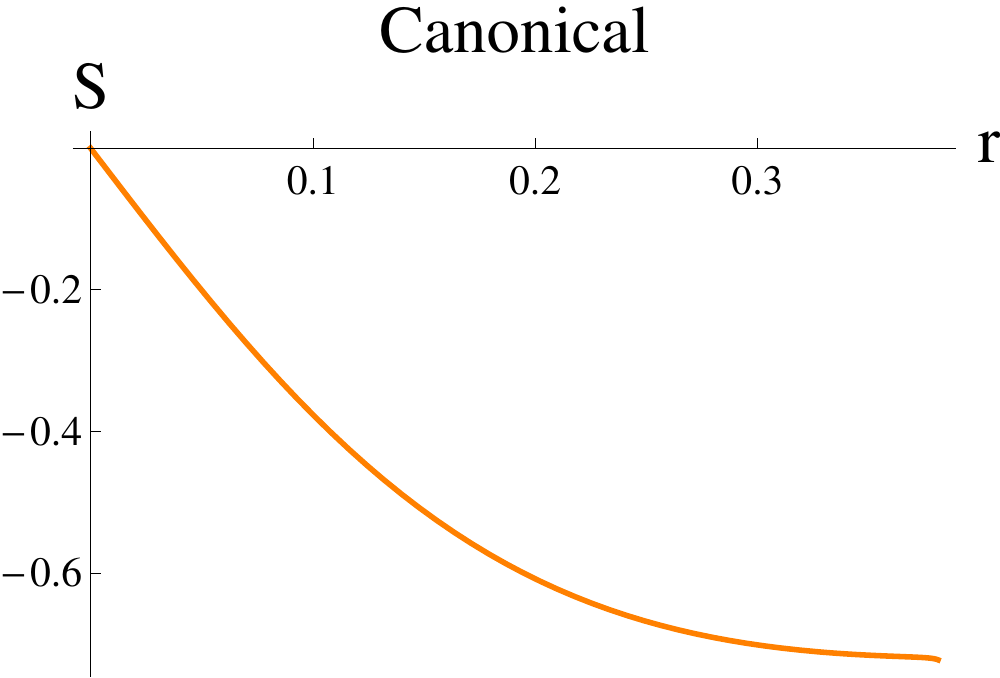}}
\centerline{
(b)\includegraphics[height=1.2in]{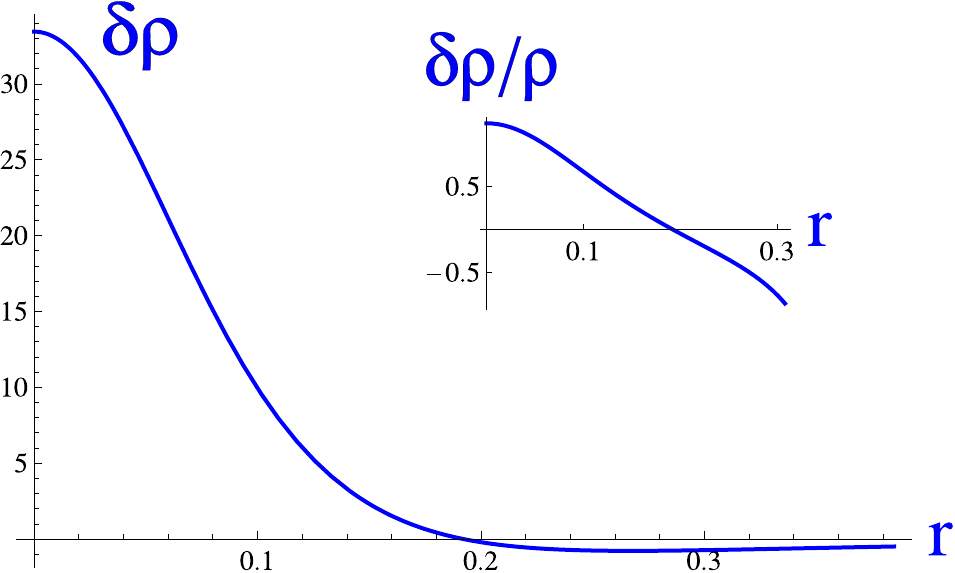}
}
\caption{Radial profile of the first order deviation in the canonical case (CEP
model) for (a) the displacement
(or velocity), and (b) the density. The
insert in (b) shows the presence of a single node. 
}
\label{Fig:cep-vit}
\end{figure}

The structure of the neutral mode is different in the canonical and
microcanonical ensembles. This has important consequences for the evolution of
the star in the collapse regime.
The important result is depicted by the spatial profile of the velocity
which is negative everywhere for the CEP model illustrated
in Figure \ref{Fig:cep-vit}-(a), whereas the velocity clearly changes sign in
the
star for the microcanonical case (see the curve $S(r)$ drawn in Figure
\ref{FigrhM}-(b)). This shows that the collapse corresponds to a pure inward
motion in the canonical ensemble while, in the microcanonical ensemble, the core
collapses (inward motion) and the halo expands (outward motion). This result has
to be completed by the spatial profile of the density deviation, which displays
only one node for the CEP model at the critical point A  (see the curve $\delta
\rho/ \rho$ in the insert of Figure
\ref{Fig:cep-vit}-(b)), so the density increases in the core and decreases in
the outer layers, whereas
in the microcanonical ensemble, the density deviation $\delta \rho/ \rho$
at the critical point A' displays two nodes (see Figure \ref{FigrhM}-(a)), so
the
density increases in the core and in the halo, while it decreases in the
intermediate region.

We note that the above results regarding the structure of the neutral modes in
the canonical and microcanonical ensembles are similar to those obtained in the
context of box-confined isothermal spheres (see \cite{sc} and references
therein). However, the box prevents the expulsion of the halo, so the box model
is limited in this sense, and the present model, which is unbounded, should be
prefered for astrophysical applications. The above
results are also in agreement with general
results of thermodynamics applied to self-gravitating systems (see
\cite{can-microcan} and Appendices A and B of \cite{sc}). Indeed, in the
canonical ensemble, the system evolves so
as to minimize its free energy at fixed mass. Therefore, one expects that  the
system collapses as a whole and ultimately forms a {\it Dirac peak} containing
all the mass.\footnote{Of course, in practice, other physical processes such as
quantum mechanics and  general relativity \cite{ac,rc} will
come into play and prevent this classical mathematical singularity to form. It
will be replaced by a quantum compact object such as a white dwarf or a neutron
star (if its mass is smaller than the Chandrasekhar \cite{chandra31} or
Oppenheimer-Volkoff \cite{ov} limit) or by a black hole.} Indeed, a Dirac peak
has an infinite
negative free energy. The
collapse of the system is
accompanied by a huge decrease of potential energy ($W$) which overcomes the
slower decrease of entropy (or increase of $-TS$). Such an evolution
is energetically favorable. On the other hand, in the microcanonical ensemble,
the system evolves so as to maximize its entropy at fixed mass and energy.
Therefore, one
expects that the system  takes a {\it core-halo} structure. Indeed, by
collapsing the core and expanding the halo we can make the entropy very large,
possibly infinite, while
conserving the energy. As the core collapses, its potential energy
decreases. Since the total energy is conserved,
the kinetic energy of the halo must increase simultaneously. As
a result, the halo overheats and is ejected at large distances. Such an
evolution
is entropically favorable.

\subsection{Weakly nonlinear regime}

Our weakly nonlinear analysis, whose relevance is confirmed by the numerical
solution of the full hydrodynamic equations, is valid during the early stage of
the collapse dynamics.  It leads to the same Painlev\'e I equation
[see equation (\ref{n25}) here and equation (75) in Paper I] in
both
canonical and microcanonical models, but the coefficients are different. In
Paper I, we obtained $\gamma\simeq 120.2$ and $K\simeq 12.3$, whereas in the
present paper we obtained  $\gamma\simeq 46.62$ and $K\simeq 1055.98$.
Therefore, the amplitude $A(t)$ increases more slowly  in the canonical model
than in the microcanonical one, compare Figure $8$ of Paper I
with Figure \ref{Fig:Painl} here.
This is related to the fact that the critical density
$\rho_0^{(c)}$ is much lower (by a factor $100$) in the CEP model with respect
to the MEP model.

\subsection{Fully nonlinear regime}

The fully nonlinear regime is marked by the collapse of the core of the
system,
the formation
of a finite time singularity, and the growth of a Dirac peak  by accretion of
the surrounding matter in the post-collapse regime. When
considering the collapse of the core, one can neglect  the pressure as
compared to the self-gravity. The core undergoes a self-similar collapse (free
fall) in both canonical and microcanonical
ensembles but the exponents are different in the two 
ensembles. For example, the density profile decreases as $r^{-24/11}$ in the CEP
model
and as $r^{-48/19}$ in the MEP model. Consequently, in the post-collapse
regime, the mass in the Dirac peak increases as $M_c(t)\sim t^{3/4}$ in the CEP
model and as $M_c(t)\sim t^{3/8}$ in the MEP model. On the other hand, in the
CEP model, the system collapses as a whole while, in the MEP model, it takes a
core-halo structure reminiscent of a red giant. The halo is heated by the energy
released by the collapsing core and, when considering the evolution of the
halo, one can consider that the pressure force overcomes the gravitational
attraction. Therefore, the canonical ensemble may be relevant to describe
the life and death of supermassive stars which collapse (implode) without
exploding (hypernova phenomenon) while the microcanonical ensemble
may be relevant to describe the life and death of less massive stars which
present a more complex evolution marked by the  collapse (implosion) of the core
and the
explosion of the halo (supernova phenomenon). The final fate of a star is to
become a neutron star if its mass is below the Oppenheimer-Volkoff limit or a
black hole if its mass is above the Oppenheimer-Volkoff limit.

{\it Remark:} In previous works on the statistical mechanics of
self-gravitating systems \cite{box,lbw}, the collapse of the system in the
microcanonical ensemble was associated with the
so-called gravothermal catastrophe. The gravothermal
catastrophe is not like an avalanche (or a free fall). During the gravothermal
catastrophe the system takes a
core-halo structure but remains in hydrostatic equilibrium \cite{lbe,inagakilb}.
Its evolution is due to the temperature gradient between the core and the halo
and the fact that the core has a negative specific heat $C=dE/dT<0$
\cite{lbw,thirring}. Therefore, by losing heat the core grows hotter and evolves
away from equilibrium. On the other hand, the halo does not explode and even
barely expands. The evolution of the system consists just in a core collapse.
This description applies to globular clusters. During the gravothermal
catastrophe their central parts collapse and get hotter while their
outer parts are left behind.  In our model, which rather applies to gaseous
stars
described by fluid equations, we are in the opposite regime (see also
\cite{sc}).\footnote{This is necessary to account for the very different
timescale governing the collapse of globular clusters and stars. The timescale
of the gravothermal catastrophe is of the order of the age of the Universe
while the timescale of star collapse (e.g. supernova) is of the
order of a few days.} There is no gradient of temperature but the system is not
in
hydrostatic equilibrium. At low energies and low temperatures, the
pressure cannot balance the gravitational attraction and the star collapses.
The core experiences a free fall and the halo expands
because it is heated by the energy released by the collapsing core (we
have adopted a rough energetic constraint where the temperature is uniform but
increases with time). We have suggested that this simple model could be related
to the onset of red giant structure and to supernova explosions. We note that
Lynden-Bell and Wood \cite{lbw} and Thirring \cite{thirring} have
also related the gravitational instability resulting from the  negative specific
heat of self-gravitating systems to the onset of red giant
structure and to 
supernova
explosions (see \cite{ac} for additional comments). Probably, a realistic model
of stars should take into account both
energy transfers by temperature gradients as in \cite{lbe,inagakilb} {\it and}
deviation from hydrostatic equilibrium as in our model.

\section{Conclusion}
\label{sec:discussion}

Presently, theories of supernova explosion focus on physical phenomena such as
the emission of neutrinos, or complex 3D effects which we do not consider at all
in our work.
We focus on an entirely different aspect of the physics of supernovae, namely
the fluid mechanical part, without considering the immensely complex set of
possible nuclear reactions in the core.
We show that implosion \textit{and} explosion taking place at the death of a
massive star may occur simultaneously. This yields an alternative explanation to
the yet unsolved problem of supernova description where the two steps process
makes, we believe, an unsatisfactory explanation. Using a simple model which has
no aim to reproduce the complex reality of what happens inside a star, we point
out first that the huge difference of time scales between the long life of a
star and its abrupt death can be understood in the light of a catastrophe-like
theory which includes dynamical aspects. This is performed by sweeping slowly a
saddle-center bifurcation. Starting from the stable equilibrium state and
approaching the saddle-center bifurcation, the weakly nonlinear
analysis leads to a
universal (Painlev\'{e} I) equation followed by a self-similar collapse more
rapid than the growing explosion of the outer shell.

It is important to point out that the Painlev\'{e} analysis gives 
access to the sign for the velocity field at the critical point, contrary to
what happens in ``classical" transitions from a linearly stable to a linearly
unstable situation (where the unstable mode may have either positive or negative
amplitude). As we have shown, this sign may change as a function of the radius.
This fair property of the definite sign of the growing Painlev\'{e} solution
comes from the fact that in the case of a saddle-center bifurcation, the two
stable and unstable  equilibrium states (a center and a saddle respectively)
merge at the critical point, beyond which no equilibrium state exists (neither
stable nor unstable) that makes the difference with the ``classical" case.

Our study  illustrates once more (see \cite{can-microcan}) that 
a change from canonical to microcanonical description, not looking very
important at first, does deeply change the outcome of the transition from stable
to unstable state. In the case we have studied, the canonical model collapses
without producing any outgoing flow of matter, although the microcanonical model
shows a core collapse together with an explosive outer shell. The former case
(Paper I) could reproduce what happens in the case of supermassive stars which
die via hypernovae showing very intense and directive gamma ray bursts, but no
explosion of matter (or a very faint one) and often leads to the formation of
black holes. In the present paper, on the other hand, we show that it is
possible to reproduce what
happens for massive stars which die via supernovae showing explosion of matter
and often leading to the formation of a neutron star resulting from the core
collapse. Therefore, our simple model opens up the way to a new
understanding of the explosion of stars, based on fluid mechanics, catastrophe
theory, and bifurcation properties of their equilibrium state. It also provides
a nice illustration of the property of inequivalence between canonical and
microcanonical ensembles for systems with long-range interactions.

The assumption of a uniform temperature inside the star implies physically that heat conduction is very fast so that temperature is made uniforme on a time scale much shorter than the one of the physical process we consider. This could be due, for instance, to heat transfer by photons,
moving a priori very fast in the star, even though this motion is a kind
of Brownian motion, not a straight trajectory.  Another physical
possibility is given by the well-known Laplace equilibrium in the
atmosphere of the Earth: Laplace assumed, rightly, that, because of very
fast vertical motions, the air reaches rapidly an isentropic equilibrium,
where the entropy per unit mass is constant (notice that the word entropy was
absent in Laplace's work, but he understood that fast exchanges like in
sound waves are such that there is no irreversible exchange of heat so
that the relationship between pressure and volume is given by the relation  $P V^{\gamma}$
constant, where $P$ is the  pressure,  $V$ the specific volume, and $\gamma$ the ratio of heat
conductivity at constant pressure and volume). Therefore, if fast vertical
motion (likely turbulent) is present in the star, it could be closer to
reality to take, instead of a uniform temperature and a global energy
conservation, a constraint of Laplace equilibrium, namely a uniform
entropy per unit mass together with a conserved total energy. Such
an
equilibrium with a non uniform temperature is what is expected to represent
the present state inside the Sun, with a temperature increasing toward the
center. While being not much heavier to treat numerically, this description would be hardly tractable analytically and this is why we considered a simpler isothermal model. However, we expect that, qualitatively, the results should be comparable.

In the description of the halo expansion after the explosion, we made rough
approximations. Nevertheless, the points we made clear seem important.
In a first stage, we assumed that the halo expands
self-similarly powered by the rise of temperature accompanying the contraction
of the core. In a second stage, we assumed that the free expansion stage is  an
isentropic process with two constraints, the conservation of mass and energy. 
This stage of free expansion, which has not been much studied, reveals itself to
be especially interesting because neglecting the gravity with respect to the
pressure forces, we find that no self-similar solution exists, contrary to the
free fall of dense molecular gas (where the opposite was assumed). Then, we
point out that when  both pressure and gravity are negligible, another type of
solution appears, of Burgers-type, which is a prototype for creating shocks.
Such a scenario could happen in the process of remnant expansion, but is not the
common one found in the literature which invokes Sedov-Taylor self-similar
solutions where the shock is due to the interaction between the remnant and the
interstellar matter. Our argument relies on the fact that the mean-free path  in
interstellar matter may be as large as the size of a galaxy, that makes
 such event unrealistic. In our rough description shocks are formed naturally
inside the remnant, they propagate inside this matter, the role of the
interstellar medium being ignored.

\section*{Acknowledgement}
Two of us (P.H. Chavanis and Y. Pomeau)  greatly acknowledge ``la fondation des Treilles'' which helped finance a  colloquium organized in September 2014  in their beautiful  mediterranean domain of Tourtour (83690, France)  where  our  collaboration started on this subject.

\appendix

\section{Useful relations in original variables}
\label{app:A}

We regroup in this Appendix some useful relations that are needed in our
theoretical study. We write the equations in terms of the original
(dimensional) variables.

\subsection{Newton's law}
\label{sec_gauss}

Integrating the Poisson equation (\ref{e3}) for a spherically symmetric
distribution of matter, we obtain  Newton's law
\begin{eqnarray}
{\Phi}_{,r}(r,t)=\frac{GM(r,t)}{r^2},
\label{he4}
\end{eqnarray}
where
\begin{eqnarray}
M(r,t)=\int_0^r \rho(r',t) 4\pi {r'}^2\, dr'
\label{he5}
\end{eqnarray}
is the mass contained within the sphere of radius $r$. The density is
\begin{eqnarray}
\rho(r,t)=\frac{M_{,r}(r,t)}{4\pi {r}^2}.
\label{he5b}
\end{eqnarray}
Applying Newton's law at the edge of the star, we get
\begin{equation}
{\Phi}_{,r}(R(t),t)=\frac{GM}{R(t)^2}\quad {\rm and}\quad 
{\Phi}(R(t),t)=-\frac{GM}{R(t)},
\label{he6}
\end{equation}
where $M$ is the total mass of the star (to get the second relation we have
assumed that the space is empty outside the star so that Newton's law can be
easily integrated for $r\ge R(t)$). For a steady state, using equation
(\ref{he2}), the foregoing relations from equations (\ref{he4}) and (\ref{he6})
 imply
\begin{eqnarray}
{h}_{,r}(r)=-\frac{GM(r)}{r^2},\qquad {h}_{,r}(r_0)=-\frac{GM}{r_0^2}.
\label{he7}
\end{eqnarray}

\subsection{Gravitational energy}
\label{sec_pot}

The gravitational energy of the star is given by
\begin{eqnarray}
\label{e6}
W(t)=\frac{1}{2}\int\rho\Phi\, d{\bf r}.
\end{eqnarray}
Using Poisson's equation (\ref{e3}), integrating by parts, and using
equation (\ref{he6}) valid for a spherically symmetric distribution of
matter, we find that the gravitational energy is given by
\begin{eqnarray}
\label{e6b}
W(t)&=&\frac{1}{8 \pi
G} \int \Phi \Delta \Phi \, d{\bf r} \nonumber\\ 
&=&-\frac{GM^2}{2R(t)}-\frac{1}{8 \pi
G} \int (\nabla \Phi)^2 \, d{\bf r}.
\end{eqnarray}
Alternatively, using  equation (\ref{he5b}), we can write
\begin{equation}
\label{ew}
W(t)=\frac{1}{2}\int_0^{R(t)}\rho\Phi \, 4\pi r^2\,
dr=\frac{1}{2}\int_0^{R(t)}\Phi M_{,r}\, dr.
\end{equation}
Integrating equation (\ref{ew}) by parts and using equation (\ref{he6}), we
obtain
\begin{eqnarray}
W(t)=-\frac{GM^2}{2 R(t)}-\frac{1}{2}\int_0^{R(t)}M \Phi_{,r} \, dr.
\end{eqnarray}
For a steady state, using equation (\ref{he2}), the foregoing equation becomes
\begin{eqnarray}
W=-\frac{GM^2}{2 r_0}+\frac{1}{2}\int_0^{r_0}M h_{,r} \, dr.
\end{eqnarray}
Therefore, at equilibrium, the energy (\ref{ae6}) can be written as
\begin{eqnarray}
E=\frac{3}{2}N k_B T-\frac{GM^2}{2 r_0}+\frac{1}{2}\int_0^{r_0}M h_{,r} \, dr.
\end{eqnarray}

\subsection{Virial theorem}

For a self-gravitating gas in a steady state, the scalar virial theorem writes
\begin{eqnarray}
3\int P\, d{\bf r}+W=0.
\label{he9}
\end{eqnarray}
Using equation (\ref{he9}), the energy (\ref{ae6}) takes the form
\begin{eqnarray}
\label{he10}
E=\frac{3}{2}N k_B
T-3\int P\, d{\bf r}.
\end{eqnarray}

\subsection{Radial displacement}

We consider a spherically symmetric evolution of the system and define the
radial displacement $S(r,t)$ by 
\begin{equation}
\delta u=S_{,t},
\end{equation}
where $u(r,t)$ is the radial
component of the velocity field. In the linearized equations, recalling that the
perturbations evolve with time as $e^{\lambda t}$, we get
\begin{equation}
\delta u=\lambda
S.
\end{equation}
The linearized continuity equation (\ref{l4}) may be written as
\begin{equation}
\delta \rho+\frac{1}{r^2}(r^2 \rho S)_{,r}=0.
\label{rs1}
\end{equation}
Multiplying equation (\ref{rs1}) by $4\pi r^2$ and integrating between $0$ and
$r$, we get
\begin{equation}
S(r)=-\frac{\delta M(r)}{4\pi\rho(r) r^2}.
\label{rs2}
\end{equation}
This relation is valid for $r<r_0$. It is undetermined at $r=r_0$ where
$\rho=0$. However, coming back to equation (\ref{rs1}), and expanding the
derivative, we obtain
\begin{equation}
S(r_0)=-\frac{\delta\rho(r_0)}{\rho_{,r}(r_0)}.
\label{rs3}
\end{equation}

\section{Useful relations in scaled variables at the critical point}
\label{app:B}

In this Appendix, we regroup some useful relations in scaled variables that we
apply at the critical point. In all the subsequent formulae, we suppress the
hats (in the final equations) in order to simplify the notations.

Writing equation (\ref{marg3}) in scaled variables, we
get
\begin{equation}
j_{,r}(r)=-\delta \Phi^{(c)}_{,r}(r)=-\frac{\delta M^{(c)}(r)}{r^2}.
\label{ur1}
\end{equation}
Since $\delta M^{(c)}(r_c)=0$ because the 
total mass is conserved, the foregoing equation gives  $j_{,r}(r_c)=0$. Knowing
$j(r)$ we obtain $\delta M^{(c)}(r)$ by the relation
\begin{equation}
\delta M^{(c)}(r)=-r^2 j_{,r}(r).
\label{ur2}
\end{equation}
Writing equation (\ref{he5b}) in perturbed form and introducing the scaled
variables, we get
\begin{equation}
\delta\rho^{(c)}(r)=\frac{\delta M^{(c)}_{,r}}{4\pi r^2}=-\frac{1}{4\pi}\Delta
j,
\label{ur3}
\end{equation}
where we have used equation (\ref{ur2}) to obtain the last equality. Writing
equation (\ref{rs2}) in scaled variables with ${\hat S}=S/T^{1/2}$, 
we get
\begin{equation}
S^{(c)}(r)=-\frac{\delta M^{(c)}(r)}{4\pi\rho^{(c)}(r)
r^2}=\frac{j_{,r}(r)}{4\pi \rho^{(c)}(r)},
\label{ur4}
\end{equation}
where we have used equation (\ref{ur2}) to obtain the last equality. At the edge
of the star, writing equations (\ref{marg10}) and (\ref{rs3}) in scaled
variables, we
get
\begin{equation}
S^{(c)}(r_c)=-\frac{\delta\rho^{(c)}(r_c)}{\rho^{(c)}_{,r}(r_c)}
=-\frac{2j(r_c)}{\rho^{(c)}_{,r}(r_c)}.
\label{ur5}
\end{equation}

\section{A self-similar solution for the expansion of the halo}
\label{sec_a}

In this Appendix, we construct a self-similar solution describing the expansion
of the halo according to the model developed in Section \ref{subsec:phcsol}.

\subsection{Euler equations}

We consider the Euler equations
\begin{eqnarray}
\label{a1}
\frac{\partial\rho}{\partial t}+\nabla\cdot (\rho {\bf u})=0,
\end{eqnarray}
\begin{eqnarray}
\label{a2}
\frac{\partial {\bf u}}{\partial t}+({\bf u}\cdot \nabla){\bf
u}=-\frac{1}{\rho}\nabla P,
\end{eqnarray}
with a polytropic equation of state of the form
\begin{eqnarray}
\label{a3}
P=K(t)\rho^{\gamma},
\end{eqnarray}
where $K(t)$ is a given function of time. We assume that $K(t)\ge 0$ in order to
have a positive pressure. We assume that
$\gamma>0$ so that the pressure
force leads to an expansion of the halo: $-(1/\rho)P'(\rho)d\rho/dr>0$
(recalling that $d\rho/dr<0$). Finally,
we neglect the
self-gravity of the halo
in equation (\ref{a3}), an approximation whose validity will be discussed in
Section
\ref{sec_val}.

\subsection{Scaling ansatz}

We look for a self-similar solution of equations (\ref{a1})-(\ref{a3})
of the form
\begin{eqnarray}
\label{a4}
\rho({\bf r},t)=\frac{M}{R(t)^3} f\left\lbrack \frac{\bf r}{R(t)}\right
\rbrack,\qquad
{\bf u}({\bf r},t)=H(t){\bf r}.
\end{eqnarray}
We have assumed that the velocity field is proportional to the radial
distance ${\bf r}$  with a proportionality factor $H(t)$. Defining
\begin{eqnarray}
\label{a4b}
{\bf x}=\frac{{\bf r}}{R(t)},
\end{eqnarray}
we can rewrite equation (\ref{a4}) as
\begin{eqnarray}
\label{a4c}
\rho({\bf r},t)=\frac{M}{R(t)^3} f({\bf x}),\qquad
{\bf u}({\bf r},t)=H(t)R(t){\bf x}.
\end{eqnarray}
In the
foregoing equations $R(t)$ is the typical
size (radius) of the halo and $f({\bf x})$ is the
invariant density profile. We assume that the density profile contains all the
mass ($\int \rho({\bf r},t)\, d{\bf r}=M$) so that  $\int f({\bf x})\,
d{\bf x}=1$.

The continuity equation (\ref{a1}) can be rewritten
as
\begin{eqnarray}
\label{a5}
\frac{\partial\ln\rho}{\partial t}+\nabla\cdot {\bf u}+ {\bf
u}\cdot \nabla\ln\rho=0.
\end{eqnarray}
From equation (\ref{a4}), we obtain
\begin{eqnarray}
\label{a6}
\frac{\partial\ln\rho}{\partial t}=-\frac{\dot R}{R}{\bf x}\cdot \nabla_{\bf
x}\ln f-3\frac{\dot R}{R},\nonumber\\
 \nabla\ln\rho=\frac{1}{R}\nabla_{\bf x}\ln
f,\qquad \nabla\cdot {\bf u}=3H.
\end{eqnarray}
Substituting the foregoing relations into equation (\ref{a5}), we get
\begin{eqnarray}
\label{a7}
\left (H-\frac{\dot R}{R}\right )\left (3+{\bf x}\cdot \nabla_{\bf x}\ln f\right
)=0.
\end{eqnarray}
This equation must be satisfied for all ${\bf x}$. This implies
\begin{eqnarray}
\label{a8}
H(t)=\frac{\dot R}{R}.
\end{eqnarray}
We note the formal analogy with the Hubble constant in cosmology. We have ${\bf
u}({\bf r},t)=[\dot R/R(t)]{\bf r}=\dot R {\bf x}$.

Using equation (\ref{a4}), the left hand side of
the Euler equation (\ref{a2}) can be written as
\begin{eqnarray}
\label{a9}
\frac{\partial {\bf u}}{\partial t}+({\bf u}\cdot \nabla){\bf
u}=(\dot H+H^2){\bf r}=\frac{\ddot R}{R}{\bf r}=\ddot R {\bf x}.
\end{eqnarray}
For an equation of state of the form of equation (\ref{a3}), the
pressure term in the right hand side of equation (\ref{a2}) is given by
\begin{eqnarray}
\label{a10}
-\frac{1}{\rho}\nabla P=-K(t)\gamma\rho^{\gamma-2}\nabla\rho.
\end{eqnarray}
With the scaling ansatz from  equation (\ref{a4}),   we obtain
\begin{eqnarray}
\label{a11}
-\frac{1}{\rho}\nabla
P=-K(t)\gamma\frac{M^{\gamma-1}}{R^{3\gamma-2}}f^{\gamma-2}\nabla_{\bf x} f.
\end{eqnarray}
Substituting equations (\ref{a9}) and (\ref{a11}) into the Euler equation
(\ref{a2}), and assuming that $f$ depends only on $x=|{\bf x}|$, we get
\begin{eqnarray}
\label{a12}
\ddot
R=-K(t)\gamma\frac{M^{\gamma-1}}{R^{3\gamma-2}}f^{\gamma-2}\frac{f'(x)}{x}.
\end{eqnarray}
The variables of position and time separate provided that 
\begin{eqnarray}
\label{a13}
f^{\gamma-2}\frac{df}{dx}+2Ax=0
\end{eqnarray}
and
\begin{eqnarray}
\label{a14}
\ddot R=2AK(t)\gamma\frac{M^{\gamma-1}}{R^{3\gamma-2}},
\end{eqnarray}
where $A$ is a constant (the factor $2$ has been introduced for
convenience). These differential equations determine the invariant halo profile
$f(x)$ and the evolution of the halo radius $R(t)$.

\subsection{Invariant halo profile and halo radius}

The differential equation (\ref{a13}) determining the invariant profile of the
halo  can be integrated into
\begin{eqnarray}
\label{a15}
f(x)=\left\lbrack C-(\gamma-1)A x^2\right\rbrack_+^{1/(\gamma-1)},
\end{eqnarray}
where $[x]_+=x$ if $x\ge 0$ and $[x]_+=0$ if $x\le 0$. Therefore, the invariant
profile (\ref{a15}) is given by a Tsallis distribution \cite{tsallis} of index
$\gamma$ (see Section VI of \cite{sc2}). We can take $C=A$ 
without loss of generality. Denoting this constant by $Z^{1-\gamma}$, we get
\begin{eqnarray}
\label{a16}
f(x)=\frac{1}{Z}\left\lbrack 1-(\gamma-1)x^2\right\rbrack_+^{1/(\gamma-1)},
\end{eqnarray}
where $Z$ is determined by the normalization condition $\int f({\bf x})\, d{\bf
x}=1$. This yields
\begin{eqnarray}
\label{a17}
Z=\int_0^{x_{\rm max}}\left\lbrack
1-(\gamma-1)x^2\right\rbrack_+^{1/(\gamma-1)}\, 4\pi x^2\, dx,
\end{eqnarray}
where $x_{\rm max}=1/\sqrt{\gamma-1}$ if $\gamma\ge 1$ and $x_{\rm max}=+\infty$
if $1/3<\gamma\le 1$. The distribution is not normalizable when $\gamma\le
1/3$. Therefore, in the following, we assume $\gamma>1/3$. The integral can be
expressed in terms of Gamma functions leading to
\begin{eqnarray}
\label{a18}
Z=\frac{\pi^{3/2}\Gamma\left (\frac{1}{\gamma-1}\right
)}{(\gamma-1)^{5/2}\Gamma\left (\frac{3}{2}+\frac{\gamma}{\gamma-1}\right
)}\qquad (\gamma\ge 1),
\end{eqnarray}
\begin{eqnarray}
\label{a19}
Z=\frac{\pi^{3/2}\Gamma\left (\frac{1}{1-\gamma}-\frac{3}{2}\right
)}{(1-\gamma)^{3/2}\Gamma\left (\frac{1}{1-\gamma}\right )} \qquad
(1/3<\gamma\le 1).
\end{eqnarray}
On the other hand, the differential equation (\ref{a14}) determining the
evolution of the halo radius becomes
\begin{eqnarray}
\label{a20}
\ddot R=2 Z^{1-\gamma} K(t)\gamma\frac{M^{\gamma-1}}{R^{3\gamma-2}}.
\end{eqnarray}
By a proper rescaling, we can write this equation as
\begin{eqnarray}
\label{a21}
\ddot R R^{3\gamma-2}  = K(t).
\end{eqnarray}
For the isothermal equation of state $P=\rho
k_B
T(t)/m$ (corresponding to
$\gamma=1$), the invariant halo profile is the Gaussian
\begin{eqnarray}
\label{a22}
f(x)=\frac{1}{\pi^{3/2}}e^{-x^2}
\end{eqnarray}
and  the evolution of the halo radius is determined by a differential equation
of the form
\begin{eqnarray}
\label{a23}
\ddot R=\frac{T(t)}{R}.
\end{eqnarray}
For the polytropic equation of state 
$P=K(t)\rho^2$ (corresponding to $\gamma=2$), the invariant halo profile is
parabolic
\begin{eqnarray}
\label{a24}
f(x)=\frac{15}{8\pi}(1-x^2)_+
\end{eqnarray}
and the evolution of the halo radius is determined by a differential equation
of the form
\begin{eqnarray}
\label{a25}
\ddot R=\frac{K(t)}{R^4}.
\end{eqnarray}
We note that the expansion of the halo is always accelerating ($\ddot R>0$).

\subsection{Validity of the approximations}
\label{sec_val}

The Euler equations (\ref{a1}) and (\ref{a2}) are valid
provided that we can neglect the self-gravity of the halo as compared to the
pressure force. The pressure force scales as
\begin{equation}
\label{a26}
\frac{1}{\rho}|\nabla P|=\frac{1}{\rho}K(t)|\nabla \rho^{\gamma}|\sim
\frac{K(t)\rho^{\gamma-1}}{R}\sim \frac{K(t)M^{\gamma-1}}{R^{3\gamma-2}},
\end{equation}
while the gravitational force scales as
\begin{eqnarray}
\label{a27}
|\nabla \Phi|\sim
\frac{GM}{R^{2}}.
\end{eqnarray}
Therefore, when $R$ is large, the gravitational force is negligible in front of
the pressure force provided that
\begin{eqnarray}
\label{a28}
\frac{GM}{R^{2}}\ll  \frac{K(t)M^{\gamma-1}}{R^{3\gamma-2}}.
\end{eqnarray}
The validity of this approximation depends on the function $K(t)$ 
and on the value of the polytropic index $\gamma$. Some examples are given
below.

\subsection{The case $K(t)=1$}
\label{sec_k}

In this subsection, we assume that the temperature is constant:
\begin{eqnarray}
\label{k1}
K(t)=1.
\end{eqnarray}
In that case, the differential
equation  (\ref{a21}) becomes
\begin{eqnarray}
\label{k2}
\ddot R   = \frac{1}{R^{3\gamma-2}}.
\end{eqnarray}
It is similar to the fundamental equation of dynamics (Newton's equation) for a
fictive particle of unit mass and position $R(r)$ submitted to a repulsive
force of the form $F=1/R^{3\gamma-2}$. The case of an arbitrary polytropic
index $\gamma$ is treated in
Ref. \cite{prep} by developing an analogy with the Friedmann equations of
cosmology.

\subsubsection{Isothermal equation of state $(\gamma=1)$}
\label{sec_ka}

Here, we specifically consider the case $\gamma=1$
corresponding to an isothermal equation of state. In that case, the
differential
equation  (\ref{k2}) becomes
\begin{eqnarray}
\label{k3}
\ddot R  = \frac{1}{R}.
\end{eqnarray}
It can be written as
\begin{eqnarray}
\label{k4}
\ddot R=-\frac{dV}{dR}\qquad {\rm with}\qquad V(R)=-\ln R.
\end{eqnarray}
The first integral of motion is
\begin{eqnarray}
\label{k5}
E=\frac{1}{2}\left (\frac{dR}{dt}\right )^2+V(R),
\end{eqnarray}
where $E$ is a constant. The evolution of the halo radius $R(t)$ is
therefore
determined by the integral
\begin{eqnarray}
\label{k6}
t=\int_{R_0}^{R(t)}\frac{dR}{\sqrt{2(E-V(R))}},
\end{eqnarray}
where $R_0$ is its value at $t=0$. In writing equation (\ref{k6}) we have
assumed
that $R(t)$ always increases with time. Substituting the potential $V(R)$ from
equation (\ref{k4}) into
equation (\ref{k6}), we obtain 
\begin{eqnarray}
\label{k7}
t=\int_{R_0}^{R(t)}\frac{dR}{\sqrt{2(E+\ln
R)}}.
\end{eqnarray}
Making the change of
variables $x=\sqrt{E+\ln R}$ in equation (\ref{k7}), we get
\begin{eqnarray}
\label{k8}
t=\sqrt{2}e^{-E}\int_{\sqrt{E+\ln
R_0}}^{\sqrt{E+\ln
R(t)}} e^{x^2}\, dx.
\end{eqnarray}
This equation can be rewritten as
\begin{eqnarray}
\label{k9}
t=\sqrt{2} R(t)D(\sqrt{E+\ln
R(t)})  \nonumber\\
-\sqrt{2}R_0 D(\sqrt{E+\ln
R_0}),
\end{eqnarray}
where $D(x)$ is Dawson's function
\begin{eqnarray}
\label{k10}
D(x)=e^{-x^2}\int_{0}^{x} e^{t^2}\, dt.
\end{eqnarray}
It has the asymptotic behavior
\begin{eqnarray}
\label{k11}
D(x)=\frac{1}{2x}+\frac{1}{4x^3}+... \qquad (x\rightarrow +\infty).
\end{eqnarray}
Therefore, for $t\rightarrow +\infty$, we obtain
\begin{eqnarray}
\label{k12}
t\sim \frac{R}{\sqrt{2\ln R}},
\end{eqnarray}
leading to (at leading order):
\begin{eqnarray}
\label{k13}
R(t)\sim t\sqrt{2\ln t}.
\end{eqnarray}
The radius of the halo expands linearly in time with a logarithmic correction.
The velocity
of expansion
\begin{eqnarray}
\label{k14}
\dot R(t)\sim \sqrt{2\ln t},\qquad {\dot R}^2(t)\sim 2\ln
t
\end{eqnarray}
increases logarithmically in time. 

\subsubsection{Asymptotic results for an arbitrary index}
\label{sec_kb}

Here, we provide asymptotic results valid when $t\rightarrow +\infty$. For an
arbitrary index $\gamma$, the potential writes
\begin{eqnarray}
V(R)=\frac{1}{3(\gamma-1)}\frac{1}{R^{3(\gamma-1)}}.
\end{eqnarray}
We first assume $\gamma>1$. For $R\rightarrow +\infty$, the potential
$V(R)\rightarrow 0$ and the first integral of motion (\ref{k5}) reduces to $\dot
R\sim \sqrt{2E}$ (with $E>0$) leading to
\begin{eqnarray}
R(t)\sim \sqrt{2E}\, t \qquad (t\rightarrow +\infty).
\end{eqnarray}
We now
assume $1/3<\gamma<1$. For $R\rightarrow +\infty$, the potential
$V(R)\rightarrow -\infty$ and the first integral of motion (\ref{k5}) reduces
to $\dot R\sim 
\sqrt{-2V(R)}$ leading to 
\begin{equation}
R(t)\sim \left\lbrack
\frac{(3\gamma-1)^2}{6(1-\gamma)}\right\rbrack^{{1}/{(3\gamma-1)}}t^{
2/(3\gamma-1)} \;\;(t\rightarrow +\infty).
\end{equation}

For $K(t)=1$ the condition of validity of our study (\ref{a28}) takes the form
$R^{4-3\gamma}\gg 1$. Since $R(t)\rightarrow +\infty$ for $t\rightarrow
+\infty$, the foregoing asymptotic behaviors are valid provided that
$\gamma<4/3$.

\subsection{The case $K(t)=t^a$ (post-collapse)}
\label{sec_b}

In this subsection, we assume that the temperature evolves with time as a power
law:
\begin{eqnarray}
\label{b1}
K(t)=t^a.
\end{eqnarray}
In that case, the differential
equation  (\ref{a21}) becomes
\begin{eqnarray}
\label{b2}
\ddot R R^{3\gamma-2}  = t^a.
\end{eqnarray}
For the sake of generality, we let the value of $a$ arbitrary (positive or
negative).  For $a>0$ the temperature increases with time up to infinity. This
is the situation corresponding to the post-collapse regime considered in Section
\ref{sec:post-coll}
where $a=3/8$ and $\gamma=1$ (isothermal gas). For $a<0$ the
temperature decreases with time up to zero.

\subsubsection{Solution $R(t)= A t^q$ with $q>0$}
\label{sec_ut}

We consider a solution of equation (\ref{b2}) of the
form 
\begin{eqnarray}
\label{b3b}
R(t)= A t^q
\end{eqnarray}
with $q>0$ (and, of course, $A>0$). In that case, the halo radius increases with
time up to infinity. Substituting this ansatz into
equation (\ref{b2}) we get 
\begin{eqnarray}
\label{b3}
Aq(q-1)t^{q-2}A^{3\gamma-2}t^{(3\gamma-2)q}= t^a,
\end{eqnarray}
implying
\begin{eqnarray}
\label{b4}
q=\frac{a+2}{3\gamma-1}
\end{eqnarray}
and
\begin{eqnarray}
\label{b5}
A^{3\gamma-1}=\frac{1}{q(q-1)}.
\end{eqnarray}
Considering equation (\ref{b5}), and recalling that $q>0$ and $A>0$, we see that
a
necessary condition for the existence of a solution is that $q>1$. Considering
equation (\ref{b4}) with $q>1$, and recalling that $\gamma>1/3$, we find that
the solution exists provided that  $a>-2$ and
$1/3<\gamma<(a+3)/3$. 

The
condition of validity of our study (\ref{a28}) takes the form
\begin{eqnarray}
\label{b6}
\frac{1}{t^{2q}}\ll \frac{t^a}{t^{(3\gamma-2)q}}\qquad {\rm for}\qquad
t\rightarrow +\infty.
\end{eqnarray}
When $\gamma>1/3$
this requires $\gamma<(3a+8)/6$.  This condition is always satisfied when the
solution
exists. When $a=0$ the solution exists and is
valid provided that
$1/3<\gamma<1$. When $\gamma=1$ the solution exists and is valid provided that
$a>0$. When $\gamma=1$ and $a=3/8$ (see Section  \ref{sec:post-coll}), we get 
$R=(16/\sqrt{57})t^{19/16}$, which is close
 to the law $R\sim
vt$ corresponding to a constant kinetic energy (see Section \ref{sec:free
exp}).

{\it Remark:} There exist solutions where the radius increases with time ($q>0$)
while the temperature decreases with time ($a<0$).

\subsubsection{Solution $R(t)= A t^q$ with $q<0$}

We consider a solution of equation (\ref{b2}) of the
form of equation (\ref{b3b}) with $q<0$ (and, of course, $A>0$). In that case,
the
halo radius decreases with
time up to zero. Substituting this ansatz into
equation (\ref{b2}) we get equation (\ref{b3}) implying equations (\ref{b4})
and (\ref{b5}). Considering
equation (\ref{b4}) with $q<0$, and recalling that $\gamma>1/3$, we find that
the solution exists provided that $a<-2$. 

The
condition of validity of our study (\ref{a28}) takes the form of
equation (\ref{b6}). When $\gamma>1/3$
this requires $\gamma<(3a+8)/6$. This condition is never fulfilled when the
solution
exists. When $a=0$ the solution does not exist. When $\gamma=1$ the solution
exists provided that $a<-2$ but it is not valid.

\subsubsection{Asymptotic solution $\dot R(t)\rightarrow v$ with $v>0$}
\label{sec_cst}

We consider an asymptotic solution of equation (\ref{b2}) of the form 
\begin{eqnarray}
\label{b7b}
R(t)= vt+\epsilon(t)
\end{eqnarray}
with
$v>0$ and $|\epsilon(t)|\ll v t$
for
$t\rightarrow +\infty$. This means that the velocity $\dot R$ of the halo 
(or its kinetic energy $\propto {\dot R}^2$) tends to a constant for large
times. 
Substituting this ansatz into
equation (\ref{b2}) we get for $t\gg 1$:
\begin{eqnarray}
\label{b7}
\ddot \epsilon \sim \frac{1}{v^{3\gamma-2}} t^{a-3\gamma+2}.
\end{eqnarray}
After two integrations, we obtain (the constants of integration can be taken
equal to zero without restriction of generality)
\begin{eqnarray}
\label{b8}
\epsilon(t) \sim \frac{1}{v^{3\gamma-2}}
\frac{t^{a-3\gamma+4}}{(a-3\gamma+3)(a-3\gamma+4)}.
\end{eqnarray}
The velocity of the halo is
\begin{eqnarray}
\label{b8b}
\dot R \simeq v+ \frac{1}{v^{3\gamma-2}}
\frac{t^{a-3\gamma+3}}{a-3\gamma+3}.
\end{eqnarray}
Note that the terminal velocity $v$ cannot be determined by this asymptotic
approach as it depends on the initial condition. The condition $|\epsilon(t)|\ll
v t$ for
$t\rightarrow +\infty$ impose $\gamma>(a+3)/3$. Since our approach assumes
$\gamma>1/3$, we find that the solution exists (i) for any $\gamma>1/3$
when $a<-2$; (ii) for $\gamma>(a+3)/3$ when $a>-2$.

The condition of validity of our study (\ref{a28}) takes the form
\begin{eqnarray}
\label{b10}
\frac{1}{t^{2}}\ll \frac{t^a}{t^{3\gamma-2}}\qquad {\rm for}\qquad
t\rightarrow +\infty.
\end{eqnarray}
This requires $\gamma<(a+4)/3$. We note in that case that $\epsilon(t)<0$ so
that asymptotically $R(t)\lesssim vt$. In conclusion, the solution exists and
is valid provided that (i) $-3<a<-2$ and $1/3<\gamma<(a+4)/3$; (ii) $a>-2$
and $(a+3)/3<\gamma<(a+4)/3$.
When $a=0$ the solution exists and  is valid
provided that $1<\gamma<4/3$. When
$\gamma=1$ the solution exists and is valid provided that $-1<a<0$.

{\it Remark:} There exist solutions where the radius increases with time
while the temperature decreases with time ($a<0$).

\subsubsection{Asymptotic solution $R(t)\sim (t_{f}-t)^q$ with $q<0$}

We consider an asymptotic solution of equation (\ref{b2}) of the form 
\begin{eqnarray}
\label{b11b}
R(t)\sim A
(t_{f}-t)^q
\end{eqnarray}
with $q<0$ (and, of course, $A>0$). This corresponds to a future finite time
singularity in the sense
that the halo radius becomes infinite in a finite time $t_f$. Defining 
 $\tau=t_{f}-t$ and substituting this ansatz into
equation (\ref{b2}), we get for  $\tau\rightarrow 0$:
\begin{eqnarray}
\label{b11}
Aq(q-1)\tau^{q-2}A^{3\gamma-2}\tau^{q(3\gamma-2)}\sim t_{f}^a,
\end{eqnarray}
implying
\begin{eqnarray}
\label{b12}
q=\frac{2}{3\gamma-1}
\end{eqnarray}
and
\begin{eqnarray}
\label{b13}
A^{3\gamma-1}=\frac{t_{f}^a}{q(q-1)}.
\end{eqnarray}
Considering equation (\ref{b12}) and recalling that $\gamma>1/3$, we find that
the condition  $q<0$ is never fulfilled. Therefore, there is
no solution of that form.

\subsubsection{Asymptotic solution $R(t)\sim (t_{f}-t)^q$ with $q>0$}

We consider an asymptotic solution of equation (\ref{b2}) of the form of
equation (\ref{b11b}) with $q>0$ (and, of course, $A>0$). In that case,  the
halo
radius vanishes  in a finite time $t_f$. Defining 
 $\tau=t_{f}-t$ and substituting this ansatz into
equation (\ref{b2}), we get equation (\ref{b11}) for  $\tau\rightarrow 0$,
implying equations (\ref{b12}) and (\ref{b13}). Considering equation
(\ref{b13}), and recalling that $q>0$ and $A>0$, we see that a
necessary condition for the existence of a solution is that $q>1$. Considering
equation (\ref{b12}) with $q>1$, we find that the solution
exists provided that $1/3<\gamma<1$, independently of $a$.

The condition of validity of our study (\ref{a28}) takes the form
\begin{eqnarray}
\label{b1w}
\tau^{-2q}\ll \tau^{-(3\gamma-2)q}\qquad {\rm
for}\qquad \tau\rightarrow 0.
\end{eqnarray}
This requires $\gamma>4/3$.  This condition is never
fulfilled when the solution exists. When $a=0$ the solution exists provided that
$1/3<\gamma<1$ but it is not valid. There is no solution of the form
(\ref{b11b}) when $\gamma=1$.

\subsubsection{Conclusions}

Regrouping the foregoing results, and considering only solutions that satisfy
the condition of validity of our study (\ref{a28}), we come to the following
conclusions: (i) When $-3<a<-2$ the solution  of equation (\ref{b2})  behaves
asymptotically as $R\sim vt$ for $1/3<\gamma<(a+4)/3$ (see Appendix
\ref{sec_cst}); (ii) When $a>-2$ the
solution of equation (\ref{b2}) behaves asymptotically as $R\sim A t^q$ with
$q>0$ for $1/3<\gamma<(a+3)/3$ (see Appendix \ref{sec_ut}) and
as  $R\sim vt$  for  $(a+3)/3<\gamma<(a+4)/3$ (see Appendix
\ref{sec_cst}).

When $a=0$, the
solution of equation (\ref{b2}) behaves asymptotically as $R\sim A
t^{2/(3\gamma-1)}$ for
$1/3<\gamma<1$ (see Appendices \ref{sec_kb} and \ref{sec_ut}), as $R\sim
t\sqrt{2\ln t}$ for $\gamma=1$ (see Appendix \ref{sec_ka}), and as  $R\sim vt$ 
for $1<\gamma<4/3$ (see Appendices \ref{sec_kb} and
\ref{sec_cst}).

When $\gamma=1$, the
solution of equation (\ref{b2}) behaves asymptotically as  $R\sim vt$ for
$-1<a<0$ (see Appendix \ref{sec_cst}), as $R\sim
t\sqrt{2\ln t}$ for $a=0$ (see Appendix \ref{sec_ka}), and as  $R\sim A
t^{(a+2)/2}$ for $a>0$ (see Appendix \ref{sec_ut}). 

When $a=3/8$ and $\gamma=1$, which is the situation corresponding to the
post-collapse regime considered in Section
\ref{sec:post-coll}, the solution of equation (\ref{b2}) behaves
asymptotically as  $R\sim A t^{19/16}$ (see Appendix \ref{sec_ut}). 

\subsection{The case $K(t) = (t_{\rm coll}-t)^a$ (pre-collapse)}
\label{sec_c}

In this subsection, we assume that the temperature behaves
as
\begin{eqnarray}
\label{c1}
K(t)=(t_{\rm coll}-t)^a.
\end{eqnarray}
In that case,  the
differential equation (\ref{a21})  becomes
\begin{eqnarray}
\label{c2}
\ddot R R^{3\gamma-2}  = (t_{\rm coll}-t)^a.
\end{eqnarray}
Defining $\tau=t_{\rm coll}-t$, it reduces to 
\begin{eqnarray}
\label{c3}
\ddot R R^{3\gamma-2}  = \tau^a.
\end{eqnarray}
For the sake of generality, we let the value of $a$ arbitrary
(negative or positive).  When $a<0$, the temperature diverges in a finite
time $t_{\rm coll}$. This is the situation corresponding to the pre-collapse
regime
considered in Section \ref{sec:numerics} where $a=-1/24$ and $\gamma=1$
(isothermal gas). When
$a>0$, the temperature tends to zero in a finite
time $t_{\rm coll}$.

\subsubsection{Solution $R(\tau)= A (t_{\rm coll}-t)^q$ with $q<0$}
\label{sec_c1}

We consider a solution of equation (\ref{c2}) of the form
\begin{eqnarray}
\label{c4b}
R(\tau)= A \tau^q
\end{eqnarray}
with
$q<0$ (and, of course, $A>0$). In that case, the radius $R(t)$ increases and
becomes infinite at $t=t_{\rm coll}$.
Substituting
this ansatz
into
equation (\ref{c3}) we get
\begin{eqnarray}
\label{c4}
Aq(q-1)\tau^{q-2}A^{3\gamma-2}\tau^{(3\gamma-2)q}= \tau^a,
\end{eqnarray}
implying
\begin{eqnarray}
\label{c5}
q=\frac{a+2}{3\gamma-1} 
\end{eqnarray}
and
\begin{eqnarray}
\label{c5b}
 A^{3\gamma-1}=\frac{1}{q(q-1)}.
\end{eqnarray}
Considering  equation (\ref{c5}) with $q<0$, and recalling that $\gamma>1/3$, we
find that the solution exists provided that $a<-2$.

The condition of validity of our study (\ref{a28}) takes the form
\begin{eqnarray}
\label{c7}
\frac{1}{\tau^{2q}}\ll \frac{\tau^a}{\tau^{(3\gamma-2)q}}\qquad {\rm for}\qquad
\tau\rightarrow 0.
\end{eqnarray}
When $\gamma>1/3$
this requires $\gamma>(3a+8)/6$. This condition is always
fulfilled when the solution exists. When $a=0$
the solution does not exist. When $\gamma=1$ the
solution exists and is valid provided that
$a<-2$. For the case considered in Section \ref{sec:precoll-halo}, corresponding
to
$\gamma=1$ and $a=-1/24$, the condition $a<-2$ is not satisfied
so there is no solution of the form of equation (\ref{c4b}) with
$q<0$. This suggests that the radius of the halo
does not diverge at $t_{\rm coll}$ in agreement with the
numerical solution of the MEP model.

\subsubsection{Solution $R(\tau)= A (t_{\rm coll}-t)^q$ with $q>0$}
\label{sec_c1b}

We consider a solution of equation (\ref{c2}) of the form of equation
(\ref{c4b}) with
$q>0$ (and, of course, $A>0$). In that case, the radius $R(t)$ decreases and
tends to zero at  $t=t_{\rm coll}$.
Substituting
this ansatz
into
equation (\ref{c3}) we get equation (\ref{c4})  implying equations (\ref{c5})
and (\ref{c5b}).
Considering equation (\ref{c5b}), and recalling that $q>0$ and $A>0$, we see
that a
necessary condition for the existence of a solution is that $q>1$. Considering
equation (\ref{c5}) with $q>1$, and recalling that $\gamma>1/3$, we find that
the solution exists provided that  $a>-2$ and $1/3<\gamma<(a+3)/3$.

The condition of validity of our study (\ref{a28}) takes the form of
equation
(\ref{c7}). When $\gamma>1/3$
this requires $\gamma>(3a+8)/6$.  This condition is never
fulfilled when the solution exists. For the
case
considered in Section \ref{sec:precoll-halo}, corresponding to
$\gamma=1$ and $a=-1/24>-2$, the condition
$1/3<\gamma<(a+3)/3=71/72$ is not satisfied,
so there is no solution of the form of equation (\ref{c4b}) with
$q>1$  (we get $q=47/48<1$). This suggests that the radius of
the halo
does not tend to zero at $t_{\rm coll}$ in agreement with the
numerical solution of the MEP model.

\subsubsection{Asymptotic solution $R(t)\rightarrow B$}
\label{sec_c2}

We consider an asymptotic solution of  equation (\ref{c2}) of the form
\begin{eqnarray}
\label{nc8}
R(\tau)= B+ \epsilon(\tau)
\end{eqnarray}
with $|\epsilon(\tau)|\rightarrow
0$ when
$\tau\rightarrow 0$. In that case, the radius $R(t)$ reaches a
finite value $B$ at $t=t_{\rm coll}$. Substituting this ansatz into
equation (\ref{c3}) we get for $\tau\ll 1$:
\begin{eqnarray}
\label{c8}
\ddot \epsilon \sim \frac{1}{B^{3\gamma-2}} \tau^{a}.
\end{eqnarray}
After two integrations, we obtain 
\begin{eqnarray}
\label{c9}
\epsilon(t) \sim \frac{1}{B^{3\gamma-2}}
\frac{\tau^{a+2}}{(a+2)(a+1)}-v\tau,
\end{eqnarray}
where $v$ is a constant of integration (the other constant of integration can be
taken
equal to zero without restriction of generality). This solution exists provided
that $a>-2$. We note that
$B$ and $v$ cannot be
determined by this asymptotic approach since they depend on the
initial condition.  The velocity of expansion is
\begin{eqnarray}
\label{c10}
\dot R(t)=-\dot R(\tau)=-\dot\epsilon(\tau) \sim -\frac{1}{B^{3\gamma-2}}
\frac{\tau^{a+1}}{a+1}+v.
\end{eqnarray}
When $a>-1$, the velocity  $\dot R(t)$ tends to a finite value
$v$ at $t_{\rm coll}$. When
$a<-1$, the velocity $\dot R(t)$ tends to $+\infty$ as
$t\rightarrow t_{\rm coll}$.

The condition of validity of our study (\ref{a28}) takes the form
\begin{eqnarray}
\label{c11}
1\ll \tau^a \qquad {\rm for}\qquad
\tau\rightarrow 0.
\end{eqnarray}
This condition is fulfilled provided that $a<0$. In conclusion, the solution
exists and is valid provided that $-2<a<0$. When $a=0$ the solution exists but
is not valid. When $\gamma=1$ the solution exists and is valid provided that
$-2<a<0$.
For the case considered in Section \ref{sec:precoll-halo}, corresponding to
$\gamma=1$  and $a=-1/24$, we find that the solution
(\ref{nc8}) exists
and is valid. Together with the result of Appendices \ref{sec_c1} and
\ref{sec_c1b}, this strongly suggests
that the radius of the halo tends to a constant at $t=t_{\rm
coll}$ in agreement  with the
numerical solution of the MEP model. Furthermore, since 
$a=-1/24>-1$, the velocity  $\dot
R(t)$ tends to a finite value
$v$ at $t=t_{\rm coll}$.

\subsubsection{Conclusions}

Regrouping the foregoing results, and considering only solutions that satisfy
the condition of validity of our study (\ref{a28}), we come to the following
conclusions: (i) When $a<-2$ the solution  of equation (\ref{c2})  behaves
asymptotically as $R\sim A (t_{\rm coll}-t)^q$ with $q<0$ for $\gamma>1/3$ (see
Appendix
\ref{sec_c1}); (ii) When $-2<a<0$ the
solution of equation (\ref{c2}) tends to a constant $R\rightarrow B$ for
$\gamma>1/3$ (see Appendix \ref{sec_c2}).

When $\gamma=1$, the solution of equation (\ref{c2}) behaves asymptotically as 
$R\sim A (t_{\rm coll}-t)^{(a+2)/2}$ for
$a<-2$ (see Appendix \ref{sec_c1}) and tends to a constant $R\rightarrow B$
for $-2<a<0$ (see Appendix \ref{sec_c2}). 

When $a=-1/24$ and $\gamma=1$, which is the situation corresponding to the
pre-collapse regime considered in Section \ref{sec:numerics}, the
solution of equation (\ref{c2}) tends to a constant $R\rightarrow B$ (see
Appendix \ref{sec_c2}).

\end{document}